\documentclass[twocolumn,tighten]{aastex6}
\usepackage{natbib}

\newcommand{\msun}{M_\odot}
\newcommand{\lsun}{{L_\odot}}
\newcommand{\gyr}{{\rm{Gyr}}}
\newcommand{\myr}{{\rm{Myr}}}
\newcommand{\henon}{{H\'enon}}
\newcommand{\kms}{{\rm{km\ s^{-1}}}}
\newcommand{\pc}{{\rm{pc}}}

\newcommand{\fbhigh}{{f_{b, \rm{high}}}}
\newcommand{\sigmans}{{\sigma_{\rm{NS}}}}
\newcommand{\sigmabh}{{\sigma_{\rm{BH}}}}
\newcommand{\xfb}{{x_{\rm{fb}}}}

\newcommand{\rcobs}{{r_{c, \rm{obs}}}}
\newcommand{\rhlobs}{{r_{hl, \rm{obs}}}}
\newcommand{\sigmacobs}{{\Sigma_{c, \rm{obs}}}}

\newcommand{\surfdens}{{L_\odot\rm{pc}^{-2}}}

\newcommand{\nbhbh}{{N_{\rm{{BH-BH}}}}}

\newcommand{\nbhnbh}{{N_{\rm{BH-nBH}}}}

\newcommand{\nbh}{{N_{\rm{BH}}}}
\newcommand{\rgc}{{r_{\rm{G}}}}
\newcommand{\kpc}{{{\rm{kpc}}}}
\newcommand{\mtot}{{M_{\rm{tot}}}}
\newcommand{\mchirp}{{M_{\rm{chirp}}}}

\newcommand{\modelS}{{\tt S}}
\newcommand{\modelSifb}{{\tt SIwfb0.1}}
\newcommand{\modelSifbrvone}{{\tt SIwfb0.1rv1}}
\newcommand{\modelSifbrvoneKone}{{\tt SIwfb0.1rv1K1}}
\newcommand{\modelSifbrvoneKthree}{{\tt SIwfb0.1rv1K3}}

\newcommand{\modelSRone}{{\tt SR1}}
\newcommand{\modelSRtwo}{{\tt SR2}}
\newcommand{\modelSRthree}{{\tt SR4}}
\newcommand{\modelSRfour}{{\tt SR20}}
\newcommand{\modelSRoneZ}{{\tt SR1Z}}
\newcommand{\modelFzero}{{{\tt F0}}}
\newcommand{\modelFone}{{{\tt F1}}}
\newcommand{\modelFDMs}{{{\tt F0.7Ms0.1}}}
\newcommand{\modelFDq}{{{\tt F0.7q0.6}}}
\newcommand{\modelFDqRone}{{\tt F0.7q0.6R1}}
\newcommand{\modelFDqRtwo}{{\tt F0.7q0.6R3}}
\newcommand{\modelKone}{{{\tt K1}}}
\newcommand{\modelKtwo}{{{\tt K2}}}
\newcommand{\modelKthree}{{{\tt K3}}}
\newcommand{\modelKoneRone}{{\tt K1R1}}
\newcommand{\modelKoneRtwo}{{\tt K1R2}}
\newcommand{\modelKoneRthree}{{\tt K1R4}}
\newcommand{\modelKoneRfour}{{\tt K1R20}}
\newcommand{\modelKtwoRone}{{\tt K2R1}}
\newcommand{\modelKtwoRtwo}{{\tt K2R2}}
\newcommand{\modelKtwoRthree}{{\tt K2R4}}

\newcommand{\modelKthreeRone}{{\tt K3R1}}
\newcommand{\modelKthreeRtwo}{{\tt K3R2}}
\newcommand{\modelKthreeRthree}{{\tt K3R4}}
\newcommand{\modelKthreeRfour}{{\tt K3R20}}
\newcommand{\modelKoneRoneZ}{{\tt K1R1Z}}
\newcommand{\modelKtwoRoneZ}{{\tt K2R1Z}}
\newcommand{\modelKthreeRoneZ}{{\tt K3R1Z}}
\newcommand{\modelIsteep}{{\tt Is}}
\newcommand{\modelIflat}{{\tt If}}
\newcommand{\modelIsteepKone}{{\tt IsK1}}
\newcommand{\modelIsteepKtwo}{{\tt IsK2}}
\newcommand{\modelIsteepKthree}{{\tt IsK3}}
\newcommand{\modelIflatKone}{{\tt IfK1}}
\newcommand{\modelIflatKtwo}{{\tt IfK2}}
\newcommand{\modelIflatKthree}{{\tt IfK3}}
\newcommand{\modelFDMsKone}{{\tt F0.7Ms0.1K1}}
\newcommand{\modelFDMsKtwo}{{\tt F0.7Ms0.1K2}}
\newcommand{\modelFDMsKthree}{{\tt F0.7Ms0.1K3}}
\newcommand{\modelFDqKone}{{\tt F0.7q0.6K1}}
\newcommand{\modelFDqKtwo}{{\tt F0.7q0.6K2}}

\newcommand{\modelFDqKoneRone}{{\tt F0.7q0.6K1R1}}
\newcommand{\modelFDqKoneRtwo}{{\tt F0.7q0.6K1R3}}
\newcommand{\modelFDqKtwoRone}{{\tt F0.7q0.6K2R1}}
\newcommand{\modelFDqKtwoRtwo}{{\tt F0.7q0.6K2R3}}

\newcommand{\modelW}{{\tt W}}
\newcommand{\modelWKone}{{\tt WK1}}
\newcommand{\modelWKtwo}{{\tt WK2}}
\newcommand{\modelWKthree}{{\tt WK3}}
\newcommand{\modelWFDq}{{\tt WF0.7q0.6}}
\newcommand{\modelWFDqKone}{{\tt WF0.7q0.6K1}}
\newcommand{\modelWIflat}{{\tt WIf}}
\newcommand{\modelWIflatKone}{{\tt WIfK1}}
\newcommand{\modelWIflatKthree}{{\tt WIfK3}}
\newcommand{\modelWifb}{{\tt Wrv1fb0.1}}
\newcommand{\modelWifbKone}{{\tt Wrv1K1fb0.1}}
\newcommand{\modelWifbKthree}{{\tt Wrv1K3fb0.1}}

\shorttitle{Binary Black Holes in Dense Star Clusters}
\shortauthors{Chatterjee\ et al.}

\begin{document}

\title{Binary Black Holes in Dense Star Clusters:\\Exploring the Theoretical Uncertainties}

\author{Sourav Chatterjee\altaffilmark{1}, Carl L.~Rodriguez\altaffilmark{1,2}, \& Frederic A.~Rasio\altaffilmark{1}}
\affil{$^1$Center for Interdisciplinary Exploration \& Research in Astrophysics (CIERA)\\Physics \& Astronomy, Northwestern University, 
Evanston, IL 60202, USA\\sourav.chatterjee@northwestern.edu}
\affil{$^2$MIT-Kavli Institute for Astrophysics and Space Research\\77 Massachusetts Avenue, 37-664H, Cambridge, MA 02139, USA}

\begin{abstract}
Recent $N$-body simulations predict that large numbers of stellar black holes (BHs) 
could remain bound to globular clusters (GCs) at present, and merging BH--BH binaries 
are produced dynamically in significant numbers. We systematically vary ``standard" 
assumptions made by numerical simulations related to, e.g., BH formation, stellar winds, 
binary properties of high-mass stars, and IMF within existing uncertainties, and study 
the effects on the evolution of the structural properties of GCs, and the BHs in GCs. 
We find that variations in initial assumptions can set otherwise identical initial clusters 
on completely different evolutionary paths, significantly affecting their present observable 
properties, or even affecting the cluster's very survival to the present. However, these 
changes usually do not affect the numbers or properties of local BH--BH mergers. The 
only exception is that variations in the assumed winds and IMF can change the masses 
and numbers of local BH--BH mergers, respectively. All other variations (e.g., in initial 
binary properties and binary fraction) leave the masses and numbers of locally merging 
BH--BH binaries largely unchanged. This is in contrast to binary population synthesis 
models for the field, where results are very sensitive to many uncertain parameters 
in the initial binary properties and binary stellar-evolution physics. Weak winds are 
required for producing GW150914-like mergers from GCs at low redshifts. LVT151012 
can be produced in GCs modeled both with strong and weak winds. GW151226 is 
lower-mass than typical mergers from GCs modeled with weak winds, but is similar to 
mergers from GCs modeled with strong winds.
\end{abstract}


\section{Introduction}
\label{S:intro}
Our understanding of how black holes (BHs) evolve inside star clusters has a long and varied history. 
Following the classic work by \citet{1969ApJ...158L.139S}, it was suggested that old 
($\sim 12\,\gyr$) globular clusters (GCs) cannot retain a significant BH
population up to the present day. It was argued that, due to the much higher mass of the BHs compared to typical stars,
the BHs will quickly ($\lesssim 10^2\,\myr$) mass segregate to form an isolated 
subcluster that is dynamically decoupled from the GC. Due to the small size,
high density, and small number of objects in the subclusters, relaxation and
strong encounters were expected to eject the majority of BHs on a timescale of 
$\sim\,1\gyr$. Thus, at most a few BHs would remain in the old GCs observed in
the Milky Way \citep[MW, e.g.,][]{1993Natur.364..421K,1993Natur.364..423S,2000ApJ...528L..17P,2004ApJ...601L.171K}. 
Furthermore, it was argued that if significant numbers of BHs are present in today's GCs, a subset of 
them might be in accreting binary systems, and detectable as X-ray sources. 
However, observations of luminous X-ray sources in the MW GCs prior to 2012 suggested that {\em all} of these sources are 
accreting neutron stars and not BHs, consistent with the theoretical expectation at the time\citep[e.g.,][]{2004MNRAS.350..649V,2006csxs.book.....L,2010ApJ...712L..58A,2012ATel.4264....1A,2011A&A...535L...1B}.    

This classical picture started to change with recent discoveries of BH
candidates in extragalactic GCs, characterized 
by their super-Eddington luminosities and high variability on short timescales 
\citep{2007Natur.445..183M,2010ApJ...712L...1I}. More recently, with the
completion of the 
upgraded VLA, surveys combining radio and X-ray data for the MW GCs detected quiescent BHs by comparing their radio and X-ray luminosities 
\citep[e.g.,][]{2012Natur.490...71S,2013ApJ...777...69C}. Interestingly, the
MW GCs containing the 
detected BH candidates show large ranges in structural properties, indicating
that the retention of 
BHs may be quite common. Several recent or ongoing surveys promise much richer 
observational constraints on this question 
\citep[e.g.,][]{2013atnf.prop.5724S,2014cxo..prop.4353S,2014atnf.prop.6417M,2014atnf.prop.6457M}.    

To be detectable either via electromagnetic signatures or via gravitational waves (GW) from 
BH--BH mergers, the BHs must be in binary systems with very specific ranges of properties. 
Although dynamical interaction can enhance 
the production of binaries containing a BH (BBHs)
\citep[e.g.,][]{2002ApJ...574L...5K,2003ApJ...591L.131P}, 
modern simulations find that the fraction of BHs in binaries is typically low \citep[e.g.,][]{2015ApJ...800....9M,2014MNRAS.444...29L}. 
Hence, it has been argued that detection of just a few BH candidates 
in GCs indicates the existence of a much larger population of BHs that are not 
detectable 
\citep[e.g.,][]{2012Natur.490...71S,2012Natur.490...46U,2013ApJ...763L..15M,2015ApJ...800....9M}. 
Indeed, recent numerical studies have found that BH ejection is not nearly as efficient as 
was previously thought. In these studies, it has been shown that the BH subcluster does not 
stay decoupled from the rest of the cluster for prolonged periods. The same interactions that 
eject BHs from the subcluster also cause it to expand and re-couple with the rest
of the cluster, dramatically increasing the timescale for BH evaporation \citep[e.g.,][]{2013MNRAS.432.2779B,2015ApJ...800....9M,2016MNRAS.458.1450W}. 
These simulations suggest that tens to thousands of BHs may remain in today's 
GCs \citep[e.g.,][]{2008MNRAS.386...65M,2009ApJ...690.1370M,2012MNRAS.422..841A,2013ApJ...763L..15M,2015ApJ...800....9M}.    
 
Most recently, \citet{2015ApJ...800....9M} showed that the presence of a significant number of BHs 
can dramatically alter the overall dynamical evolution of GCs. 
Through repeated cycles of core-collapse and core re-expansion, 
BH-dynamics acts as a significant and persistent source of energy. On the other hand, the clustered 
environment and high frequency of strong scattering interactions, especially involving binaries, 
can change the numbers and properties of BBHs that form inside clusters \citep[e.g.,][]{2015PhRvL.115e1101R,2016PhRvD..93h4029R,2016ApJ...816...65A}.  
Both of these aspects intricately depend on several physical processes, many of which lack 
strong observational constraints. For example, the distribution of the natal kicks the BHs 
receive can control how many of them will be directly ejected from the GCs
immediately upon formation. Natal kicks also control the fraction of BHs that may 
retain their binary companions after SN. 
The high end of the stellar initial-mass function (IMF) determines how many BHs a cluster can form. 
The binary fraction and binary orbital properties can alter the binary stellar evolution of
a BBH (or their progenitors) as well as the rate at which they take part in dynamical encounters. 
In addition, the ratio between the BH mass and the average stellar mass, also
directly set by the IMF, determines the timescale for BHs to sink to the center.
Answers to a wide variety of questions, such as
how many BHs and BBHs can a GC retain at present, how many BBHs it can form 
over its whole lifetime, at what rate do BHs and BBHs get ejected from the
clusters, and how do the 
BHs affect the overall evolution of the clusters, can potentially depend 
on these initial assumptions. 
Hence, we must understand the sensitivity of our models to initial conditions and assumptions
that are poorly constrained by existing observations.

In this study, we vary the initial assumptions affecting the high-mass 
stars and BHs, that are usually considered ``standard" in theoretical studies of
clusters, within their observational uncertainties. 
In particular, we 
vary the stellar IMF, 
the birth-kick distribution for the BHs, and the  
primordial binary fraction and the binary properties 
for massive stars. We also vary the assumed prescription for mass loss via stellar winds. 
Furthermore, we vary the galactocentric distance ($\rgc$) and metallicities. 
Starting from otherwise identical initial star clusters, we 
study how varying these assumptions affect BH populations, 
and the overall evolution and final observable properties of the host clusters. 
We also study how these initial assumptions affect the number and properties 
of merging BH--BH binaries. We put these 
findings in the context of the recent landmark detections of GWs
from BH--BH mergers by Advanced LIGO \citep{PhysRevLett.116.061102,2041-8205-818-2-L22,PhysRevLett.116.241103,2016arXiv160203842T,2016arXiv160604856T}. 
 
In \S\ref{S:numerical} we describe our numerical models and we list the 
initial assumptions that we vary. In \S\ref{S:obs_derivation} we define how we evaluate 
observable cluster properties from our models. In 
\S\ref{S:results_clusprop} we show how the various assumptions affect the overall evolution and final 
observable properties of star clusters. 
\S\ref{S:results_bhbinaries} focuses on how these assumptions and resulting differences in the cluster 
evolution as a whole alter the binary properties of BHs. In \S\ref{S:results_merger} 
we focus on the numbers and properties of BH--BH mergers and put those results in the 
context of the recent discoveries of GWs \citep{PhysRevLett.116.061102,PhysRevLett.116.241103,2016arXiv160604856T}.  
We summarize our results and conclude in \S\ref{S:discussion}. 

%
%
%
%
\section{Numerical Models}
\label{S:numerical}
We use our \henon-type \citep{1971Ap&SS..14..151H} Cluster Monte Carlo (CMC) code, 
developed and rigorously tested in our group over 
the past 15 years \citep{2000ApJ...540..969J,2001ApJ...550..691J,2003ApJ...593..772F,2012ApJ...750...31U,2013ApJS..204...15P,2010ApJ...719..915C,2013ApJ...777..106C,2013MNRAS.429.2881C}. 
For a detailed description of the most recent updates and parallelization 
see \citet{2013ApJS..204...15P} and \citet{2015ApJ...800....9M}. The Monte Carlo approach is more approximate 
 than a direct $N$-body integration \citep[e.g.,][]{2010gnbs.book.....A}, but
 requires only a fraction of the computational time.  This rapidity allows us to
 fully explore the parameter space of dense star clusters.   Results from CMC
 have been extensively compared to recent state-of-the-art $N$-body simulations,
 and were
found to produce excellent agreement in all quantities of interest in this study \citep{2016MNRAS.463.2109R}. 

\subsection{Standard Assumptions}
\label{S:standard}
In order to understand the influence of each initial assumption on 
the production and subsequent 
evolution of BHs in a cluster, we anchor our numerical models using the same assumptions 
as used in \citet{2015ApJ...800....9M}. We call this our ``standard'' model and
denote it as  
{\tt S} (Table\ 1, 2). Our standard model initially has $N=8\times10^5$ stars. 
The position and velocities of the stars are set according to a 
King profile with $W_0=5$ \citep{1962AJ.....67..471K,1965AJ.....70..376K,1966AJ.....71...64K}. 
We adopt the commonly-used IMF presented in 
\citet{2001MNRAS.322..231K}, and use the central value for each slope in the different 
mass ranges between $0.1$--$100\,\msun$ to assign masses to these stars. 
In model {\tt S} we assume an initial 
binary fraction $f_b = 0.05$. This is realized by randomly selecting the appropriate number ($N_b = N\times f_b$) 
of stars, independent of their masses or positions in the cluster, and assigning
binary companions to them. The mass of the secondary ($m_s$) is drawn from
a uniform distribution with a lower limit taken from the assumed IMF,
$m_{\rm{min}} = 0.1\,\msun$, and an upper limit equal to the mass of the primary ($m_p$). 
The orbital period ($P$) is drawn from a distribution 
flat in $\log P$ between $5$ times the sum of the radii for the binary
companions, and the local hard--soft boundary given by $v_{\rm{orb}}=v_\sigma$, where, 
$v_{\rm{orb}}$ is the orbital velocity and $v_\sigma$ is the local velocity dispersion. 
Note that, although all initial binaries are locally hard, dynamical evolution can make them 
soft at a later time, either by increasing the local velocity dispersion of other
stars (typically in the core) or by moving the 
binary from its initial location to where the velocity dispersion is higher (due to mass segregation). Such soft 
binaries are maintained in all our simulations until strong scattering encounters disrupt them. 
The initial orbital eccentricities for the binaries are drawn from a thermal distribution.   

Single and binary stellar evolution is performed 
with SSE and BSE \citep{2000MNRAS.315..543H,2002MNRAS.329..897H}. We have modified 
the prescription for stellar remnant formation in SSE and BSE by using the results of 
\citet{2001ApJ...554..548F} and \citet{2002ApJ...572..407B}. 
All core-collapsed neutron stars get birth kicks 
drawn from a Maxwellian distribution with $\sigma=\sigmans=265\,\kms$. 
We assume momentum ($|\vec{p}|$) conserving 
kicks for BHs following the prescription of \citet{2002ApJ...572..407B}.
BHs formed via the direct collapse scenario do not get any natal kicks, since there is no associated
explosion or mass loss. 
BHs formed with significant fallback get natal kicks calculated by initially sampling from 
the same kick distribution as the neutron stars, but reduced in magnitude according 
to the fractional mass of the fallback ($\xfb$) material \citep[see ][for a more detailed description of our 
standard model {\tt S}]{2015ApJ...800....9M}. 

Below we describe how we vary the above-mentioned initial assumptions. In each case 
we describe only the assumptions we change relative to the baseline model {\tt S}, 
with all other initial conditions held constant. 

%
%
%
\subsection{Initial Binary Fraction and Binary Properties of Massive Stars}  
\label{S:fbhigh}
One source of uncertainty in setting up the initial conditions is whether the binary fraction 
depends on the stellar mass. 
We create two models, \modelFzero\ and \modelFone, varying  
the fraction of high-mass stars ($>15\,\msun$) that are initially in binaries ($\fbhigh$). In models 
\modelFzero\ and \modelFone, we adopt limiting initial values of $\fbhigh=0$ and $1$, respectively (Table\ 1, 2). 
We assign the binaries in these 
two models such that the overall binary fraction $f_b$ is kept fixed at $0.05$.
In model {\tt F0}, $N_b=0.05\times N$ stars are randomly chosen from all stars 
with mass $\leq 15\,\msun$ and are assigned 
as binaries. In model {\tt F1}, we first assign all stars with mass $>15\,\msun$ as binaries. 
We randomly choose 
the appropriate residual number of low-mass stars ($\leq 15\,\msun$) and assign them as 
binaries to make $f_b=0.05$. In \modelFzero\ and \modelFone, the distributions for the 
mass ratios ($q$) and the orbital properties are identical to model {\tt S}.

In addition to the binary fraction in high-mass stars, the binary orbital
properties can potentially affect both the formation and interaction rate of
BHs in a cluster. The observed distributions of initial separations and $q$ for binaries may 
have significant uncertainties and selection biases. To understand the 
effects of these initial assumptions, 
we create a set of models where we vary the initial binary properties of the high-mass ($>15\,\msun$) 
stars. In these models, we assume that the initial $\fbhigh = 0.7$ following the observational constraint 
provided in \citet{2012Sci...337..444S}. 
In addition, we choose an initial 
period distribution described by $dn/d\log P \propto P^{-0.55}$ for the high-mass stars 
\citep{2012Sci...337..444S}. The number of binaries for low-mass stars 
are again adjusted so that the overall $f_b$ is $\approx 0.05$, similar to model \modelS. 
We denote these models with ``{\tt F0.7}." Within this variant, we also consider
two different $q$ distributions for the high-mass binaries. In one, we use a uniform distribution of 
$q$ between $q = m_{\rm{min}}/m_p$ and $1$. 
These models are denoted with the string ``{\tt Ms0.1}". 
Another set of models uses the same uniform distribution in $q$, but within a much smaller range, 
$0.6$ and $1$. 
We denote this set of models with the string ``{\tt q0.6}" (Table\ 1, 2). 
%

%
%
\subsection{Natal Kick Distribution for Black Holes}  
\label{S:kick}
It has been widely accepted that the neutron stars get large natal kicks when they form via 
core-collapse SN
\citep[e.g.,][]{1993Natur.362..133C,1994Natur.369..127L,2006ApJ...639.1007W,2015A&A...573A..58Z}.
However, the magnitudes of natal kicks imparted to BHs, is still a matter of debate. 
Observational constraints come 
from modeling the kicks required to explain the positions and velocities of
known BH X-ray binaries (XRB) in the MW galactic potential. Detailed analysis of
individual BH XRBs results in widely varying constraints on their natal kick magnitudes 
\citep[e.g.,][]{1995MNRAS.277L..35B,1999A&A...352L..87N,2005ApJ...625..324W,2005ApJ...618..845G,2007ApJ...668..430D,2009ApJ...697.1057F,2012ApJ...747..111W,2014ApJ...790..119W}.  
Instead of modeling individual systems, 
\citet{2012MNRAS.425.2799R} and \citet{2015MNRAS.453.3341R}
have performed population synthesis using different assumptions 
of natal kick distributions and compared their results with the positions of the observed BHs 
in the MW. They have found that BHs may get large kicks, perhaps even as large as 
the neutron stars formed via core-collapse SN. In addition, they have not found evidence of a mass-dependent kick distribution, 
which would be expected for the widely used $|\vec{p}|$-conserving kick prescription. 
Recent theoretical study of core-collapse SN by \citet{2015ApJ...801...90P} also suggests that 
the birth kicks may not be directly correlated with the BH mass. 
In short, the distribution of formation kicks for BHs is largely 
uncertain, even at the qualitative level.  

In addition to the one used in \modelS, we create models with three 
different natal kick distributions for the BHs. These models assume that the 
kick magnitudes are independent of the BH masses and $\xfb$, and are 
drawn from a Maxellian given by $\sigma=\sigmabh$. We adopt three variations that 
are obtained by using $\sigmabh=\sigmans=265\,\kms$ (denoted by the string ``{\tt K1}"), 
$\sigmabh=0.1\sigmans$ (denoted by the string ``{\tt K2}"), and $\sigmabh=0.01\sigmans$ 
(denoted by the string ``{\tt K3}") in Table\ 1, 2.

%
\subsection{Initial Stellar Mass Function}  
\label{S:imf}
A lot of observational effort is aimed towards finding the expected IMF for stars born in 
clusters \citep[e.g.,][and the references therein]{2002MNRAS.336.1188K}. Of course, 
the number of BHs a cluster can form is directly dependent on the 
number of massive BH-progenitor stars it initially contains. This, in turn, is directly 
dependent on the stellar IMF, especially the power-law slope of the IMF near the high-end 
of stellar masses. As described earlier, our standard models use the central 
values for the IMF slopes from \citet[][]{2001MNRAS.322..231K}. However, we note that 
the best-fit power-law exponents $\alpha$ in 
$dn/dm \propto m^{-\alpha}$ throughout 
all mass ranges have large uncertainties. For example, the power-law exponent 
$\alpha_1$, for stars more massive than $1\,\msun$ is $2.3$ with $1\sigma$
error of $0.7$ \citep{2001MNRAS.322..231K}. 
We vary $\alpha_1$ within the quoted $1\sigma$ uncertainties, 
and create models with $\alpha_1=1.6$ (denoted using the string ``{\tt If}") and 
$\alpha_1=3$ (denoted using the string ``{\tt Is}") in Table\ 1, 2. 
%

%
%
\subsection{Stellar Wind Prescription}
\label{S:wind}
The details of how stars lose mass to stellar winds is complicated and hard to
model theoretically. 
Most stellar evolution prescriptions instead model the mass and
metallicity-dependent stellar winds by calibrating the wind-driven mass loss to 
catalogues of observed stars
\citep[e.g.,][]{1997A&A...317L..23D,1998A&A...329..971V,2000MNRAS.315..543H,2001A&A...369..574V}.
The assumed wind prescription dramatically
affects the mass of the BH progenitor at the time of SN, which in turn
determines the mass of the resultant BH. 
The wind prescription described in \citet{2000MNRAS.315..543H} is widely used as
part of the SSE and BSE software packages in many widely-used 
cluster dynamics codes. The majority of our models use this stellar wind prescription. For simplicity, we 
call BSE's implementation for winds as the ``strong wind'' prescription.  

Recent observations of high-mass stars suggest that the stellar winds
may not be as strong as suggested by earlier studies
\citep[e.g.,][]{2001A&A...369..574V,2008NewAR..52..419V,2010ApJ...714.1217B,2010ApJ...715L.138B,2012ApJ...759...52D,2015MNRAS.451.4086S}.
The details of this wind prescription, based on, e.g., \citet[][]{2001A&A...369..574V} is documented
in detail in \citet{2010ApJ...715L.138B}, and implemented in our code in
\citep[][]{2016PhRvD..93h4029R}. For simplicity, we call this implementation the
``weak wind'' prescription. All models adopting weak winds are denoted with the string ``{\tt W}" in Table\ 1, 2. 

%
%
\subsection{Other Assumptions}
\label{S:other_variation}
In addition to the above variations to our standard model \modelS, we also vary
the galactocentric distance 
($\rgc$), metallicity ($Z$), the initial virial radius ($r_v$), and the overall spread in the stellar IMF for a 
handful of models. 

For easier understanding of the variations of initial assumptions in specific models we name all 
models in a way that specific strings in the name would indicate specific variations. In Table\ 1 we 
summarize the specific strings in model names and what they mean. All model names are created using some 
combination of these strings indicating combinations of specific initial assumptions. 
The details of all models and initial assumptions are listed in Table\ 1.  

\begin{deluxetable}{p{0.1cm}|p{6.5cm}}
\tabletypesize{\footnotesize}
\tablecolumns{4}
\tablewidth{0pt}
\tablecaption{Naming convention for models}
\tablehead{
	  \colhead{String} &
	  \colhead{Meaning for initial property variations} \\ 
}
\startdata
\modelS & Our baseline model; we use ``standard" assumptions for BH kicks, galacto-centric distance, IMF, $f_b$, and $Z$ (\S\ref{S:standard}). \\
\hline
{\tt Rx} & Galacto-centric distance is varied; $r_{G}=x\,\kpc$ (\S\ref{S:other_variation}). \\
\hline
{\tt Z} & Metallicity is the same as in the baseline model, $Z=0.001$, independent of the galacto-centric distance of the cluster (\S\ref{S:other_variation}). In contrast, in other cases, we assume $Z$ is anti-correlated with $r_G$ and assign metallicities consistent with the observed MW GCs at a given galacto-scentric distance \citep{1994AJ....108.1292D}. \\
\hline
{\tt fb0.1} & Overall initial binary fraction is changed from our fiducial value, $f_b=0.05$, to $f_b=0.1$ 
without changing the distributions of initial binary orbital properties. \\
\hline
{\tt rv1} & Initial $r_v=1\,\pc$ in contrast to our fiducial value of $r_v=2\,\pc$. \\
\hline
{\tt Iw} & Wider range ($0.08$--$150\,\msun$) in the IMF is used relative to our fiducial range ($0.1$--$100\,\msun$). \\
\hline
{\tt Fx} & Binary fraction for high-mass ($>15\,\msun$) stars $\fbhigh=x$ while keeping the overall binary 
fraction $f_b=0.05$, the same as in the baseline model. \\
\hline
{\tt Ms0.1} & Minimum secondary mass for initial binaries is $0.1\,\msun$. In addition, the initial period distribution is taken from \citet{2012Sci...337..444S}. \\
\hline
{\tt q0.6} & Minimum secondary mass for initial binaries is determined such that the mass ratio 
$q=m_s/m_p\geq0.6$. In addition, the initial period distribution is taken from \citet{2012Sci...337..444S}. \\
\hline
{\tt Ki} & BH natal kicks are independent of fallback fraction and remnant mass. 
Index {\tt i}$=${\tt 1}, {\tt 2}, {\tt 3} denote $\sigmabh/\sigmans=1$, $0.1$, and $0.01$, respectively. \\
\hline
{\tt Is} & Steep power-law exponent is used ($\alpha_1=3$) for the IMF for stars more massive than $1\,\msun$. \\
\hline
{\tt If} & Flat power-law exponent is used ($\alpha_1=1.6$) for the IMF for stars more massive than $\msun$. \\
\hline
{\tt W} & Weak winds (Vink et~al. 2001) are assumed. \\
%
%
\enddata
\tablecomments{We give informative names to our models. The names are combinations of several strings 
where each string refers to particular initial assumptions. To aid the readers understand the initial assumptions 
for particular models directly from the model's name we list specific strings in the names of our models and their 
corresponding meaning for the initial assumptions. 
}
\label{T:names}
\end{deluxetable}

%
%
\section{Derivation of Observed Cluster Properties}
\label{S:obs_derivation}
The definitions for key structural properties in numerical models are often different from those 
defined by observers for real clusters \citep[e.g.,][]{2013MNRAS.429.2881C}. To be consistent, we 
``observe'' our simulated models to extract structural properties with definitions similar to those 
used for real observations. We use the last snapshot from all our model clusters to extract the 
observable structural properties. All relevant final structural properties, measured both using theoretical 
definitions, and observers' definitions are listed in Table\ 3.  

\subsection{Estimation of ``Observed" Structural Properties}
\label{S:derivation}
We create two-dimensional projections for each model assuming spherical symmetry. The half-light 
radius, $\rhlobs$, is then estimated by finding the projected radius containing
half of the total light. 
We obtain the observed core radius, $\rcobs$, and the observed central density, 
$\sigmacobs$, by fitting an analytic King model to the cumulative stellar luminosity at a given
projected radius including stars within a projected distance of $\rhlobs$ from the center 
\citep[][their Eq.\ 18]{1962AJ.....67..471K}. This method was suggested earlier by \citet{2015ApJ...800....9M}. 
Since, this approach avoids binning of data, this is more robust against statistical fluctuations, especially at 
low projected distances compared to the often-used method of fitting the King profile directly to the surface brightness profile (SBP). 

We estimate the observed central velocity dispersion $v_{\sigma,c,\rm{obs}}$ 
by taking the standard deviation of the magnitudes of the three-dimensional velocities of all luminous stars 
(excluding compact objects) within a projected distance of $\rcobs$. In case of binaries, we 
take into account the center of mass velocities. 

\subsection{Estimation of Dissolution Times}
\label{S:tdiss}
Depending on initial assumptions, some of our cluster models get tidally
disrupted before the integration 
stopping time of $12\,\gyr$. The basic assumptions of our Monte Carlo approach are spherical symmetry, 
and a sufficiently large $N$ to ensure that the relaxation timescale is significantly
longer than the dynamical timescale. 
Both assumptions break down for clusters that have begun to tidally disrupt, since
the tidal boundary is not spherically symmetric, and a disrupting cluster can
lose mass on a timescale $<<$ than the relaxation time.  To that
end, once $t_{r}(t)>M(t)/\dot{M}$ for a cluster, where $t_r$, and $M$ denote relaxation time and total cluster 
mass respectively, we consider the cluster to have dissolved.  For clusters that
dissolve before 12 Gyr, we list the approximate dissolution times and mark them as 
``Dissolved" in Table\ 3.

%
\section{Overall Cluster Evolution}
\label{S:results_clusprop}
%
%
%
%
%
\begin{figure}
\begin{center}
\plotone{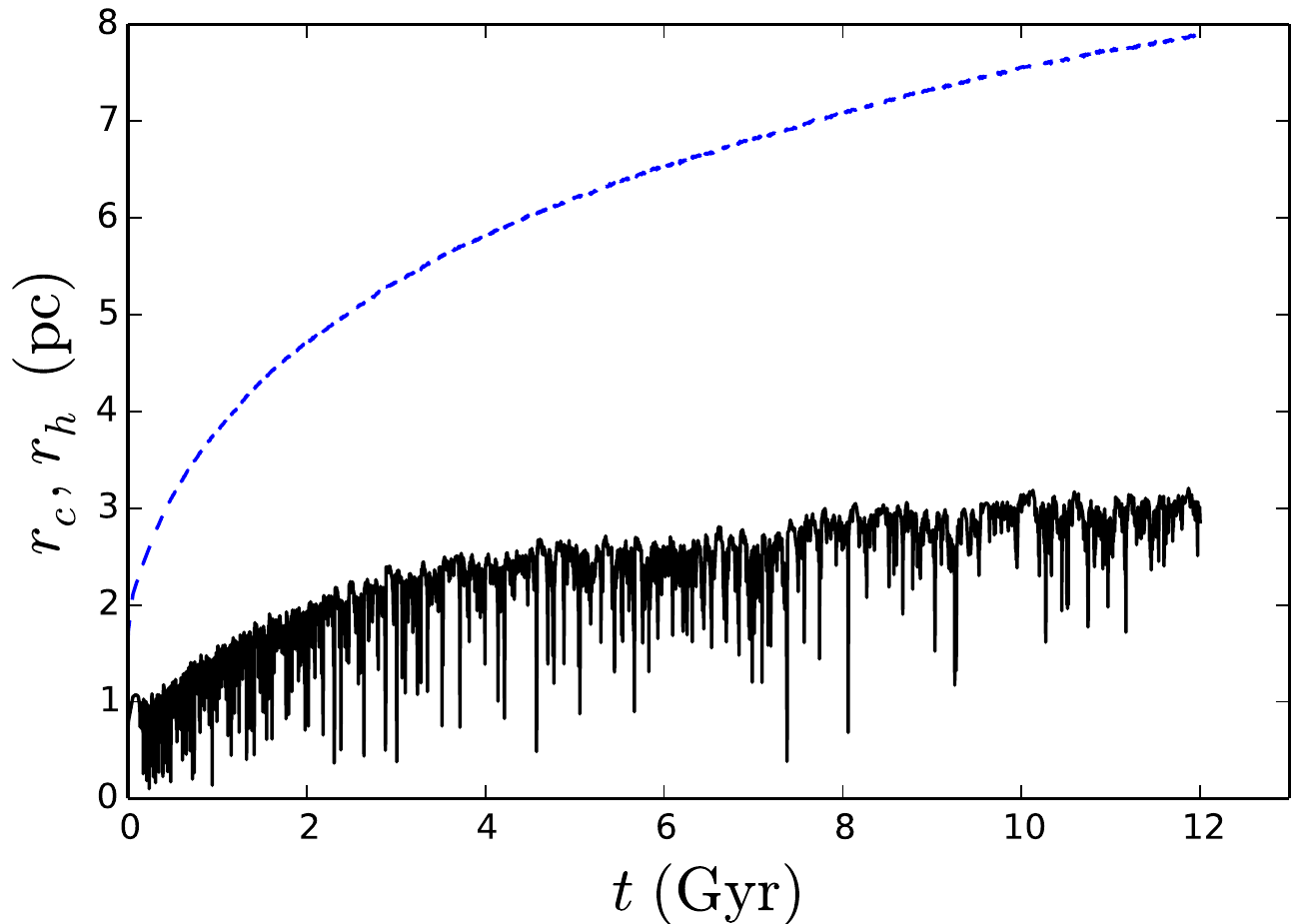}
\caption{
Evolution of the core radius ($r_c$) and the half-mass radius ($r_h$) for model {\tt S}. The solid (black) 
and dashed (blue) lines denote $r_c$ and $r_h$, respectively. The spikes in $r_c$ due to BH-driven 
core collapse continue until the end of the simulation at $12\,\gyr$. Both $r_c$ and $r_h$ expand 
all the way to the end.    
}
\label{fig:rcrh_s}
\end{center}
\end{figure} 
%
%
%
%
\begin{figure}
\begin{center}
\plotone{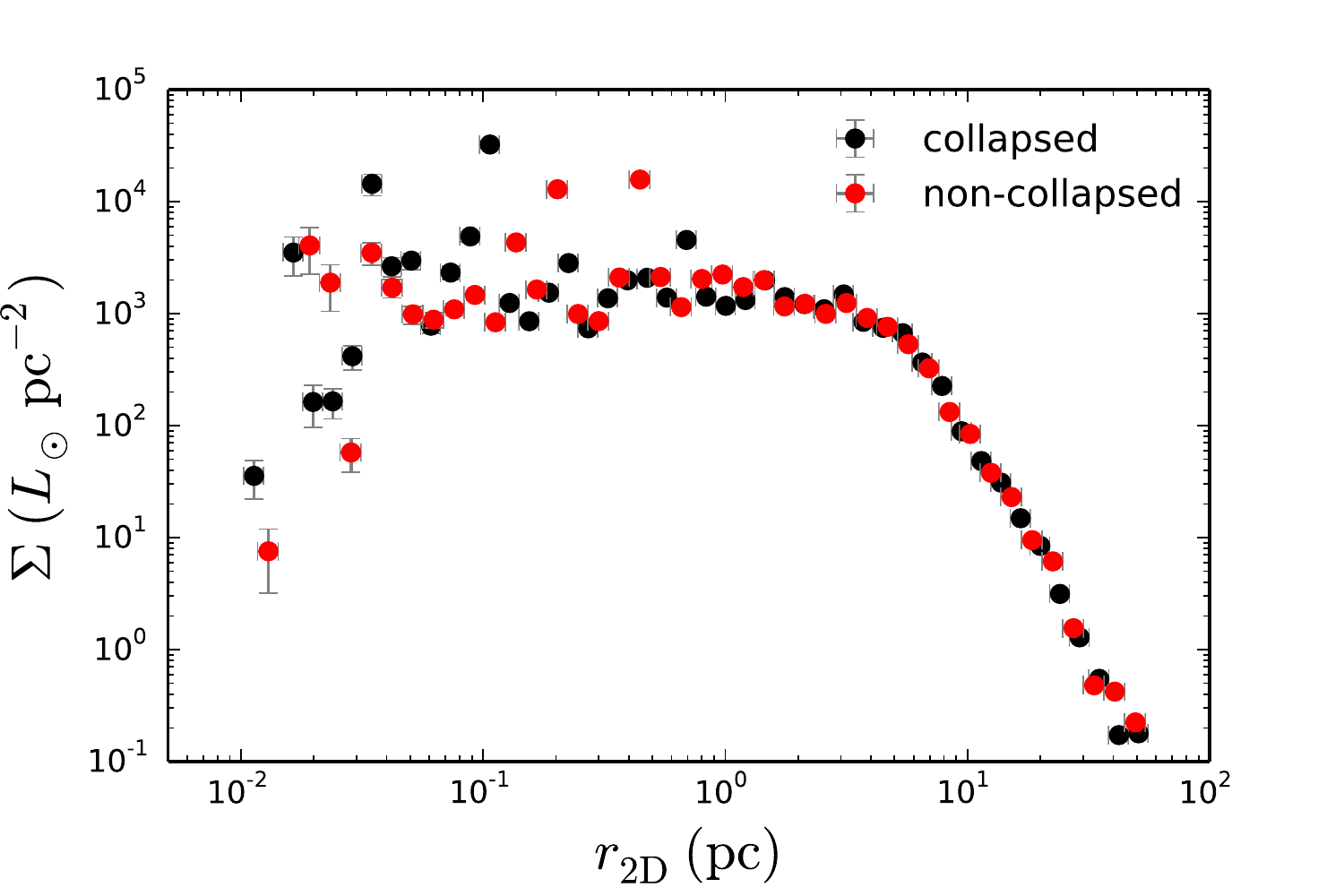}
\caption{
Comparison between the SBPs at two different times, one corresponding to 
a core-collapsed state, seen as the downward spikes in Fig.\ \ref{fig:rcrh_s} (at $t=9.27\,\gyr$; black), and 
the other corresponding to a non-collapsed state (at $t=9.29\,\gyr$; red) for model \modelS. The theoretically 
defined core radius changes from $r_c \approx 0.4$ during the collapsed state to about $3\,\pc$ out of 
that collapse. However, the observable SBP barely changes. 
}
\label{fig:comp_coll_noncoll_sbp_s}
\end{center}
\end{figure} 
%
%
%
%
\begin{figure}
\begin{center}
\plotone{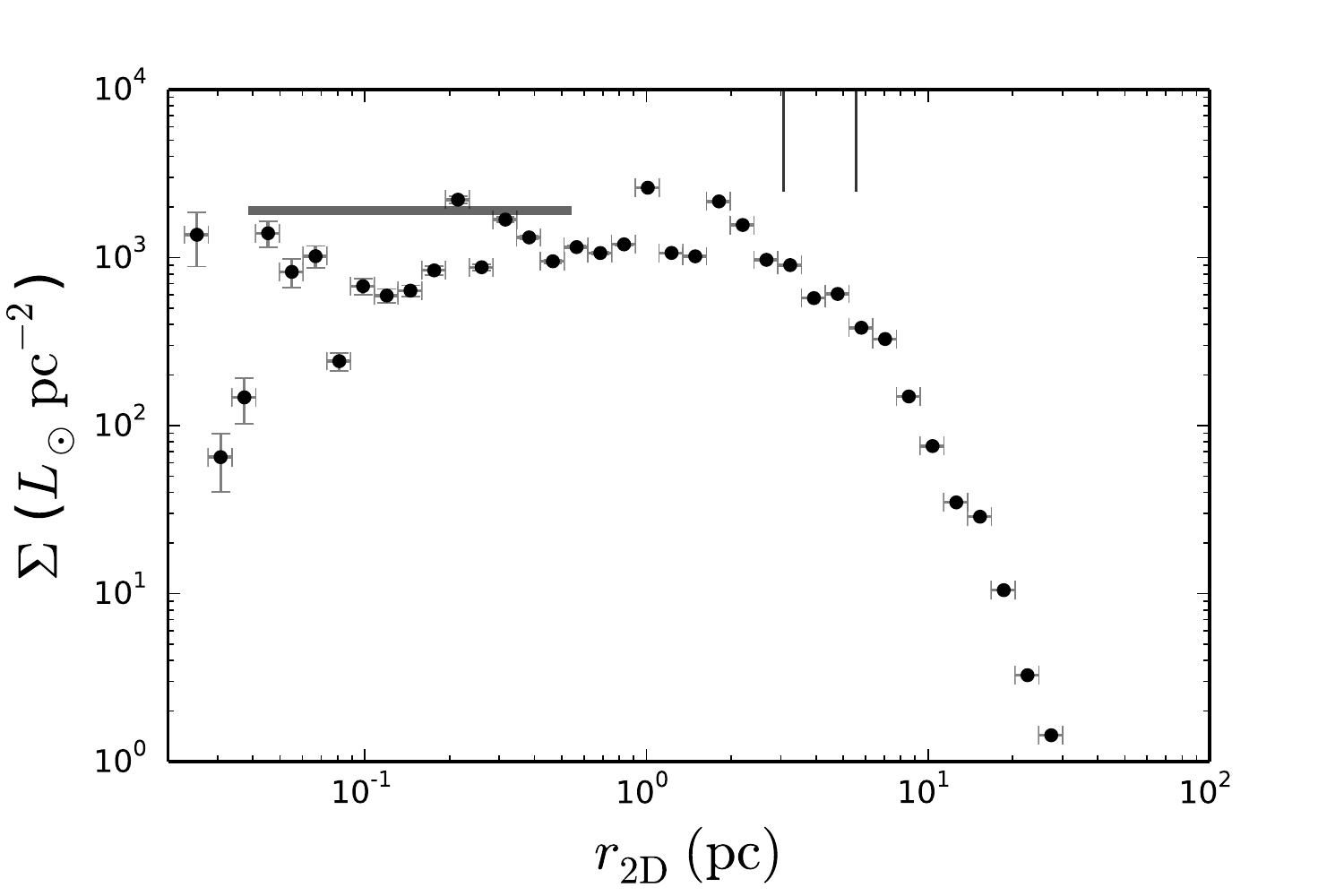}
\caption{
Final SBP at $t=12\,\gyr$ for model \modelS. The dots denote the surface luminosity density. While calculating the 
SBP, we discard stars brighter than $20\,\lsun$ to reduce noise from a small number of bright 
giant stars \citep[e.g.,][]{2006AJ....132..447N}. The horizontal line denotes the central surface 
luminosity density ($\sigmacobs$) based on the best-fit King model (\S\ref{S:obs_derivation}). 
The vertical lines 
denote the observed core radius $\rcobs$ (obtained from the King fit) and observed half-light 
radius $\rhlobs$. 
}
\label{fig:SBP_s}
\end{center}
\end{figure} 
%
%
%
%
\begin{figure*}
\begin{center}
\plottwo{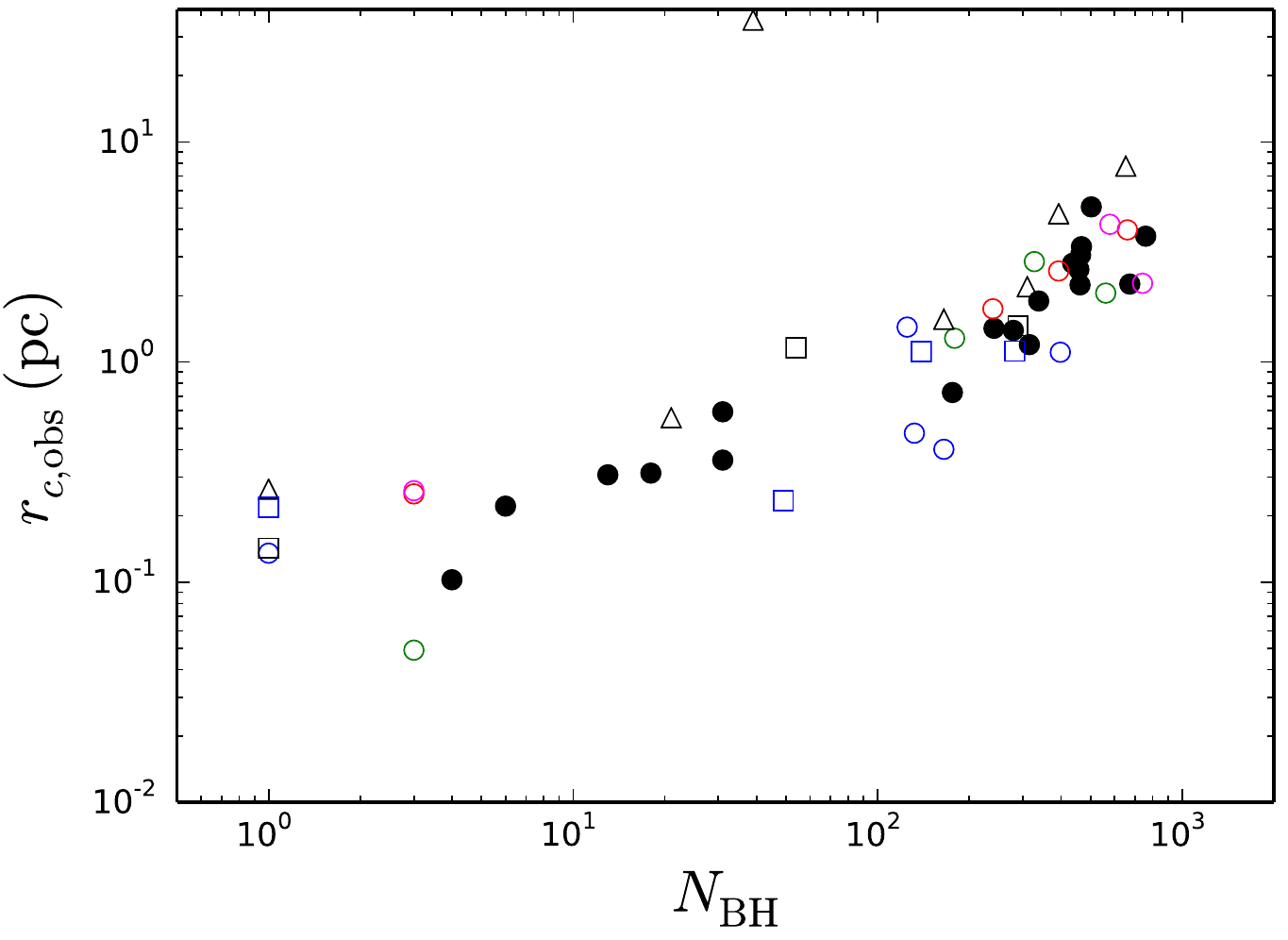}{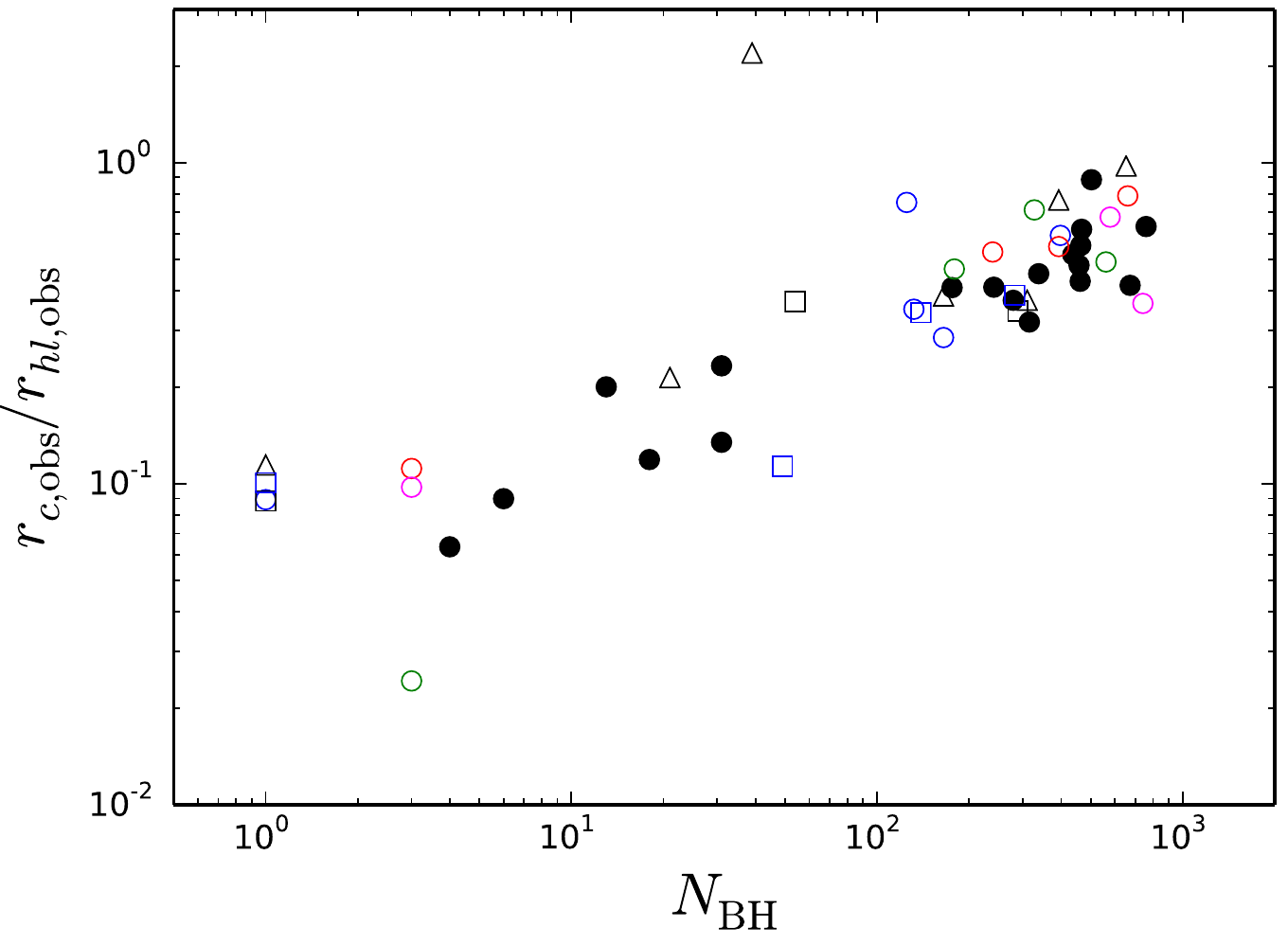}
\caption{
Final number of BHs bound to the cluster, $\nbh$ vs $\rcobs$ ({\em Left}) and $\rcobs/\rhlobs$ ({\em Right}) 
for all models that survived for at least 
$11\,\gyr$. Filled black circles represent models with $\rgc=8\,\kpc$,
$r_v=2\,\pc$ and strong winds. Blue, green, 
red, and magenta empty circles denote models with the same assumptions for $r_v$ and stellar wind, but with 
$\rgc=1$, $2$, $4$, and $20\,\kpc$, respectively. Black triangles denote models with $\rgc=8\,\kpc$, $r_v=2\,\pc$, 
and weak winds. Black and blue squares both denote models with $\rgc=8\,\kpc$, $r_v=1\,\pc$, and a wider 
IMF, but with weak and strong winds, respectively. 
In general, the larger the $\nbh$, the higher the $\rcobs$ and $\rcobs/\rhlobs$. 
Effects of all other initial assumptions are minor 
(Table\ 2). One model is a clear outlier with a very large $\rcobs$ and moderate $\nbh$: 
this is model \modelWIflatKone\ (Table\ 3), which is on the verge of dissolution. 
}
\label{fig:nbh_rcobs}
\end{center}
\end{figure*} 
%
%
%
%
\begin{figure}
\begin{center}
\plotone{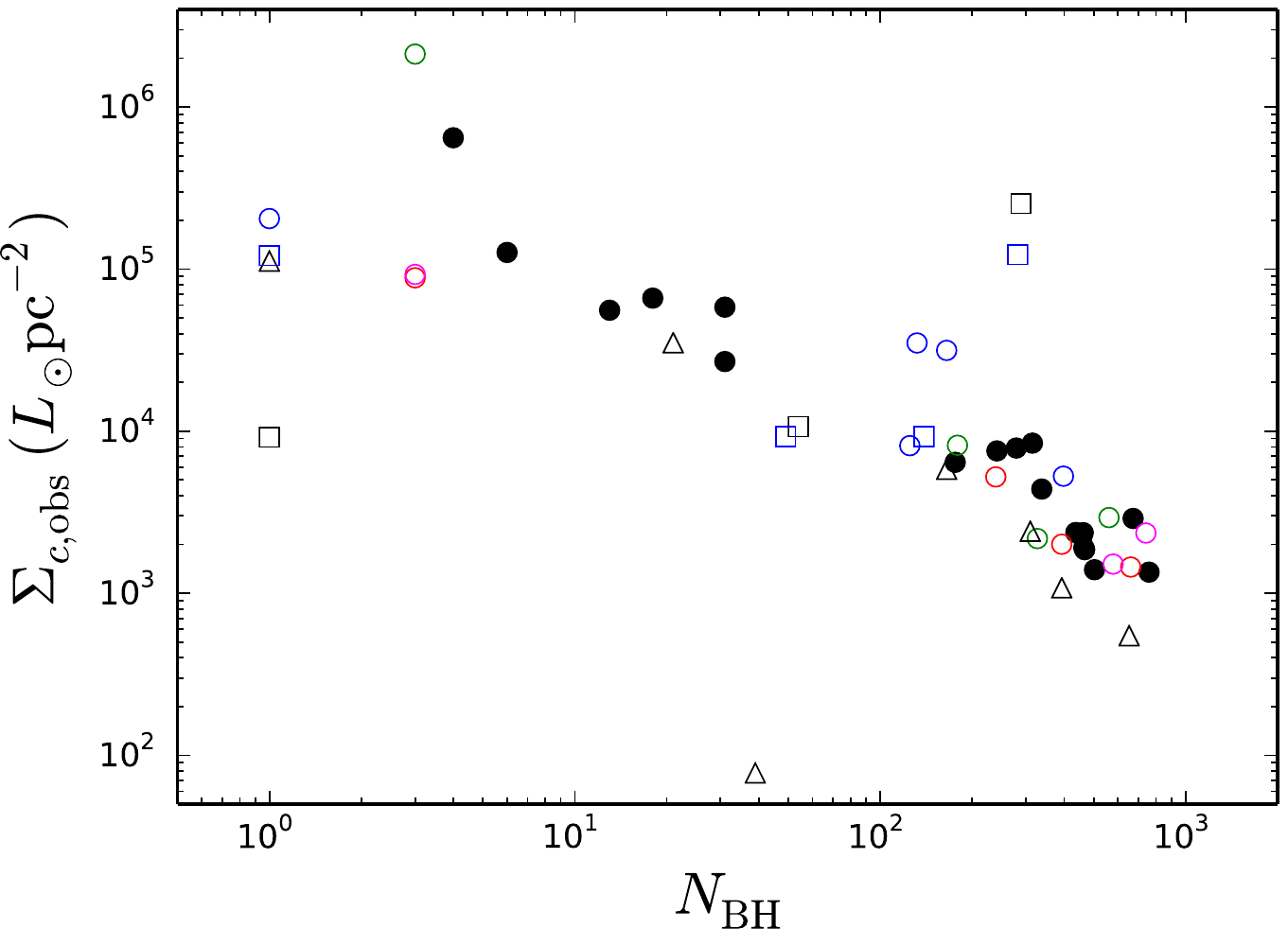}
\caption{
$\nbh$ vs $\sigmacobs$ for all models that survive for at least $11\,\gyr$ . 
Point styles are the same as in Fig.\ \ref{fig:nbh_rcobs}. The final $\sigmacobs$ 
is strongly dependent on $\nbh$. Effects of $r_v$, $\rgc$, $\fbhigh$ and winds
are minor, only effecting $\sigmacobs$ through $\nbh$ that are bound in the clusters at $12\,\gyr$. 
Model \modelWIflatKone\ is on the verge of dissolution and exhibits very low density and 
moderate $\nbh$. 
}
\label{fig:nbh_rhocobs}
\end{center}
\end{figure} 

Fig.\ \ref{fig:rcrh_s} shows the evolution of the core radius ($r_c$) and the half-mass radius ($r_h$) for our 
standard model {\tt S}. As expected, $r_c$ shows repeated downward spikes indicating collapse of the BH 
subcluster \citep[e.g.,][]{2015ApJ...800....9M}. 
Dynamical ejections and  binary formation following each BH-driven core collapse
produces energy, reversing the collapse and re-expanding the BH positions to mix with
the rest of the cluster.  Through repeated collapses, the $r_c$ increases on an average 
until the integration is stopped at $12\,\gyr$, and attains 
$r_c=2.8\pm 0.1\,\pc$. The error-bar here denotes the $1\sigma$ fluctuations during the last $1\,\myr$ 
of the cluster's evolution. $r_h$ monotonically increases as well, attaining a final value of about $8\,\pc$. 

Of course, an astronomer observing this cluster would measure different values for the structural 
quantities. Since the core collapse at any time during the evolution involves only a small number of 
the most massive BHs in the cluster, the observable SBP is insensitive to these collapses. For example, Fig.\ \ref{fig:comp_coll_noncoll_sbp_s} shows the 
SBPs for the same model cluster at two different times during its evolution, one when the cluster is 
in a core-collapsed state and the other when the cluster is out of it. 
There is no difference between the SBPs, although 
the theoretical $r_c$ changes from about $0.4$ to $2.8\,\pc$, between the 
collapsed and non-collapsed states. Thus, the core collapses driven by BH dynamics 
in these clusters are not observable.  

Fig.\ \ref{fig:SBP_s} shows the SBP including stars with luminosity 
$L_\star \leq 20 L_\odot$ for model \modelS\ at $t=12\,\gyr$. 
At this time model {\tt S} has 
$\sigmacobs \approx 2.7\times10^3\,\surfdens$, $\rhlobs=5.5\,\pc$, and $\rcobs=3\,\pc$. 
Hence, to an observer \modelS\ would appear as a cluster with low central density and puffed-up core. 

In general, if a large number of  BHs remain bound to the host cluster, dynamics involving BHs (dynamical formation 
of new binaries, binary--single and binary--binary scattering, and dynamical ejections from the cluster core) 
acts as the dominant source of energy in the core. The cluster can start the relaxation-driven 
core-contraction phase only after this source of energy is sufficiently depleted. 
The larger the number of retained BHs ($\nbh$) at a given time, 
the bigger the $\rcobs$ and $\rcobs/\rhlobs$ for the host cluster (Fig. \ref{fig:nbh_rcobs}). For example, 
excluding the disrupted clusters, the correlation coefficient between $\nbh$ and $\rcobs$ at $t=12\,\gyr$ 
is $0.82$ and that between $\rcobs/\rhlobs$ is 
$0.78$. 
Because of the large amount of energy BH-driven dynamics can inject into the clusters, the central density, 
$\sigmacobs$, also depends strongly on $\nbh$; $\sigmacobs$ and $\nbh$ 
are anti-correlated with a correlation coefficient of $-0.3$ (Fig.\ \ref{fig:nbh_rhocobs}).  

Note that due simply to the differences in the initial assumptions affecting the high-mass stars, 
clusters with very similar initial conditions attain widely varying final properties spanning $\sim4$ orders of magnitude in $\nbh$,
$\gtrsim2$ orders of magnitude in $\rcobs$, and 
$\gtrsim4$ orders of magnitude in $\sigmacobs$ at $t=12\,\gyr$ 
(\S\ref{S:numerical}; Tables\ 2 \& 3).

We find that the most dramatic differences in the overall final properties of a star
cluster come from the differences in $\nbh$. 
The variations in initial assumptions can change the evolution of $\nbh$ in different ways, and as a result, 
change the overall evolution of the cluster and its structural properties (e.g.\ $\rcobs$, $\rhlobs$, and 
$\sigmacobs$).  
We find that the most important 
assumption that determines the evolution of $\nbh$ in a cluster 
is the natal kick distribution 
for the BHs. Hence, we start by discussing the effects of initial assumptions related to BH-formation kicks. 

%
%
\subsection{Effects of the Natal Kick Distribution for BHs}
\label{S:results_kicks}
%
%
%
\begin{figure*}
\begin{center}
\plotone{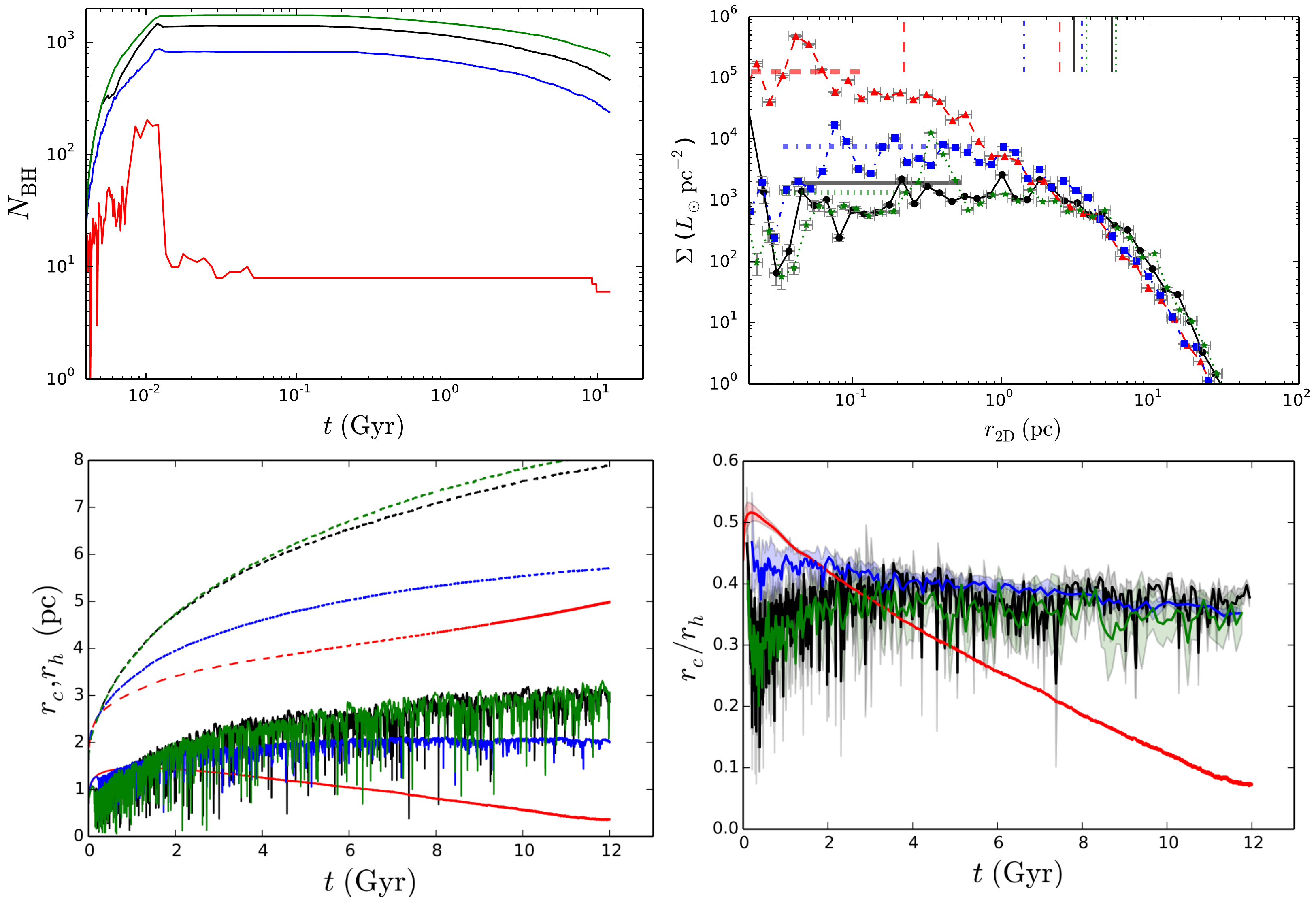}
\caption{
Comparison between model clusters \modelS\ (black), \modelKone\ (red), \modelKtwo\ (blue), and \modelKthree\ (green) 
(Tables\ 2, 3) identical except the assumption for the kick distribution during BH-formation. 
{\em Top-Left:} Evolution of the bound $\nbh$ for the four models.  
{\em Bottom-Left:} Evolution of $r_c$ and $r_h$. Solid and dashed lines denote $r_c$ and $r_h$, respectively. 
{\em Bottom-Right:} Evolution of $r_c/r_h$. To reduce scatter we take running averages for $r_c/r_h$. The lines denote 
the mean and the shaded regions show $1\sigma$ spread. 
{\em Top-Right:} Final SBPs for the four models. The horizontal and vertical lines 
denote $\sigmacobs$, $\rcobs$, and $\rhlobs$ for each model. 
Clearly, the bound $\nbh$ depends very strongly on the assumed distribution 
of formation kicks for the BHs. As a result, the overall evolution of the cluster also alters dramatically.
For example, \modelKone\ with $\sigmabh=\sigmans$ have only a handful of retained BHs. This cluster 
evolves to become a dense compact cluster and reaches the 
binary-burning phase near $t=12\,\gyr$, with a SBP typical of the so-called core-collapsed observed GCs. 
In contrast, all other models that retain significantly larger numbers of BHs at $t=12\,\gyr$, evolve very 
differently. \modelS\ and \modelKthree\ keep expanding until $t=12\,\gyr$, whereas model \modelKtwo\ 
ceases to expand at around $t=8\,\gyr$. 
}
\label{fig:kick_comp_clusprop}
\end{center}
\end{figure*} 

The assumed natal kick distribution for BHs directly controls $\nbh$ 
at a given time, and through that it controls the evolution of the cluster. 
For example, Fig.\ \ref{fig:kick_comp_clusprop} shows the evolution of 
$\nbh$, $r_c$, $r_h$, $r_c/r_h$, and the final SBPs for four identical models, all at a fixed $\rgc=8\,\kpc$. 
These models vary only in the assumed distribution of natal kicks for the BHs. 
We find that if the BHs are given full 
NS kicks, i.e., $\sigmabh=\sigmans$, most of the BHs are ejected from the 
cluster within $t\approx20\,\myr$, immeditally after formation. 
Without the source of energy from BH dynamics at the center, the cluster starts contracting due to two-body 
relaxation after the initial expansion from mass loss via stellar evolution. At about $11\,\gyr$ the cluster reaches the 
so-called binary-burning phase, when core-contraction is arrested due to extraction of binding energy 
from binary orbits via super-elastic scattering encounters \citep[e.g.,][]{2003gmbp.book.....H}. 
Based on the final SBP, this cluster would appear as a high-density core-collapsed cluster  
\citep[Fig.\ \ref{fig:kick_comp_clusprop}; see also ][]{2013MNRAS.429.2881C}. 
In contrast, clusters modeled with other natal kick distributions, where the BHs essentially receive much lower 
kicks compared to the neutron stars formed via core-collapse SN, do retain 
large numbers of BHs all the way through $12\,\gyr$. 
Due to the energy produced from BH dynamics, each of these 
clusters continues to expand till the end. \modelKtwo\ with $\sigmabh=0.1\sigmans$ contains about 
200 BHs at $t=12\,\gyr$, a sufficient number to keep the cluster in an expanded state. However, the rate of expansion 
is lower compared to models \modelS\ and \modelKthree\, where at $12\,\gyr$, $\nbh$ are much higher, with values
$464$ and $759$, respectively. While near the end there are indications that the model cluster \modelKtwo\ would start contracting 
after it ejects some more BHs, models \modelS\ and \modelKthree\ are still expanding at $t=12\,\gyr$,
and appear as puffy, low-density clusters from their final SBPs. The same
initial cluster can evolve to a final observational state with 
$\rcobs$ and $\sigmacobs$ varying by orders of magnitude, simply
because of changes in the assumed natal kick distribution for its BHs (Table\ 3).

The significant number of retained BHs can also expand the cluster closer to its
tidal radius, making the cluster more prone to disruption 
from Galactic tides. For example, with otherwise the exact same properties, at $\rgc=1\,\kpc$, clusters with 
relatively larger $\nbh$ expand more (e.g.\ \modelSRoneZ and \modelKthreeRoneZ),
and dissolve much earlier than $12\,\gyr$. 
On the other hand, with the same initial conditions, model clusters with relatively lower 
$\nbh$ are safe from tidal 
disruption even at $\rgc=1\,\kpc$ (e.g., \modelKoneRone, \modelKtwoRone; Table\ 3). 

In general, the larger the value of $\nbh$, the lower the final total mass of the cluster. 
This is because higher $\nbh$ expands the cluster more and a more expanded cluster 
loses more mass due to galactic tides. The exact
amount of expansion, mass loss, and the final cluster mass also depends 
on the metallicity ($Z$) of the cluster, since the metallicity controls the
wind-driven mass loss and the resulting mass of the BH population. Higher $Z$
leads to the formation of lower-mass BHs, and as a result the overall expansion due to 
BH ejections is reduced. The adopted wind prescription yields a similar effect. 
While the strong wind prescription leads to a higher mass loss at early times 
compared to the weak wind prescription, the latter leads to the formation of more 
massive BHs in a cluster than the former. As a result, the same initial cluster under the weak wind assumption 
eventually undergoes more expansion when the energy production in the center is dominated by 
BH dynamics. 
This increased expansion leads to an increased rate of star loss in models with weak winds compared to those with 
strong winds which reflects in the final $N$ in these clusters (Table\ 3). 

%
%
%
%
\subsection{Effects of the Assumed IMF}
\label{S:results_IMF}
%
%
%
%
\begin{figure*}
\begin{center}
\plottwo{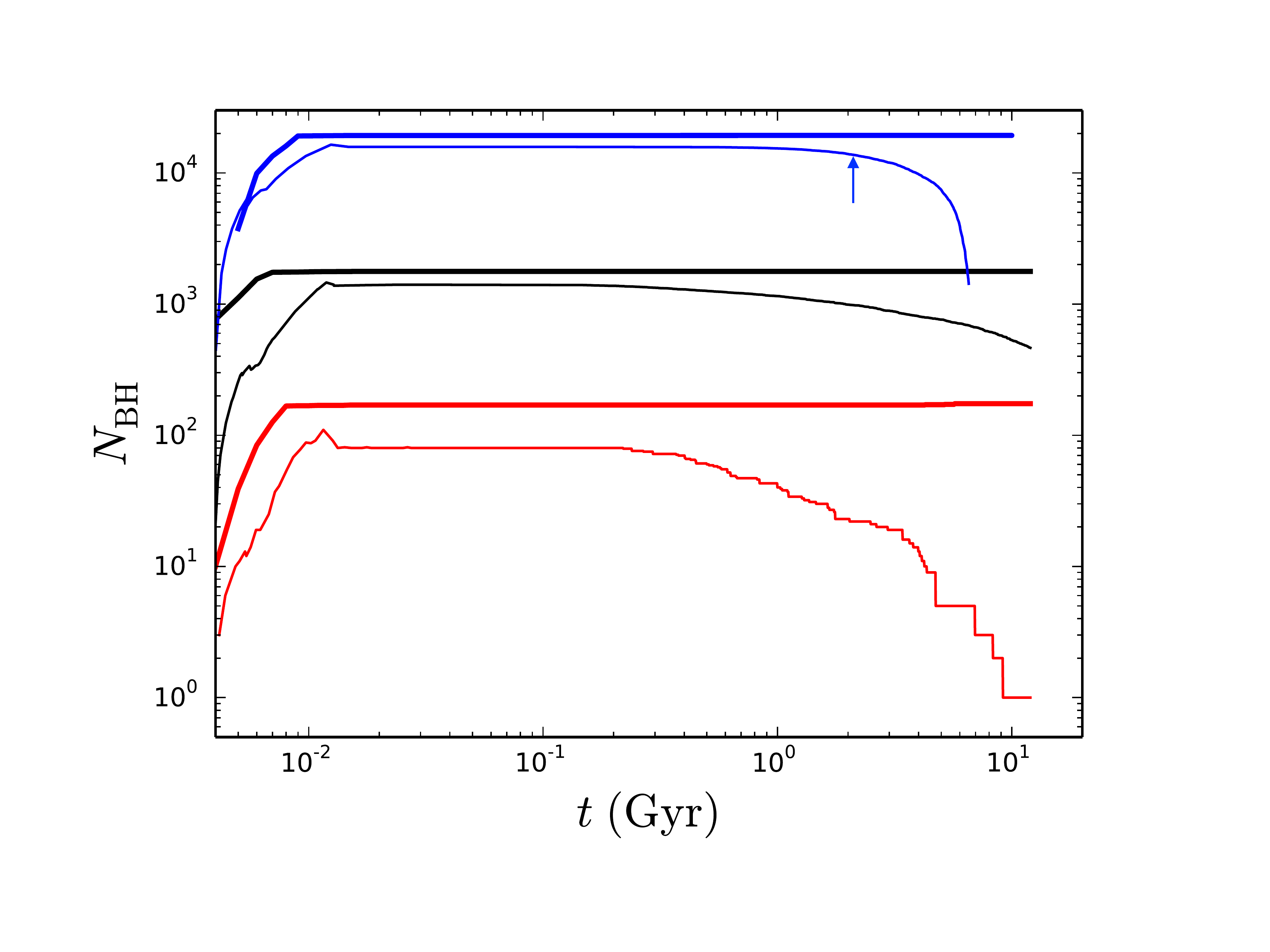}{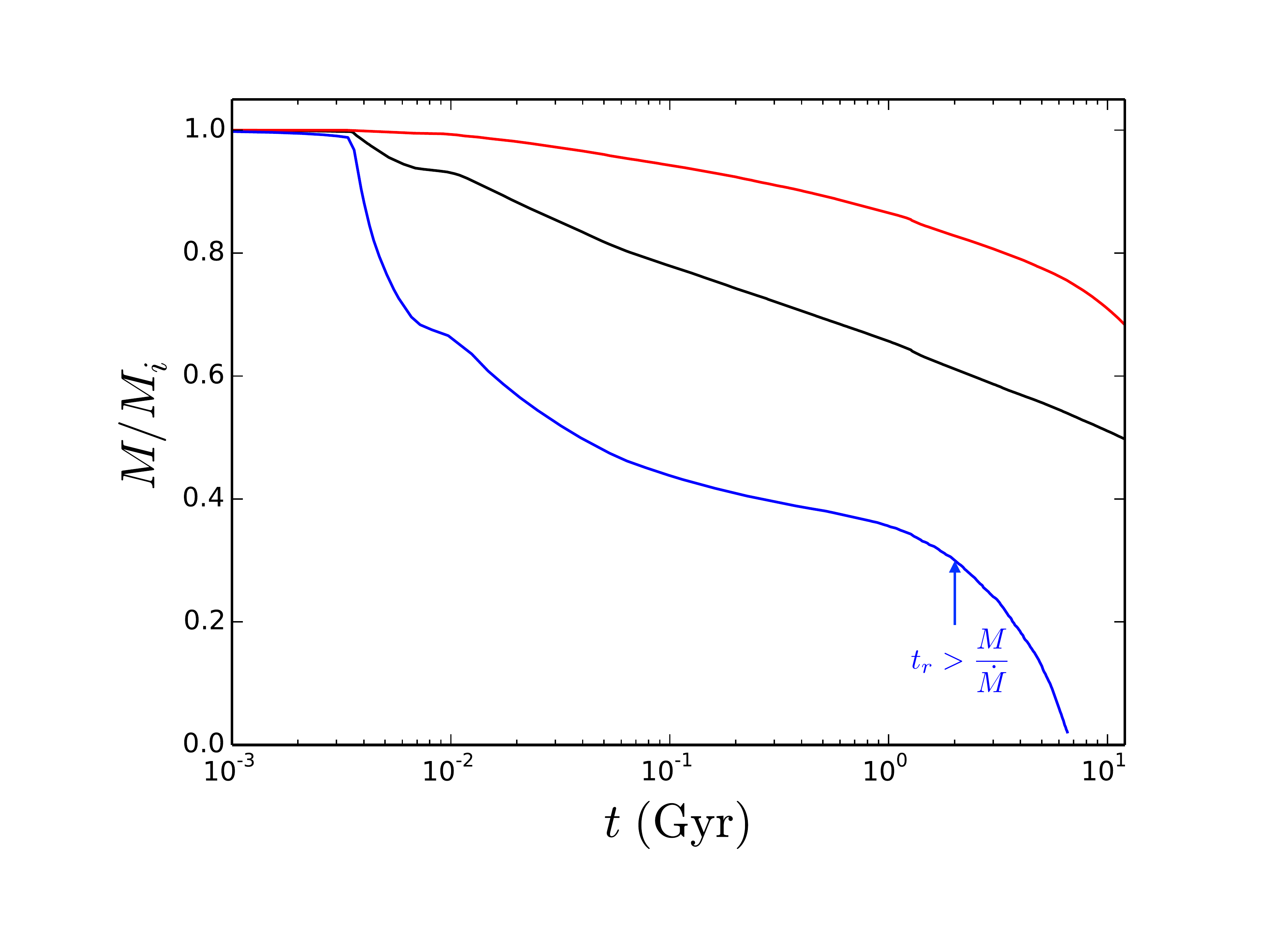}
\caption{
Comparison between models with varying IMFs. Each model 
has a different $\alpha_1$, where $\frac{dn}{dm_\star}\propto m_\star^{-\alpha_1}$ for $m_\star>1\,\msun$. 
Black (model \modelS), red (model \modelIsteep) and blue (model \modelIflat) lines denote $\alpha_1 = 2.3$, 
$3$, and $1.6$, respectively \citep[see \S\ref{S:numerical};][]{2001MNRAS.322..231K}. 
{\em Left:} Evolution of the total number of BHs formed (thick lines) and total number of BHs retained 
by the cluster (thin lines). {\em Right:} Evolution of the cluster mass
normalized to the initial cluster mass. 
Starting from otherwise identical initial conditions, the three clusters meet with very different fates. The cluster 
model with $\alpha_1=1.6$ 
produces many more BHs compared to other models, expands enormously, and gets disrupted  
around $t=2\,\gyr$. The approximate disruption time (\S\ref{S:tdiss}) is marked by arrows.  
}
\label{fig:imf_m_nbh}
\end{center}
\end{figure*} 
%
%
%
%
\begin{figure}
\begin{center}
\plotone{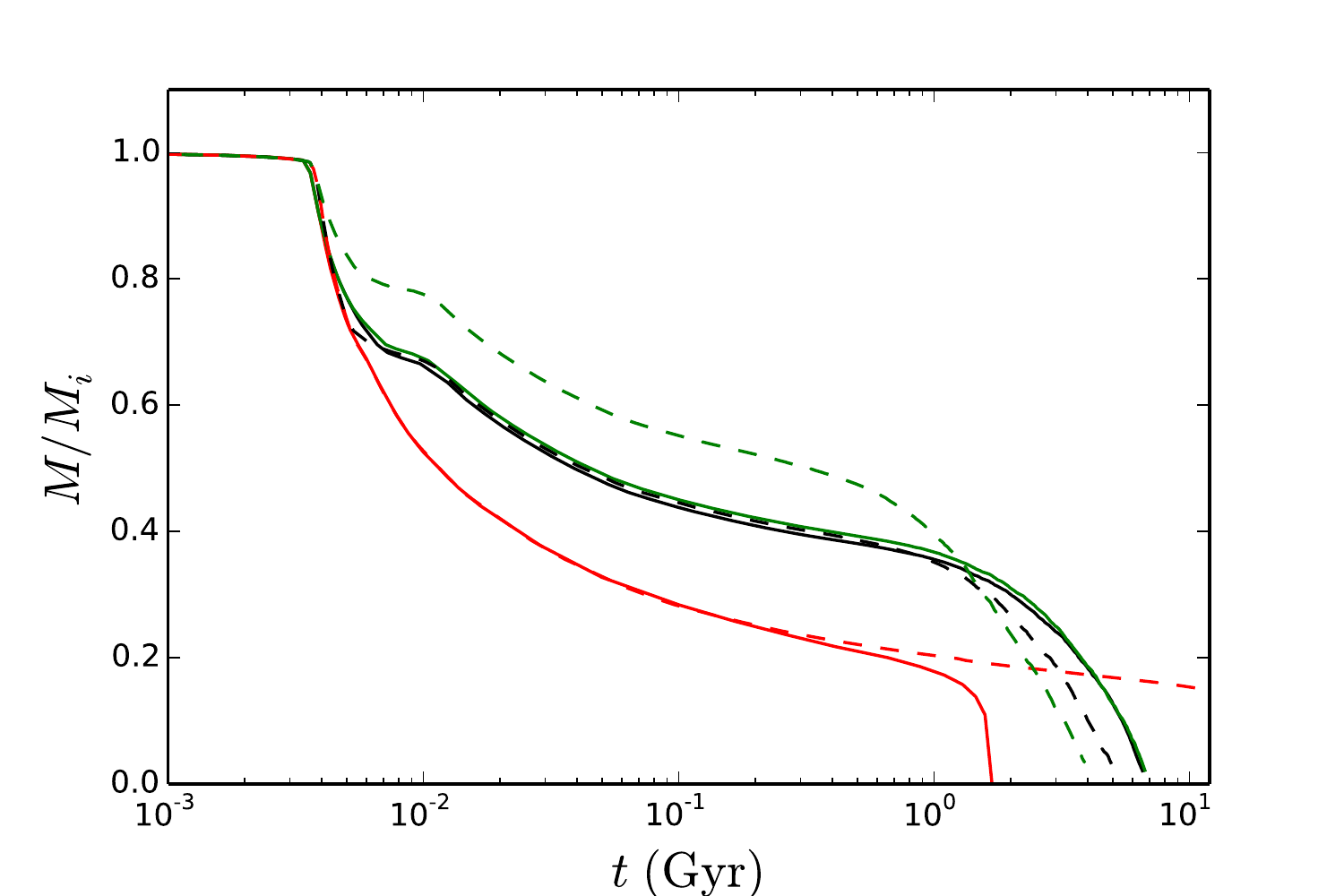}
\caption{
Evolution of the star cluster mass normalized by the initial mass of the cluster 
for models with $\alpha_1=-1.6$. Black, red, and green lines denote models 
with our standard kick prescription (models \modelIflat, \modelWIflat), and 
fallback-independent kick prescriptions with 
$\sigmabh=1$ (models \modelIflatKone, \modelWIflatKone), 
and $0.01\sigmans$ (models \modelIflatKthree, \modelWIflatKthree), respectively. 
Solid and dashed lines denote models with strong and weak winds, respectively.
All of these clusters get disrupted before $t=12\,\gyr$. 
Models that retain significant numbers of retained BHs, weak winds lead to earlier dissolution 
of the cluster relative to strong winds. 
Weak winds lead to formation of higher-mass BHs, which lead to faster expansion 
of the clusters, and correspondingly higher tidal mass loss compared to the strong wind case. 
For models 
with $\sigmabh=\sigmans$, this is reversed, since in this case with very low values of $\nbh$, 
mass loss from stellar evolution and compact object formation is the dominant
effect.   
}
\label{fig:flat_imf_m_kicks_standard_vink}
\end{center}
\end{figure} 

While for a given IMF the BH natal kick distribution is the dominant factor
controlling BH retention and overall star cluster 
evolution, any change to the IMF, especially for the high-mass stars, can also
bring about dramatic differences in how the cluster evolves. 
By directly controlling both the number of high-mass stars formed in a cluster 
(hence the number of BHs, stellar evolution driven mass loss), and the average stellar mass, 
variations in IMF can control the evolution of a cluster and even its survival 
\citep{1990ApJ...351..121C,2011ApJ...741L..12B}. 

We have tested 
the effects of variation in the exponent $\alpha_1$ of the IMF for stars more massive than $1\,\msun$ within the quoted uncertainty 
$2.3\pm0.7$ in \citet{2001MNRAS.322..231K}. Clusters evolve very differently
depending on the choice of $\alpha_1$ (Fig.\ \ref{fig:imf_m_nbh}). We first focus on our standard set of models 
(models \modelS, \modelIsteep, and \modelIflat; Table\ 2). While models with $\alpha_1=2.3$ and $3.0$ 
evolve normally and survive until $t=12\,\gyr$, the model with $\alpha_1=1.6$ 
dissolves at around $2\,\gyr$ (see \S\ref{S:tdiss} for how dissolution times are estimated). 
At first, mass is lost primarily via stellar
winds. Cluster model \modelIflat\ loses slightly more mass 
compared to the other model clusters with steeper $\alpha_1$. Dramatic differences appear during the stage when the clusters lose mass 
via compact object formation. As expected, the steeper the high-end of the IMF,
the lower the mass loss from compact object formation. 
This episode of quick mass loss ends by the time all the BHs are formed and many of the BHs are ejected due to their birth kicks. 
Following this episode, mass loss slows down and is driven by dynamical ejections of BHs
from the core and mass loss through the tidal boundary. At this stage, the number of retained BHs in a cluster becomes 
very important. A larger value of $\nbh$ leads to more dynamical ejections, which in turn leads to faster 
cluster expansion and higher tidal mass loss rate. Eventually, if the cluster expands too
much, it gets disrupted. 

\begin{figure}
\begin{center}
\plotone{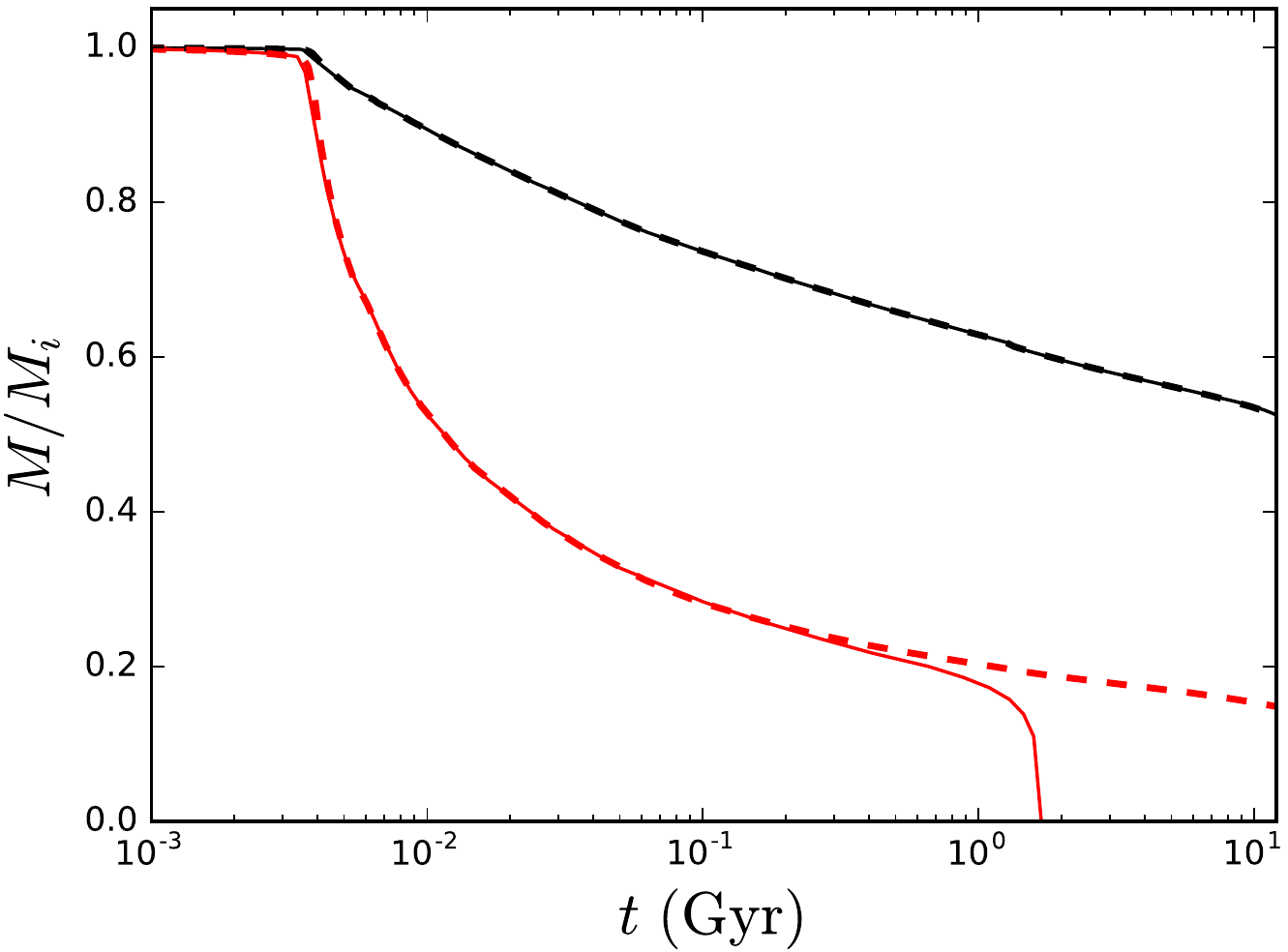}
\caption{
Similar to Fig.\ \ref{fig:flat_imf_m_kicks_standard_vink}, but for models 
\modelKone\ (black solid), \modelWKone\ (black dashed), \modelIflatKone\ (red solid), 
and \modelWIflatKone\ (red dashed). Since in all models the assumption of $\sigmabh=\sigmans$ 
ejects almost all BHs formed in these clusters, dynamical effects of BHs become unimportant. The 
dominant effect is from mass loss via compact object formation and immediate ejection of most 
BHs. In models with $\alpha_1=1.6$, the number of high-mass stars is significantly larger compared 
to that in models with $\alpha_1=2.3$. This leads to a much higher mass loss in models 
\modelIflatKone\ and \modelWIflatKone\ compared to models \modelKone\ and \modelWKone. 
}
\label{fig:flat_imf_m_nskicks_vink}
\end{center}
\end{figure}

The details of this process depends both on the kick distribution for the BHs as
well as the wind-driven mass loss (Fig.\ \ref{fig:flat_imf_m_kicks_standard_vink}). 
During the initial stages, clusters modeled with weak winds lose less mass than those modeled 
with strong winds. However, 
lower wind-driven mass loss leads to the formation of more massive BHs. 
These higher-mass BHs segregate more rapidly
in the cluster potential. In addition, higher-mass BHs inject more energy
into the cluster via dynamics 
and ejections than their lower-mass counterparts. As a result, once the cluster 
is sufficiently old for BH-dynamics to dominate energy production, 
the clusters modeled with weak winds expand faster and get disrupted earlier 
than those modeled with strong winds. The
higher the number of retained BHs, the bigger the 
difference between the clusters modeled with strong and weak winds (Fig.\ \ref{fig:flat_imf_m_kicks_standard_vink}). 
This trend is reversed in the models \modelIflatKone\ and \modelWIflatKone, both modeled with 
the highest natal kicks for the BHs we consider (Table\ 2). Since in these models almost all BHs are ejected during formation, 
the above-mentioned difference due to BH-dynamics is not relevant. Instead, since in the high-kick case, 
the weak-wind model loses less mass than the strong-wind model, \modelIflatKone\ is disrupted 
earlier than \modelWIflatKone. To further illustrate this point, Fig.\ \ref{fig:flat_imf_m_nskicks_vink} shows 
the effects of $\alpha_1$ on models with $\sigmabh=\sigmans$. Since most BHs are ejected during formation from all 
of these models (\modelKone, \modelWKone, \modelIflatKone, \modelWIflatKone) the evolution really depends only 
on the mass loss from compact object ejections during formation, i.e., the number of high-mass stars formed 
in the cluster. 

%

%
%
%
\subsection{Effects of Other Assumptions}
\label{S:results_other_clusprop}
%
%
%
%
\begin{figure*}
\begin{center}
\plotone{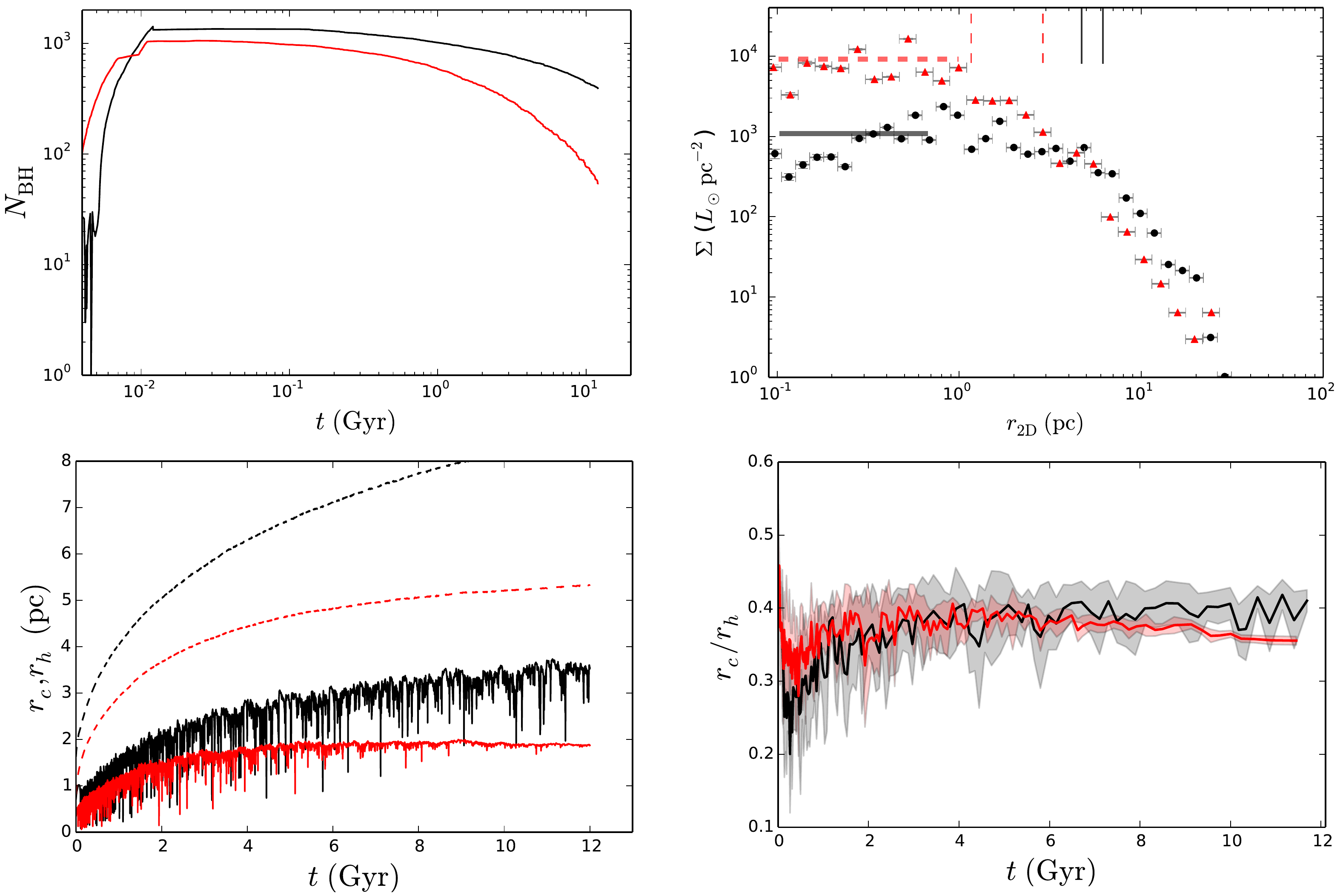}
\caption{
Same as Fig.\ \ref{fig:kick_comp_clusprop}, but comparing between models 
\modelW\ (black) and \modelWifb\ (red; Table\ 2). The lower relaxation timescale and 
average stellar mass for model \modelWifb\ compared to model \modelW\ leads to faster 
mass segregation, and as a result faster dynamical processing of BHs. 
The retained number of BHs $\nbh$ in model cluster \modelWifb\ is significantly lower compared 
to $\nbh$ in model cluster \modelW.  
As a result, model cluster 
\modelWifb\ ceases to expand
by $t=12\,\gyr$, and begins the relaxation-driven slow-contraction phase. In contrast, model cluster 
\modelW\ keeps expanding till the end. \modelWifb\ appears as a higher-density and more compact cluster 
compared to \modelW.  
}
\label{fig:rv_imfspread_comp_clusprop}
\end{center}
\end{figure*} 

The process and timescale for energy production from BH dynamics depend critically on the mass segregation 
timescale ($t_S$) in a cluster \citep[e.g.,][]{2013MNRAS.432.2779B,2013ApJ...763L..15M,2015ApJ...800....9M}. 
The mass segregation timescale of a massive object depends on the relaxation 
timescale ($t_r$), and the ratio of its mass to the average stellar mass ($<m>$) in its neighborhood, 
$t_S \propto t_r \frac{<m>}{m_i}$ \citep[e.g.,][]{2004ApJ...604..632G}.
 Thus, $\nbh$ 
and as a result, the overall cluster evolution depends on assumptions, such as: initial virial 
radius ($r_v$), and the IMF, that can affect 
either $t_r$ or $\frac{<m>}{m_i}$. 
Fig.\ \ref{fig:rv_imfspread_comp_clusprop} shows the difference in 
the evolution of two clusters (\modelW\ and \modelWifb) with different initial $r_v$ and the overall spread in the 
IMF (Table\ 2). The initial average stellar mass for models \modelW\ and \modelWifb\ are 
$<m>=0.66\,\msun$ and $0.62\,\msun$, respectively. The initial half-mass relaxation times for the models are $5.2\,\gyr$ and 
$1.9\,\gyr$. Due to these differences, \modelWifb\ processes
through its BHs much faster than \modelW. 
Furthermore, \modelWifb\ retains significantly fewer $\nbh$, and evolves to a
much denser and more compact final cluster than \modelW.

We have also tested how changes in the binary fraction and binary properties of 
high-mass stars affect the overall cluster evolution. We find that with a fixed overall 
$f_b$, variations simply in $\fbhigh$ and the distributions of orbital and companion properties for the high-mass binaries 
do not significantly affect the overall evolution of the clusters and their global observable properties. 

%
%
\section{Black Hole Binaries}
\label{S:results_bhbinaries}
%
%
%
%
\begin{figure}
\begin{center}
\plotone{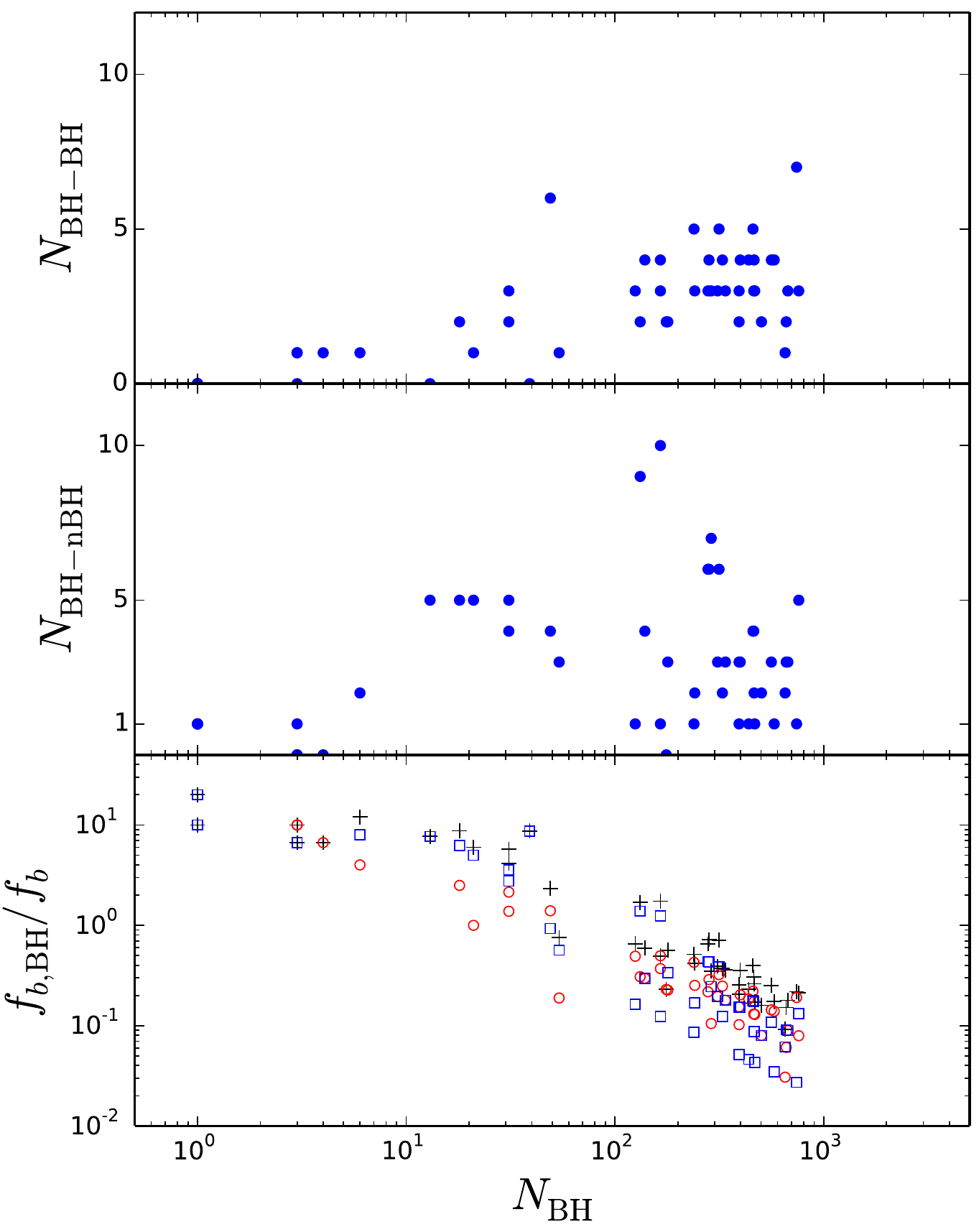}
\caption{
{\em Top:} The number of bound BHs $\nbh$ vs the number of bound BH-BH binaries $\nbhbh$. 
{\em Middle:} $\nbh$ vs the number of bound BH-nBH binaries $\nbhnbh$. 
{\em Bottom:} Binary fraction in BHs ($f_{b,\rm{BH}}$) normalized by the overall binary fraction $f_b$. 
Plusses (black), circles (red), and squares (blue) denote the normalized binary fraction for all 
BH binaries, BH-BH binaries, and BH-nBH binaries, respectively. A larger fraction of BHs are in binaries 
relative to the overall binary fraction for models with low $\nbh$. High $\nbh$ 
leads to BH-driven core-collapse which typically disrupts and ejects BH binaries. As $\nbh$ decreases, 
BH-driven core-collapses stop occuring, reducing the efficiency of BBH destruction and ejection. 
BHs being the most massive objects, typically get exchanged into 
binaries via binary-mediated scattering encounters. As a result, the fraction of 
BHs in binaries increases as $\nbh$ decreases.   
}
\label{fig:nbh_nbbh}
\end{center}
\end{figure} 

To be detectable, either by observation of electromagnetic signals generated via 
accretion from a non-BH companion in a BH--non-BH binary (BH--nBH),  
or by detection of GWs from a BH--BH merger, BHs 
must be in binary systems. Here we explore the effects of our various initial assumptions on the number and 
properties of BBHs. Dense star clusters are expected to be efficient factories of dynamically created binary BHs 
\citep[e.g.,][]{2010MNRAS.407.1946D,2011MNRAS.416..133D,2002ApJ...574L...5K,2003ApJ...591L.131P} although 
at any given time the binary fraction for BHs inside the cluster remains low due to an ongoing competition between dynamical 
creation, and disruption and ejection of binary BHs \citep{2015ApJ...800....9M}. Table\ 4 lists 
all our models and the corresponding numbers of single and binary BHs, retained and ejected from 
each cluster over its lifetime. 

%
%
%
\subsection{Black hole binaries retained in clusters}
\label{S:results_bhbinaries_retained}
In general, we find that the number of BBHs (of any kind) 
in each of our model clusters is much lower compared to the total number of BHs (Table\ 4). 
Moreover, the BBH numbers are {\em not} strongly dependent on $\nbh$ (Fig.\ \ref{fig:nbh_nbbh}). 
For example, depending on the assumptions in our models, we find a 
large range in the values of $\nbh$, varying from $0$ to $935$ at $t=12\,\gyr$. In contrast, at $t=12\,\gyr$ 
the variation in the numbers of BH--BH ($\nbhbh$) and BH--nBH ($\nbhnbh$) binaries 
are $0$ to $7$ and $0$ to $17$, respectively. The correlation coefficient between $\nbh$ and 
$\nbhbh$ at $t=12\,\gyr$ for all model clusters that did not get disrupted earlier than $12\,\gyr$ is $0.56$ and that between 
$\nbh$ and $\nbhnbh$ is $0.07$ (Table\ 4). 

\begin{figure}
\begin{center}
\plotone{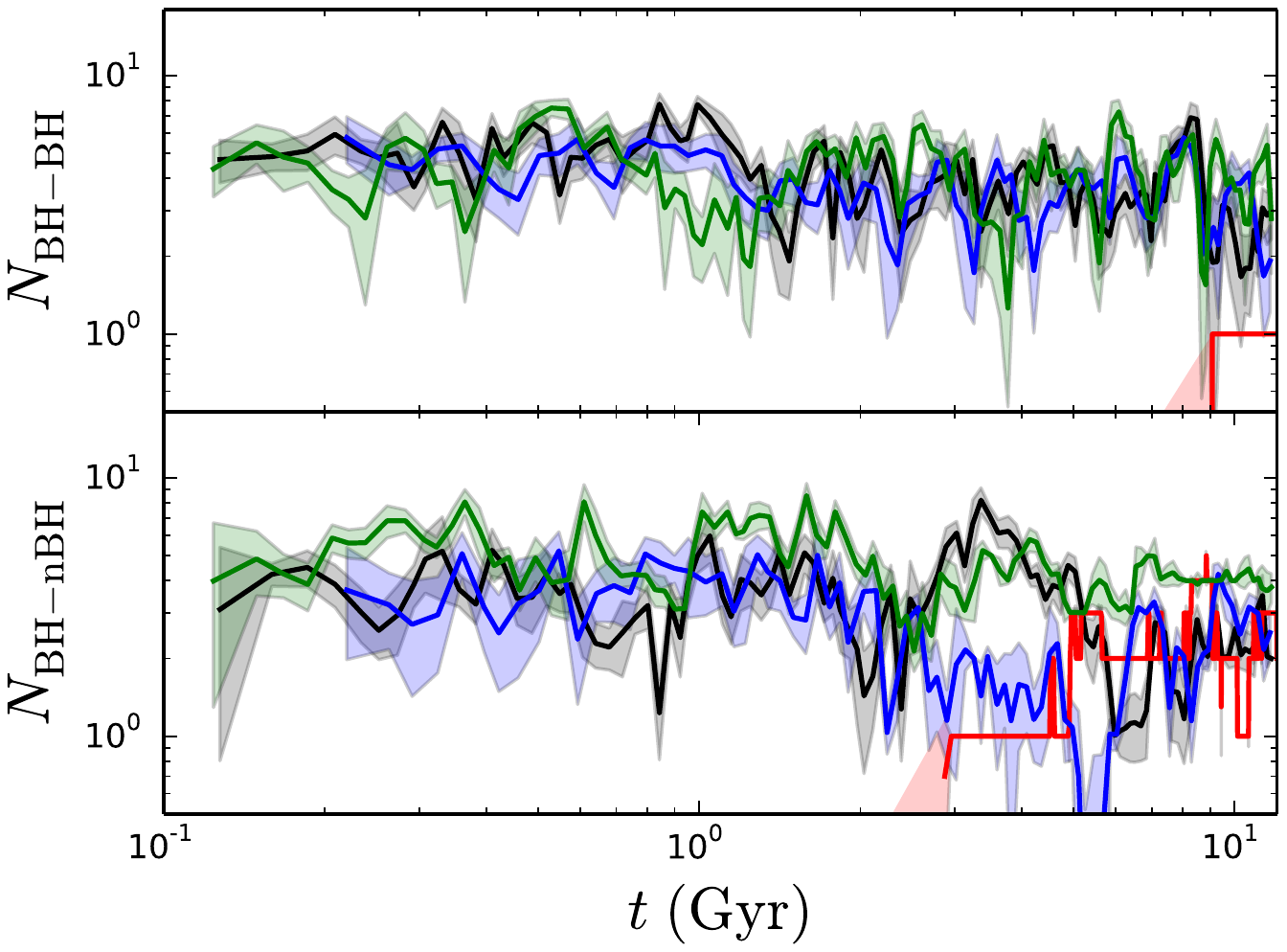}
\caption{
Evolution of the number of binary BHs bound to the cluster for different assumed natal 
kick distributions for BHs. Top and bottom panels show BH-BH and 
BH-nBH binary numbers. 
Both $\nbhbh$ and $\nbhnbh$ show large scatters over time. This is a direct consequence of the continuous 
disruption and ejection of existing binaries and dynamical formation of new ones at any given 
time in clusters. To reduce scatter we have under-sampled and show the mean (lines) and $\pm$ one standard 
deviation (shaded region). Black, red, blue, and green denote 
models \modelS, \modelKone, \modelKtwo, and \modelKthree, respectively (Table\ 2). Both 
$\nbhbh$ and $\nbhnbh$ remain low independent of the assumed 
distribution of natal kicks for BHs. This indicates that the softening of orbits for massive binaries 
due to mass loss via winds and compact object formation is responsible for dynamical disruption of most 
primordial binary orbits, and that this process does not depend on the natal kick distribution. Even with very low 
adopted natal kicks for BHs, $\sigmabh=2.65\,\kms$, \modelKthree\ contains low numbers of binary BHs.     
}
\label{fig:t_nbbh_kicks}
\end{center}
\end{figure}

\begin{figure}
\begin{center}
\plotone{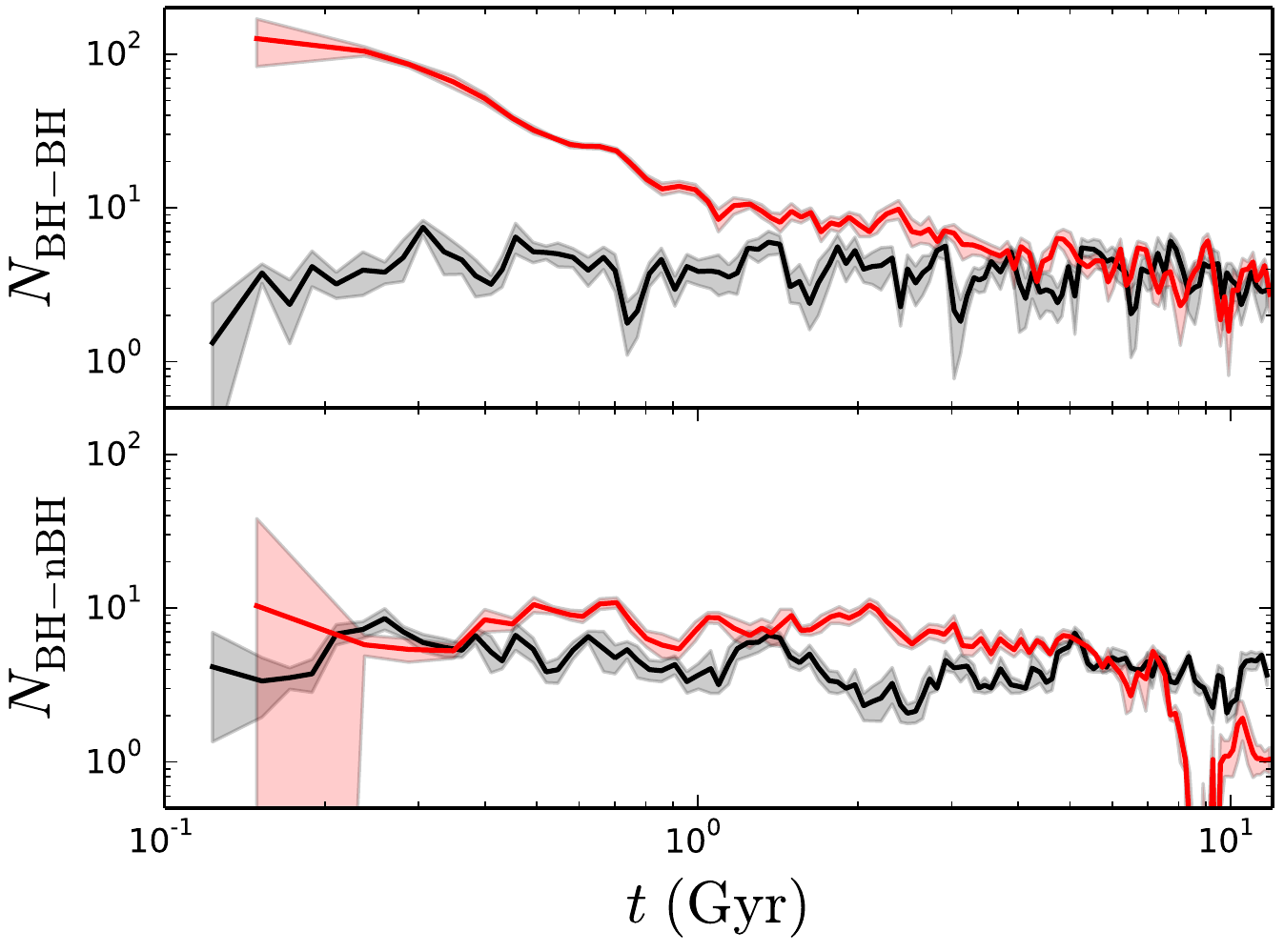}
\caption{
Same as Fig.\ \ref{fig:t_nbbh_kicks} but showing a comparison between the evolution for two 
identical initial models differing only by the initial binary fraction in high-mass ($>15\,\msun$) stars, $\fbhigh$. 
Black and red denote models \modelFzero\ with initial $\fbhigh=0$ and \modelFone\ with initial $\fbhigh=1$, 
respectively (Table\ 2). In both cases, the overall binary fraction $f_b$ is kept fixed at $0.05$. 
Independent of the initial $\fbhigh$, the final retained $\nbhbh$ converge to a low value. The final 
retained $\nbhnbh$ depends on the efficiency with which interactions involving BHs 
and binaries with non-BH components can produce BH-nBH binaries via exchange. 
The cluster with $\fbhigh=1$ has fewer low-mass binaries than the cluster with $\fbhigh=0$. As 
a result, BH-nBH binary formation is less effective in \modelFone\ compared to that in \modelFzero\ 
at late times. 
}
\label{fig:t_nbbh_fbhigh}
\end{center}
\end{figure}

\begin{figure}
\begin{center}
\plotone{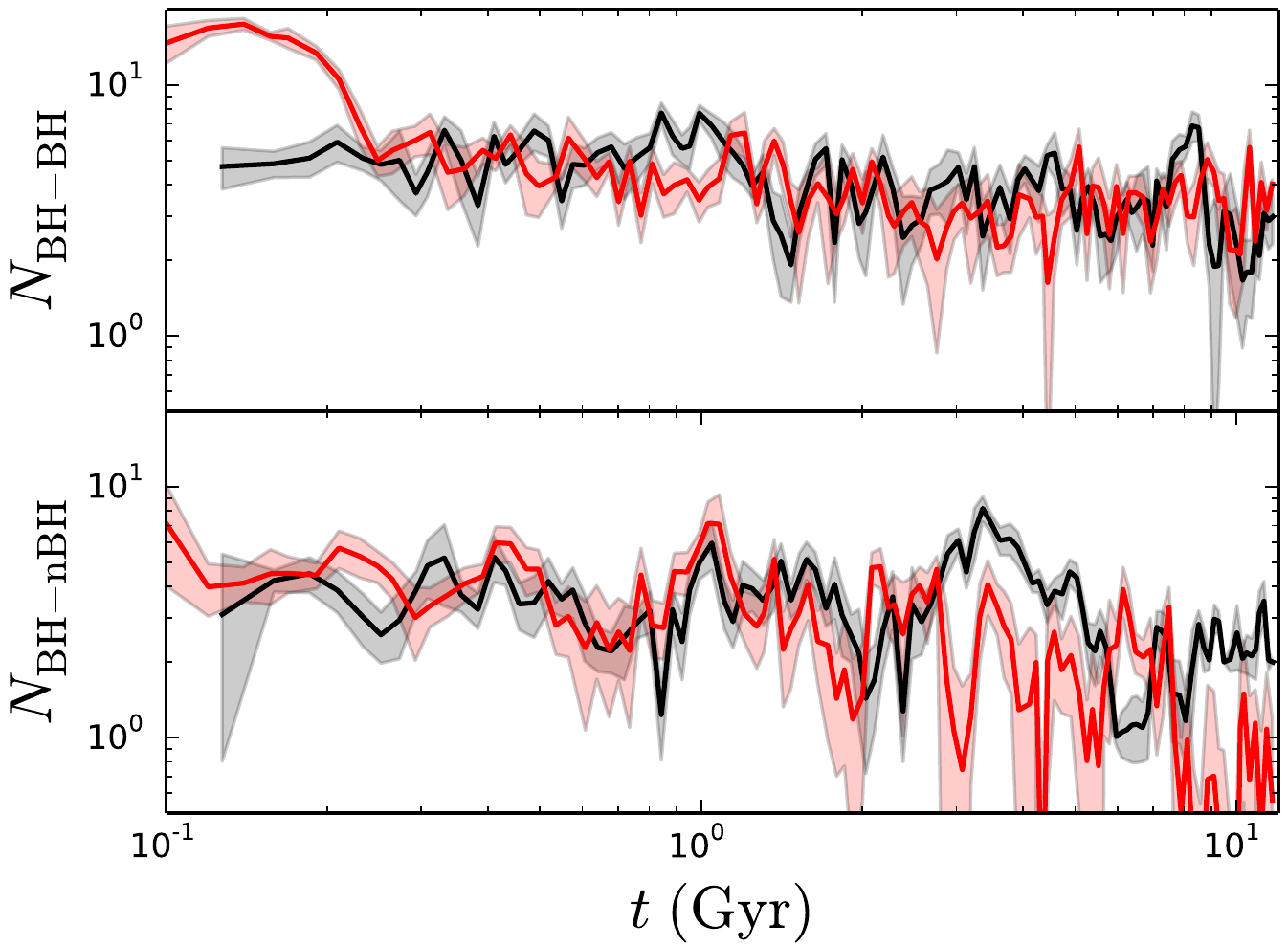}
\caption{
Same as Fig.\ \ref{fig:t_nbbh_kicks} but compares between the evolution for 
two identical initial models differing by the assumed prescription for mass loss via stellar 
winds. Black and red denote models \modelS\ with the strong and \modelW\ with the 
weak wind prescriptions, respectively. The values of $\nbhbh$ shows no difference between 
the two models. The value for $\nbhnbh$ for the weak wind case is slightly lower compared to 
that for the the strong wind case. Weak winds lead to more massive BHs, which in turn lead to more expanded 
clusters with lower central densities (e.g., Table\ 3). 
As a result, in the weak wind case, at late times formation of BH--nBH binaries via binary-mediated 
exchange interactions between single BHs and non-BH binaries is less efficient. 
}
\label{fig:t_nbbh_winds}
\end{center}
\end{figure}

\begin{figure}
\begin{center}
\plotone{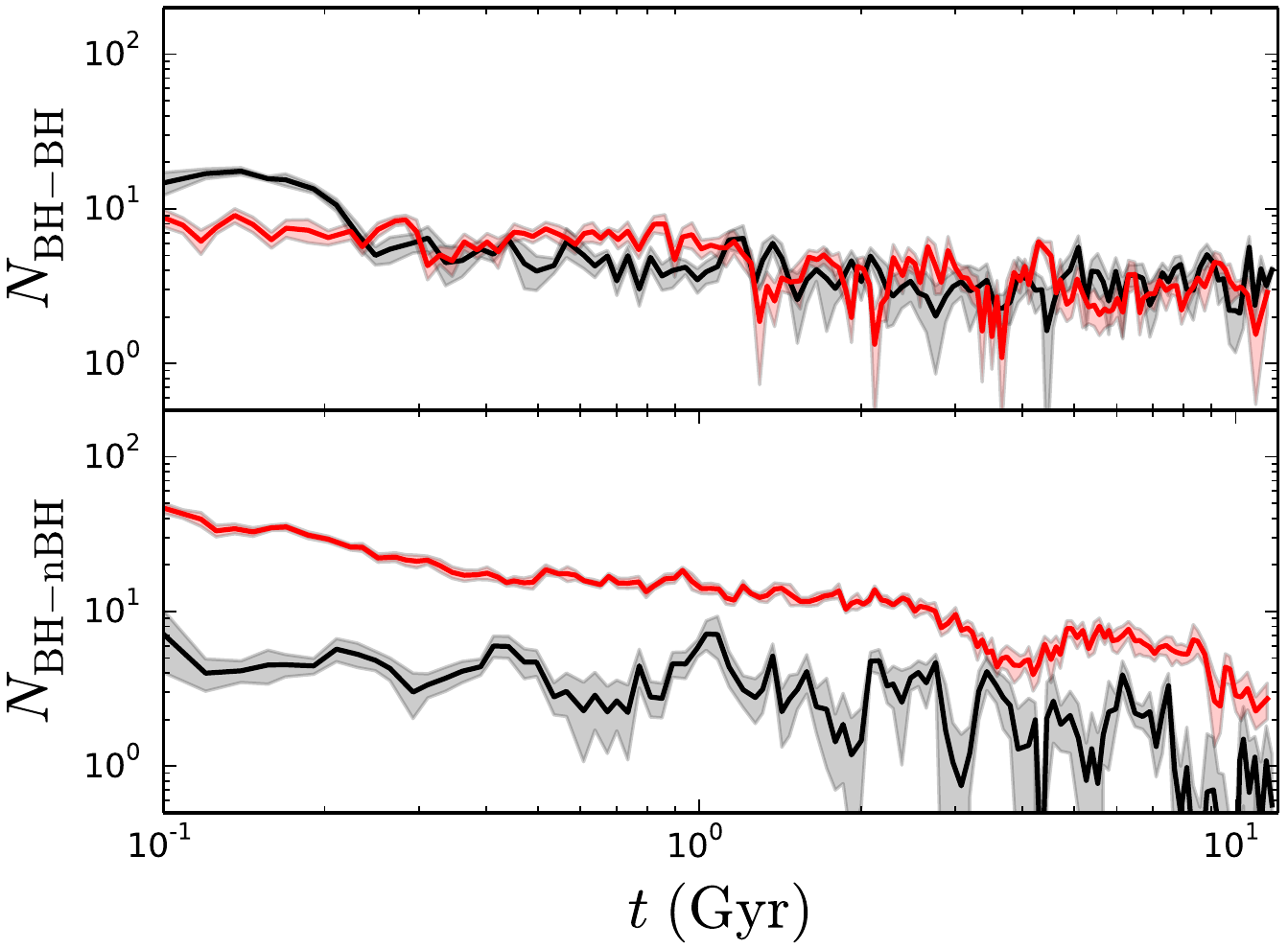}
\caption{
Same as Fig.\ \ref{fig:t_nbbh_kicks} but showing a comparison between the evolution for two 
models differing by the overall binary fraction, $r_v$, and the IMF spread. Black and red denote 
models \modelW\ and \modelWifb, respectively (Table\ 2). The retained $\nbhbh$ for both 
models converge within $t\approx200\,\myr$. The model cluster \modelW\ with a lower initial $f_b$ 
retains a lower $\nbhnbh$ compared to \modelWifb. 
(Tables\ 2, 4).  
}
\label{fig:t_nbbh_fbfrac}
\end{center}
\end{figure}

These general trends are understandable from 
the basic process of how BBHs are created and dynamically processed 
inside dense star clusters. 
Almost all primordial binaries containing massive 
stars are disrupted inside a cluster. This is mainly due to orbital expansion via mass loss, which makes these binaries 
susceptible to disruption via subsequent strong scattering encounters (Fig.\ \ref{fig:t_nbbh_kicks}). Thus, most BHs in dense star clusters 
are singles. Almost all BBHs in dense star clusters, as well as BBHs that are ejected 
from them after $t\sim10^2\,\myr$, are binaries that are dynamically created in the dense cores 
of these clusters, and are {\em not} primordial. 
Note that even if an old star cluster typical of the present-day GCs now appears to have 
a low central density, its BHs likely have been processed in dramatically 
higher density environments as a result of repeated past BH-induced core collapse episodes 
(e.g., Figs.\ \ref{fig:rcrh_s}, \ref{fig:SBP_s}, \& \ref{fig:kick_comp_clusprop}). 
Both the values of $\nbhbh$ and $\nbhnbh$ in a cluster 
show large fluctuations over time and can vary between zero to 
$\sim 10$ (e.g., Figs.\ \ref{fig:t_nbbh_kicks}--\ref{fig:t_nbbh_fbfrac}). 
This is reflective of the high frequency of dynamical processes that create, modify, 
disrupt, and eject the BBHs in a dense star cluster, especially during a BH-driven core collapse. 
In these dense environments BBHs continuously form via 
three-body encounters, change via swapping of partners in binary-mediated 
scattering, get disrupted via strong scattering, and get ejected due to recoil 
from strong scattering encounters \citep[e.g.,][]{1975MNRAS.173..729H}. 
The dynamical processes including disruption of primordial binaries, creation of new ones, and dynamical ejections 
essentially erase the binary properties the high-mass stars were born with. 
This in fact is the main difference between BBHs formed in a dense star cluster 
and BBHs that are born in isolation in the field. In the latter case, all BBHs are formed from 
primordial binaries, hence, the BBH properties as well as numbers relative to single stars 
are directly related to the assumptions of initial binary properties for the high-mass stars. 

While large numbers of BHs are still present in a star cluster, 
the core of the cluster is dominated by the BHs due to mass segregation. Internal 
dynamics determines how many of these BHs can hold onto their companions once formed 
via dynamics. 
As $\nbh$ decreases, the BH-driven core collapses become less pronounced. 
As a result, a higher fraction of BHs can remain in binaries, and the ejection rate 
of BBHs decreases. Hence, although $\nbh$ decreases, the fraction of 
BHs in BH--BH binaries increases, thus keeping $\nbhbh$ largely unchanged (Fig\ \ref{fig:nbh_nbbh}). 
On the other hand, while $\nbh$ is large, the very central regions are dominated by the (mostly) single BHs and lower-mass 
stars are driven out. 
A large $\nbh$ also keeps the cluster puffed up, lowering the rate of encounters 
between BHs and binaries with non-BH members. 
Only after a cluster is sufficiently depleted of the BHs, can the rate of formation of BH--nBH binaries 
via exchange interactions involving a BH and a binary with non-BH members increase. However, 
at this stage the maximum $\nbhnbh$ is likely limited by the low number of remaining BHs.

The lack of a strong correlation between $\nbh$ and $\nbhnbh$ has some interesting implications. 
For example, 
major efforts are now underway to detect BH candidates in GCs 
\citep[e.g.,][]{2012Natur.490...71S,2013ApJ...777...69C,2013atnf.prop.5724S,2014atnf.prop.6457M,2014atnf.prop.6417M}. For 
detection via electromagnetic signals (e.g., by comparing the radio and X-ray luminosities) there must be a BH with a non-BH accreting 
companion. For simplicity, if we treat $\nbhnbh$ as a proxy for the absolute upper limit on the number of BH candidates 
that may be detectable via electromagnetic signatures, the lack of correlation between $\nbh$ and $\nbhnbh$ poses a serious challenge 
in inferring the number of total BHs in the GC from the discovery of BH candidates in that GC. 
Note, however, that creation of accreting BHs in star clusters is likely a complex 
process which requires that the binary is not disrupted for a sufficient time to allow accretion. 
Even when this is satisfied, the duty cycle may be low for such accreting binaries 
\citep{2004ApJ...601L.171K}. We encourage a more detailed study on this topic. 

%
%
%

We now focus our attention on understanding the detailed evolution of BBHs inside a cluster and the 
effects of various initial assumptions through selected example 
models (Figs\ \ref{fig:t_nbbh_kicks}--\ref{fig:t_nbbh_fbfrac}). 
Since we have shown that the assumed BH natal-kick distribution
can bring dramatic changes to the overall cluster evolution, we first investigate 
the effects of BH formation kicks on the 
evolution of BBHs that are retained in the cluster (Fig.\ \ref{fig:t_nbbh_kicks}). 
The number of BBHs in the cluster 
is quite insensitive to the details of the kick distribution except for 
the case with $\sigmabh=\sigmans$ (e.g., model \modelKone). In the high-kick cases, the large 
formation kicks essentially eject most of the BHs from the cluster during formation. The natal kicks are 
also large enough to disrupt all binaries during BH formation. Hence, not surprisingly, in the 
high-kick models, the values for $\nbhbh$ as well as $\nbhnbh$ are always low. 
Interestingly though, 
the number of BBHs is low even in our lowest 
kick models. For example, \modelS, which assumes a fallback-dependent momentum-conserving kick prescription 
and \modelKthree, which assumes that $\sigmabh=2.65\,\kms$, typically much lower 
compared to orbital speeds of the massive binaries, both show low numbers of BBHs. As discussed earlier, 
the combined 
mass loss from stellar winds and compact object formation for high-mass stars 
expands the binary orbits and make them 
dynamically soft. Thus the majority of the high-mass binaries are disrupted 
independent of the magnitude of the SN kicks.   
%
%
%
To further investigate this we compare two of our models that are identical in all 
aspects except the fraction of high-mass 
stars that are initially in binaries. To illustrate the limiting cases, Fig.\ \ref{fig:t_nbbh_fbhigh} 
shows the evolution of retained BBHs for models \modelFzero\ with initial $\fbhigh=0$ 
and \modelFone\ with initial $\fbhigh=1$ (\S\ref{S:fbhigh}; Table\ 2). Although initially 
the values of $\nbhbh$ are vastly different between the models, within about $3\,\gyr$, they converge 
to essentially the same steady value in both models. This further highlights that the number and properties 
of BH-BH binaries that would be retained in a cluster at late times are set by the internal dynamics and 
overall cluster properties, and {\em not} on the details of the initial binary orbital properties, or binary fraction in 
high-mass stars. 
The number of BH-nBH binaries, $\nbhnbh$, for the model with 
initial $\fbhigh=1$ is slightly lower compared to that in the model with initial $\fbhigh=0$ for $t>5\,\gyr$. 
This is due to the fact that to keep the overall $f_b$ fixed at $0.05$, the $\fbhigh=1$ case 
started with fewer low-mass stars in binaries compared to the $\fbhigh=0$ case.
At late times, all BH--nBH binaries are 
dynamically formed via exchange interactions where a BH inserts itself into a
binary by ejecting a non-BH component. Because of this, the relatively small number of low-mass 
stellar binaries in the $\fbhigh=1$ case (compared to $\fbhigh=0$ case) reduces 
the efficiency of new BH--nBH binary formation. 

Independent of how many high-mass stars were born in binaries, cluster dynamics regulates 
the total number of BBHs retained in old GC-like clusters. Other variations in the 
binary orbital properties and $\fbhigh$ \citep[e.g.,][]{2012Sci...337..444S} show the same general result. 
All knowledge of the primordial binary fraction and binary orbital properties of BHs or their progenitors 
is lost due to dynamical encounters in a star cluster. If massive stars were initially mostly in binary systems, 
these binaries would take part in strong dynamical encounters with other stars. These frequent scattering encounters either disrupt 
the binaries or eject them from the cluster. On the other hand, if initially BHs or their progenitors are not in binaries, new binaries 
can form dynamically. Through both channels the final number of BBHs
attains small but similar values in all GCs (e.g., Table\ 4).

%
%
%
%

%
%
%

Similar trends are found by comparing models
with differing prescriptions for 
stellar winds (Fig.\ \ref{fig:t_nbbh_winds}). While the values of $\nbhbh$ are very similar between clusters 
\modelS\ and \modelW, the lower final $\rho_c$ in \modelW\ results in production of a lower number of 
BH--nBH binaries compared to 
\modelS\ at late times. This is consistent with our understanding that most BH--nBH binaries retained in 
old clusters typical of the GCs are dynamically created. 
The dynamical age, mass-segregation timescale, and the overall binary fraction can also moderately affect 
$\nbhnbh$. The first two control $\nbh$ in the cluster, and as a result the cluster's dynamical properties 
and overall evolution. The initial binary fraction controls the number of binaries with 
no BH components that can take part in exchange 
interactions with single BHs and BH--BH binaries in old clusters. Fig.\ \ref{fig:t_nbbh_fbfrac} compares 
$\nbhbh$ and $\nbhnbh$ for two models \modelW\ and \modelWifb\ that differ in their initial $r_v$, and $f_b$ (Table\ 2). 
While $\nbhbh$ converges 
to similar values within $200\,\myr$, $\modelWifb$ with a higher $f_b$ creates and retains a larger number of BH--nBH binaries 
compared to \modelW. 
In all cases, the fraction of BHs that are in binaries increases only after $\nbh$ is sufficiently reduced 
(e.g., Fig.\ \ref{fig:nbh_nbbh}; also Table\ 4).


%
%
%
\subsection{Black hole binaries ejected from clusters}
\label{S:results_bhbinaries_ejected}
%
%
%
%
\begin{figure*}
\begin{center}
\plotone{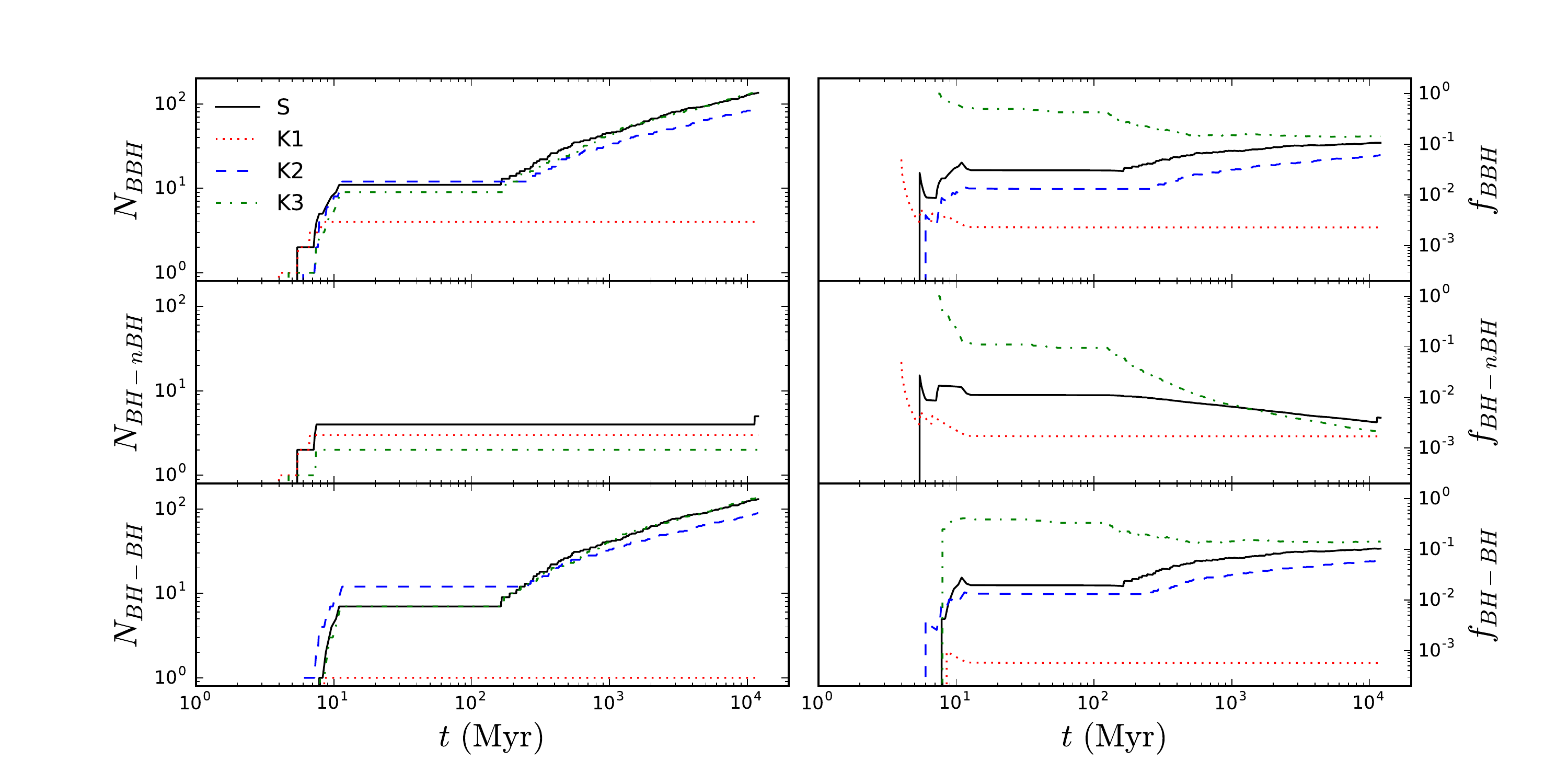}
\caption{
Evolution of the number (left) and fraction (right) of ejected binary BHs. Top to bottom, panels show 
evolution of all binary BHs, BH-nBH binaries, and BH-BH binaries ejected from the cluster vs the cluster 
age. Black (solid), red (dotted), blue (dashed), and green (dash-dot) denote models {\tt S}, {\tt K1}, {\tt K2}, 
and {\tt K3} (Table\ 2; \S\ref{S:numerical}), respectively. The number of ejected BH-nBH binary stays 
low in all cases. No BH-nBH binaries were ejected in {\tt K2} likely due to statistical fluctuations.  
}
\label{fig:ejected_bbh_kicks}
\end{center}
\end{figure*} 
Figure\ \ref{fig:ejected_bbh_kicks} shows the evolution of the number and fraction of BH binaries 
ejected from model clusters {\tt S}, {\tt K1}, {\tt K2}, and {\tt K3}. There are some noticeable trends 
in these models which are actually common to all our models. In model {\tt K1}, where we apply high natal kicks to BHs, ejected binary 
BHs of all types, BH-BH or BH-nBH, have very low values. High natal kicks not only disrupt binaries during 
BH formation, these high kicks also eject most of the BHs from the cluster during formation 
(e.g., Figure\ \ref{fig:kick_comp_clusprop}), hence, they never get a chance to dynamically acquire companions. 
As a result, most of the ejected BHs are singles. Also, the handful 
of ejected BH binaries leave the cluster at very early times, essentially with the rest of the BHs due to natal kicks. 
In contrast, in the other models with either fallback-dependent natal kicks (e.g., {\tt S}) or scaled-down natal 
kicks for BHs (e.g., {\tt K2}, {\tt K3}), a significant number of BHs are retained even after all BHs are formed. In these cases, there are 
three distinct evolutionary stages. The first sharp rise between few to $\sim 10\,\myr$ in the number 
of ejected BH binaries is coincident with BH formation. These are essentially BH binaries that are ejected 
from the cluster primarily due to formation kicks. 
By this time all BHs are already formed. Mass loss from stellar evolution, compact object formation and ejection 
due to natal kicks have already taken place resulting in significant expansion of the cluster (e.g., Fig.\ \ref{fig:rcrh_s}). 
This stage is followed by a flat part, between $\sim 10$ and $200\,\myr$. 
During this time the most massive BHs that are retained in the cluster are 
in the process of mass segregation. As a result, few BHs, single or binary, 
are ejected between $\sim10$--$200\,\myr$. A corresponding flat part can also be seen in the evolution of 
$\nbh$ (e.g., Figure\ \ref{fig:kick_comp_clusprop}). By $\sim 200\,\myr$ the heaviest retained BHs are mass segregated 
and the BH-driven core-collapse episodes start. During this stage BHs are ejected from the cluster via strong 
dynamical encounters. BH binaries are also ejected steadily. BH-BH binaries dominate all ejected BBHs 
in this stage. Most of the BH-nBH binaries are ejected at the early stages primarily 
due to BH natal kicks. This is an expected consequence of BH mass segregation. BHs dominate the population in the 
central part where most of the strong binary-mediated encounters take place. Thus, single BHs can easily form new BH-BH 
binaries. Recoils can eject them from the cluster. In contrast, while a significant number of BHs are retained 
in the cluster, it is hard to dynamically form or eject a BH-nBH binary simply because the BHs and binaries with non-BH members 
do not interact as often or as energetically. Thus BH-nBH binaries are not created in large numbers, and when they do form, 
they are typically retained in the cluster. 

%
%
%
\begin{figure*}
\begin{center}
\plotone{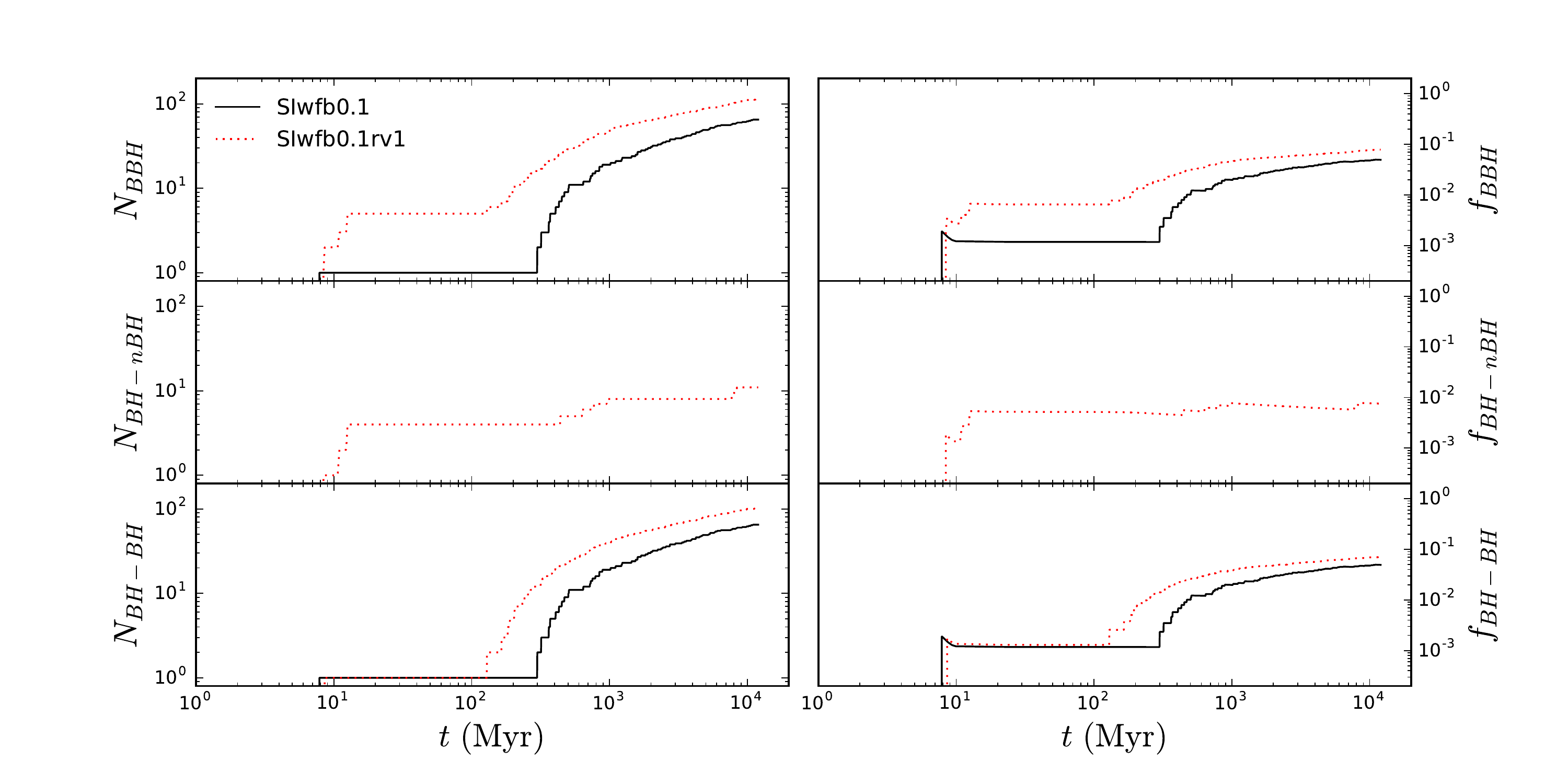}
\caption{
Same as Figure\ \ref{fig:ejected_bbh_kicks}, but for models {\tt SIwfb0.1} (black solid) and {\tt SIwfb0.1rv1} (red dotted) illustrating 
differences resulting from the difference in the initial virial radius of the cluster. The qualitative stages in the evolution of the number 
and fraction of ejected binary BHs are very similar to those shown in Figure\ \ref{fig:ejected_bbh_kicks}. Dynamical ejection of BH-BH 
binaries start earlier in the initially more compact cluster ({\tt SIWfb0.1rv1}) as expected.  
}
\label{fig:ejected_bbh_rvcomp}
\end{center}
\end{figure*} 
Interestingly, the final ejected binary fraction attains very similar values and does not seem to depend on the natal kick 
distribution, provided that the cluster retains significant numbers of BHs after formation and the BHs get a chance to 
dynamically evolve inside the cluster for a significant time (Figure\ \ref{fig:ejected_bbh_kicks}). 
Figure\ \ref{fig:ejected_bbh_rvcomp} shows 
a similar figure, but comparing models {\tt SIwfb0.1} and {\tt SIwfb0.1rv1} (Tables\ 2, 3). 
Note that in these models the IMF range is 
different from model {\modelS}. The qualitative stages for the evolution of the number and fraction of BH binaries are very 
similar. In addition, the stage of cluster dynamics driven BH binary ejection starts earlier in the model that is initially 
more compact ($r_v=1\,\pc$) with respect to the other ($r_v=2\,\pc$). Note that the number of BH binaries ejected 
via cluster dynamics is about an order of magnitude higher compared to those ejected early due to natal kicks in all cases 
(Figures\ \ref{fig:ejected_bbh_kicks}, \ref{fig:ejected_bbh_rvcomp}). This is true in all of our models except in the high-kick 
cases (models denoted by {\tt K1} in their names; Tables\ 1, 2), where few BHs remain bound to the 
cluster after their formation. 

\begin{figure}
\begin{center}
\plotone{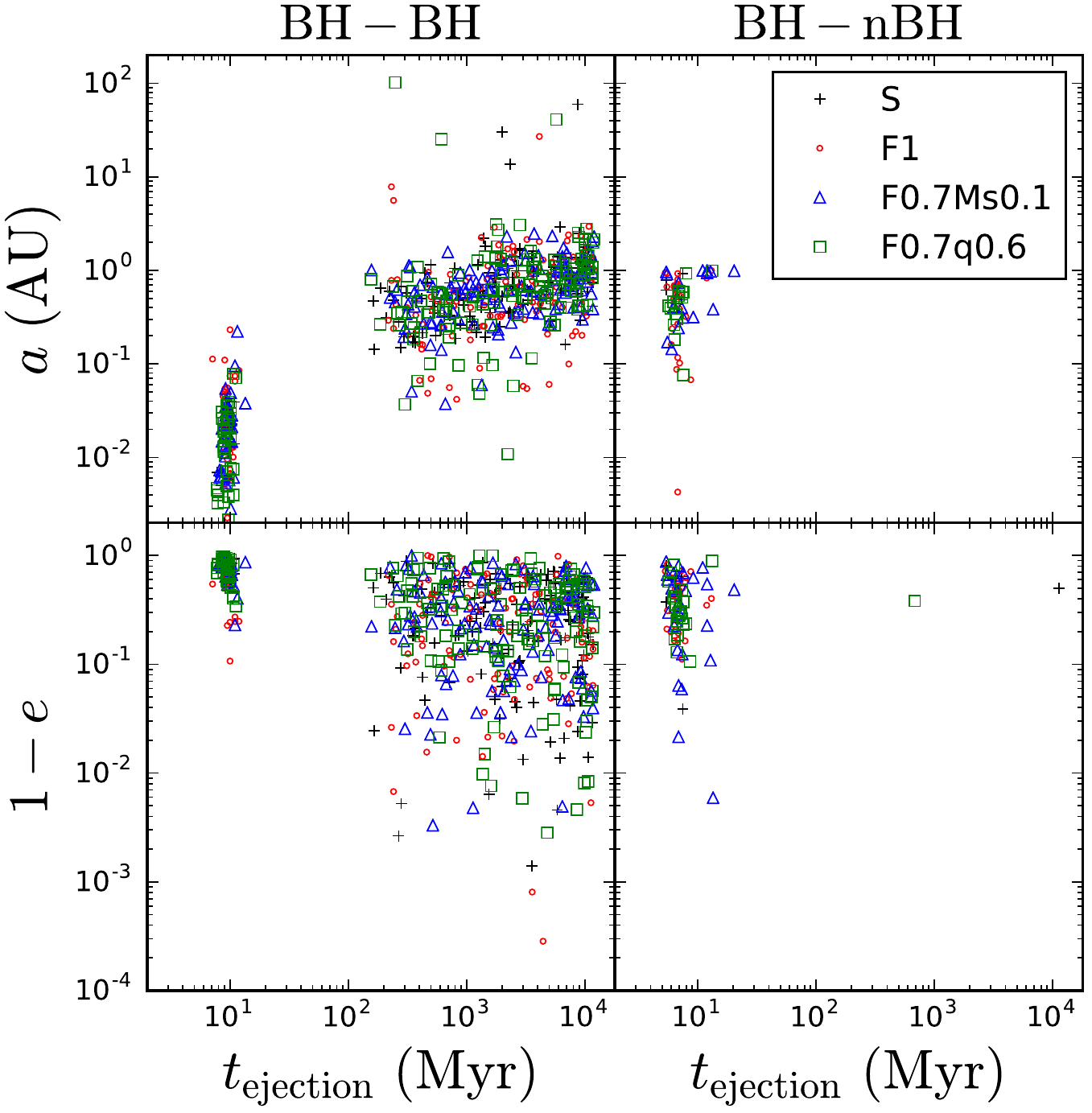}
\caption{
Semimajor axis (top) and eccentricity (bottom) at the time of ejection as a function of the ejection time 
for ejected binary BHs from models \modelS (black plus), \modelFone (red circle), \modelFDMs (blue triangle), 
and \modelFDq (green square; Tables\ 1, 2). 
Left and right panels are for ejected BH-BH and BH-nBH binaries, respectively. Independent of the initial assumptions for high-mass 
binary fraction ($\fbhigh$), and the binary orbital properties, the ejected binary BHs, especially those ejected at late times due to 
stellar dynamics, have very similar orbital properties. This bolsters the claim that the orbital properties of ejected binary BHs are set 
by the dynamical processes inside the cluster center and not by the initial assumptions of orbital properties of the BH-progenitor stars. Since 
stellar dynamics created these BH binaries, their eccentricities are also typically high. 
}
\label{fig:ejected_bbh_aet}
\end{center}
\end{figure}

Since almost all BH binaries are created via dynamical processes, their orbital properties are also 
set by these dynamical processes, and not the initial assumptions for the binarity or the binary orbital properties of high-mass stars. 
Figure\ \ref{fig:ejected_bbh_aet} compares the semimajor axes and eccentricities of 
ejected BH binaries from four different models, namely \modelS, \modelFone, \modelFDMs, and \modelFDq, with widely different assumptions for the initial binary fraction of the high-mass stars, 
and the initial distributions of their orbital properties (Table\ 2). Independent of these initial assumptions, the ejected 
BH binaries, attain very similar orbital properties at the time of ejection from the respective clusters. The general trend for the 
dynamically ejected BH-BH binaries is an increase of the semimajor axes with time (Figure\ \ref{fig:ejected_bbh_aet}). This is 
imprinted from the chaotic and repeated dynamical encounters these binaries experience prior to ejection. Once a hard 
binary forms, subsequent dynamical encounters typically make it harder transferring the excess energy to the center-of-mass 
velocities of the interacting stars, popularly known as ``Heggie's Law" \citep{1975MNRAS.173..729H}. 
Since the recoil speed is set by the binary orbital energy, the recoil speed of the increasingly hardening binaries 
increase after each such encounter, until it becomes sufficient to eject the binary from the cluster. 
Since recoil speeds are set by the binary orbital energy, the orbital energy and hence the semimajor axes of ejected binaries 
are set by the instantaneous escape speed of the cluster \citep[see ][for a recent in-depth discussion]{2016PhRvD..93h4029R}. 
As the cluster evolves, the escape speed from the cluster's center becomes lower due to the decrease in the cluster mass and 
the overall expansion of the core (e.g., Figures\ \ref{fig:rcrh_s}, \ref{fig:kick_comp_clusprop}, \ref{fig:imf_m_nbh}, \ref{fig:rv_imfspread_comp_clusprop}). Hence, the semimajor axis of typical binaries that are dynamically ejected from the cluster 
center, increases with time. Note that a competing effect could have been in action. The heavier BHs get ejected earlier. Thus, 
at earlier times, for a given energy of ejection (and thus binary orbital energy), the binary orbits could be larger. This latter 
effect does not seem to dominate in our models. 
%
%
%
%

%
%
\section{Binary Black Hole Mergers}
\label{S:results_merger}
\begin{figure*}
\begin{center}
\plotone{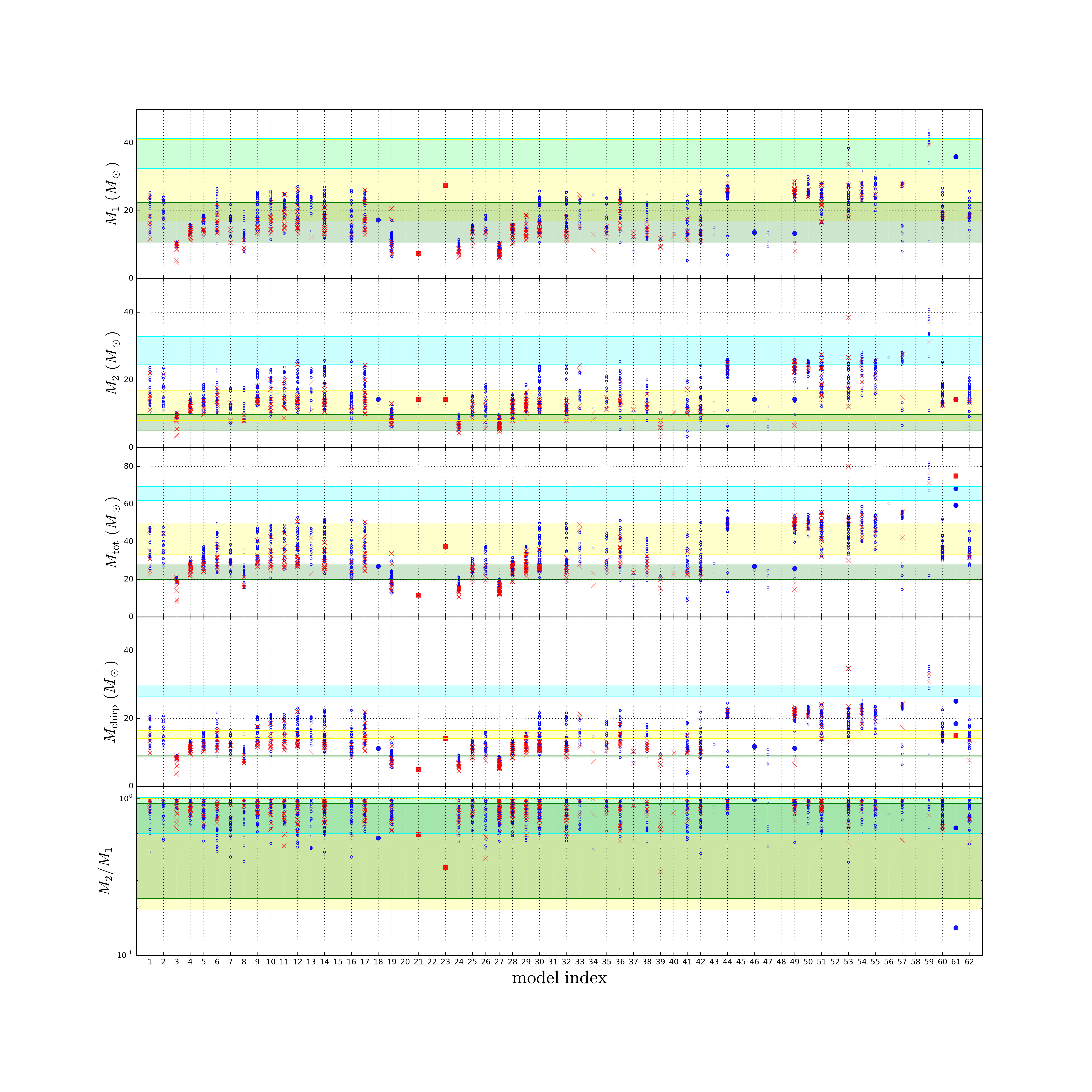}
\caption{
Top to bottom, the component masses ($M_{1,2}$), total mass ($\mtot$), chirp mass ($\mchirp$), and mass ratio ($M_2/M_1$)  
for BH-BH mergers within $0\geq z\geq0.2$ (blue) and $0.2\geq z\geq 1$ (red) for each model identified 
by the serial numbers on the horizontal axis. Crosses and dots denote in-cluster mergers after being ejected from the 
clusters. Circles and squares denote mergers within the clusters. Cyan, yellow, and green shaded regions show GW150914, 
LVT151012, and GW151226 properties for reference \citep[e.g.,][]{2016arXiv160604856T}. 
Weak winds help creating relatively more massive mergers. Assumptions of BH's natal kicks 
and IMF affect number of mergers by changing the number of BHs that clusters can retain for dynamical 
processing, and the number of BHs produced for a given $N$, respectively. 
}
\label{fig:index_bbh_masses}
\end{center}
\end{figure*} 
\begin{figure*}
\begin{center}
\plotone{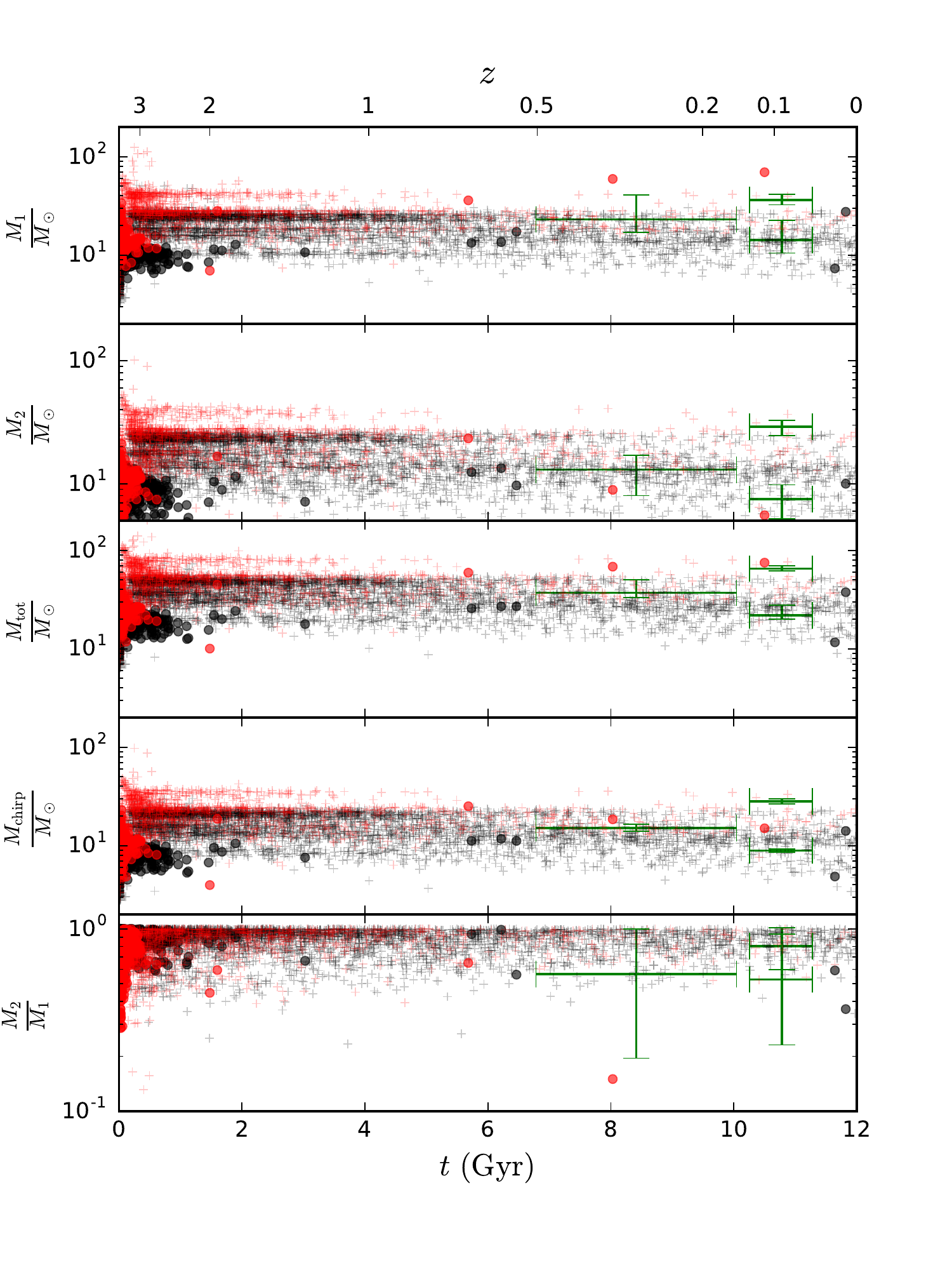}
\caption{
Top to bottom, the component masses ($M_{1,2}$), total mass ($\mtot$), chirp mass ($\mchirp$), and 
mass ratio ($M_2/M_1$) for BH-BH mergers as a function of merger time. 
The top axis shows the lookback redshift $z$ assuming 
that $t=12\,\gyr$ is equivalent to $z=0$. Plusses, and circles denote mergers that happen after the BH--BH binaries are ejected 
from the host clusters, and BH--BH mergers that happen while the binaries are still bound to the host clusters. 
The green errorbars show the detected events GW150914, LVT151012, and GW151226 \citep[e.g.,][]{2016arXiv160604856T}.  
Black and red denote BH--BH mergers from models assuming the 
strong and weak wind prescriptions, respectively. Most BH--BH mergers in the local universe happen after the binaries are ejected 
from the host clusters. BH--BH binaries that merge inside clusters typically have lower masses 
compared to those that merge after being ejected. Weak winds are necessary for producing BH--BH mergers 
as massive as GW150914 for the metallicities we consider. Most mergers in our models are closer in properties to LVT151012. 
}
\label{fig:t_bbh_masses}
\end{center}
\end{figure*} 

The BH--BH merger rates and properties of the BHs during merger 
in the local universe have been studied by multiple teams, either 
exploring the contributions from star clusters \citep[e.g.,][]{2010MNRAS.402..371B,2014MNRAS.441.3703Z,2015PhRvL.115e1101R,2016PhRvD..93h4029R,2016ApJ...824L...8R,2016ApJ...816...65A,2016MNRAS.458.1450W,2016arXiv160906689C} 
or from isolated binary stellar evolution in the field \citep[e.g.,][]{2014ApJ...789..120B,2012ApJ...759...52D,2013ApJ...779...72D,2014ApJ...789..120B,2015ApJ...806..263D,2015A&A...574A..58K,2015arXiv151004615B}. 
Here we focus on understanding the uncertainties associated with the BH--BH merger rates
formed in clusters similar to the present-day GCs \citep[e.g.,][]{2016PhRvD..93h4029R}. 
We estimate BH--BH merger times using the quadrupole approximate GW orbital evolution 
equations \citep{1964PhRv..136.1224P}. These equations are applied within BSE for tracking 
mergers of BH--BH binaries retained in the clusters. We use the same equations to track mergers of the ejected 
BH--BH binaries using their properties at the time of ejection. We ignore the ejected non-BH binaries at early times 
which might evolve into BH--BH binaries post ejection since such systems are extremely rare. 
As in earlier sections, we investigate how variations in initial assumptions 
affect the number of BH--BH mergers, merger times, and the properties of the binaries. 
The numbers of BH--BH mergers and the total masses of these mergers 
within two values of redshift ($z$) are listed in Table\ 4. Here we assume that each cluster is 12 Gyr old
at the present day, and the redshift correponds to the lookback time of each
merger.\footnote{We assume $h=0.677$, $\Omega_M=0.309$, $\Omega_K=0.$, and $\Omega_L=0.691$.} 
To highlight the assumption relating $t$ and $z$, we will refer to our model clusters as GCs in this context. 
For a detailed analysis of how the assumption of a fixed formation time for all clusters 
affect the BH--BH mergers within a given redshift see \citet{2016arXiv160906689C}. 

Fig.\ \ref{fig:index_bbh_masses} shows the BH masses for all BH--BH mergers within 
$z=0.2$ and $1$ from all of our models, identified by the index numbers listed in Tables\ 2--4. 
\citet{2015PhRvL.115e1101R,2016PhRvD..93h4029R} found that the BH--BH mergers 
in the local universe ($z<1$) is dominated by mergers that occur {\em after} the
BH--BH binaries are ejected from their host GCs. This same result is generally found in our models independent of the 
details of the initial assumptions (Table\ 4). Now we highlight the assumptions that 
affect the number and properties of BH--BH mergers originating from GCs and occurring within some small $z$. 

The largest effect arises from the assumed
distribution of natal kicks for the BHs. 
We have already seen that independent of initial assumptions, primordial binaries involving 
BH-progenitor stars are generally disrupted (\S\ref{S:results_bhbinaries}). 
All GCs modeled with an adopted BH natal kick distribution given by $\sigmabh=\sigmans$ and no scaling based on
fallback, eject almost all of their BHs due to the high natal kicks before they can dynamically form
binaries. 
The handful of BH--BH binaries these GCs 
produce are all dynamically formed via binary-mediated strong scattering encounters within clusters at 
later times. As a result, when these GCs do produce BH--BH binaries that would merge in the local universe 
the mergers usually happen while the BH--BH binary is still bound to the GC. 

The only other assumption explored in this study that can significantly impact the 
production of merging BH--BH binaries from GCs is the IMF. 
Any variations in
the IMF for high-mass stars alters the number of mergers at $z<1$ in a
predictable manner.  IMFs that produce more high-mass stars, produce more BHs,
and the number of BH--BH mergers in the local universe increases accordingly. 
For example, the IMF with $\alpha_1=1.6$, for a given initial $N$ forms much higher numbers of BHs compared 
to models with larger $\alpha_1$. Even models with the largest natal kicks for the BHs 
may produce significant numbers of BH--BH mergers if $\alpha_1=1.6$ is assumed. 
Of course, for the same reason, a combination of flat high-end for the IMF and 
low natal kicks for BHs can significantly increase the number of mergers within $z<1$.  
Nevertheless, note that we find that all clusters modeled with $\alpha_1=1.6$ get disrupted long before 
they are $12\,\gyr$ old, a typical age for the MW GCs, due to a combination of mass loss 
and energy production via BH dynamics from the large numbers of high-mass stars that such 
an IMF produces. As a result, this extreme low value of $\alpha_1$ seems very unlikely in reality. 
Similarly, if the IMF mass range is extended to lower-mass stars, the number of BHs produced, and 
as a result, the number of BH--BH mergers is reduced. 

While the distribution of natal kicks and the assumed IMF may change the actual number of BH--BH mergers 
from GCs within a given $z$, the masses of the binary mergers depend on neither. 
Instead, the mass distribution of the merging BHs is primarily
determined by 
the assumed stellar winds for a given metallicity $Z$, which we controlled by
using either the weak or strong wind prescriptions in this study (\S\ref{S:wind}).
Fig.\ \ref{fig:t_bbh_masses} 
shows the component masses ($M_{1,2}$), total mass ($\mtot$), chirp mass ($\mchirp$), and the 
mass ratios ($M_2/M_1$) of all BH--BH mergers within $12\,\gyr$ from all our models 
as a function of their merger times. We divide all model GCs in two groups 
based on the two wind prescriptions we consider. 
GCs modeled with weak winds produce 
more massive BH--BH binaries that would merge within a Hubble time compared to those modeled with strong winds. In particular, 
GCs modeled assuming weak stellar winds are the only ones that can produce BH--BH mergers with masses 
similar to GW150914 for the metallicities we consider. 
Nevertheless, we find that the majority of BH--BH binaries merging within $z=0.2$ even from 
model GCs assuming weak winds and a broader IMF have typically lower BH masses than 
the component masses of GW150914 (Table\ 2). 
Using an even broader IMF would not increase the typical intrinsic mass distributions for BH--BH binaries merging in 
the local universe for the metallicity we consider. While high progenitor mass, low metallicities, and weak winds 
help enhance production of high-mass BH--BH binaries, the majority of them merge at high redshifts 
(Fig.\ \ref{fig:t_bbh_masses}) assuming that the parent clusters had formed $12\,\gyr$ ago. Most 
recently \citet{2016arXiv160906689C} find that GW150914's $\mchirp$ is within $2\sigma$ of 
the {\em intrinsic} distributions for $\mchirp$ of BH--BH binaries formed in clusters with metallicities lower than 
$0.05Z_\odot$ and merging within $z=0.2$. Nevertheless, detection of BH--BH mergers 
as massive as GW150914 may not be as uncommon due to selection effects. For example, \citet{2016ApJ...824L...8R} find 
that accounting for selection effects, $\mchirp$ of GW150914 is within $1\sigma$ of detected BH--BH binaries 
merging within a redshift of 0.5. 

The BH masses of LVT151012 are more common in the BH--BH mergers found in our models. 
GCs modeled with both strong and weak winds can produce BH--BH mergers similar in properties to LVT151012. 
The $\mchirp$ of GW151226 is within the range of masses for BH--BH binaries merging within a redshift of 
$0.2$ from models with strong wind, but is a bit lower compared to those from models with weak winds. Nevertheless, 
note that many of these conclusions depend on the assumption of the formation time and metallicity 
of the parent clusters \citep{2016arXiv160906689C}. 
 
Very few ($\sim1\%$) of the BH--BH mergers in the local universe are expected to happen while the binary is 
still bound to the host GC. 
Most BH--BH binaries that merge while bound to the GC do so at early times, 
and typically have lower masses than those that merge after ejection. Since, the higher the 
mass of the BHs, the earlier it is ejected from the cluster \citep[e.g.,][]{2015ApJ...800....9M}, 
the mergers after ejection are typically more massive compared to those 
that happen within a cluster. 

%
\section{Summary and Discussion}
\label{S:discussion}
In this study we have explored how identical initial clusters can diverge in their evolution depending on various
ill-constrained initial assumptions affecting the high-mass stars, including
the high-end slope ($\alpha_1$) of the IMF, the
distribution of natal kicks for BHs, the binary fraction and the binary orbital properties for massive stars, and
the prescriptions adopted for stellar winds. Using Monte Carlo simulations we have studied the effects of each of these assumptions
from three different perspectives: (1) how they affect the global evolution of star clusters and their
final
observable properties; (2) how they affect the binary fraction of BHs and the numbers of different
types of BH binaries; and (3) how these assumptions affect the number and properties of merging BH--BH binaries that could be
detected as GW sources.

We find that, even when we start from the same initial cluster model, variations in
these initial assumptions can alter the number of retained BHs, $\nbh$, at any given time
by orders of magnitude (e.g., Tables\ 3, 4). This difference in
$\nbh$ can then change the overall dynamical evolution of the host star
cluster, dramatically altering its lifetime and, if it survives to the present, its final observable
properties (Fig.s\ \ref{fig:nbh_rcobs}--\ref{fig:rv_imfspread_comp_clusprop}).

In contrast, the total number and properties of BH--BH and BH--nBH binaries retained inside old star clusters
are not significantly affected by varying any of these assumptions. Through dynamical interactions in the dense
cluster cores or kicks at compact object formation, all primordial binaries containing
BH progenitors are quickly disrupted. On the relaxation timescale, binaries are reformed dynamically and eventually both
the number and properties of retained binaries with at least one BH component
are determined entirely by stellar dynamical processes. As $\nbh$ decreases, the fraction of BHs in binaries increases, so that
$\nbhnbh$ and $\nbhbh$ in old clusters do not strongly correlate with $\nbh$.
This has important implications for inferring $\nbh$ in clusters where BH XRB candidates have been identified.
In general we find that observable cluster properties
including $\rcobs$ and $\sigmacobs$ are better indicators of $\nbh$ compared to $\nbhnbh$ 
(Figs.\ \ref{fig:nbh_rcobs}, \ref{fig:nbh_rhocobs}, \& \ref{fig:nbh_nbbh}).

The number of BH--BH mergers from clusters in the local universe is only affected by
the assumptions made about BH natal kicks and the stellar IMF (Fig.\ \ref{fig:index_bbh_masses}, Table\ 4).
This is very different from the situation encountered in binary population synthesis studies for field binaries, where a myriad of
assumptions (e.g., about the initial properties of binaries) and uncertain stellar evolution parameters 
(e.g., the infamous common-envelope efficiency $\alpha_{\rm CE}$) can change results by orders of magnitude.
Instead, far more robust theoretical predictions can be made for merging BH--BH binaries that are dynamically
produced in dense star clusters.

If BHs receive natal kicks as large as
neutron stars formed in core-collapsed SN, the expected number of BH--BH mergers becomes very small, simply
because almost all BHs are ejected from the cluster at birth. Note, however, that with such large kicks the number
of BH--BH binaries generated in star clusters is significantly higher than in the
field, where BHs can never again acquire a binary companion if natal kicks disrupt the primordial binaries they were born in
\citep[see][for comparative rate estimates between star clusters and the field assuming high and low natal kicks for BHs]{2016PhRvD..93h4029R}.
Variations in the assumed IMF slope change the number of BH--BH mergers in clusters
in a predictable way, simply by changing the relative proportion of high-mass stars (see \S\ref{S:results_merger}, Fig.\ \ref{fig:index_bbh_masses}).
The component masses and total masses (or equivalently the chirp masses) of merging
BH--BH binaries are affected only by the assumed prescription for stellar-wind mass loss. Other parameters such as
$\fbhigh$ or any of the initial properties of massive binaries (e.g., the distributions of mass ratios and orbital periods)
have no effect at all (Fig.\ \ref{fig:index_bbh_masses}, Table\ 4).

In the context of GW detection, we therefore expect the BH--BH merger rates and properties estimated in 
\citet[][]{2015PhRvL.115e1101R,2016PhRvD..93h4029R}
for dynamically produced binaries to be quite robust, and in particular to remain unaffected by any change in
assumptions concerning the initial distributions of binary properties or the binary stellar evolution.
In contrast, the BH--BH merger rates estimated for isolated field binaries
\citep[e.g.,][]{2015arXiv151004615B,2013ApJ...779...72D,2015ApJ...806..263D}
depend directly and sensitively on many assumptions concerning the details of stellar evolution (mass transfer,
tidal dissipation effects, common envelopes) and on the assumed distributions of initial
properties for high-mass binaries. In addition, the assumption of truly isolated binary evolution
may be suspect because most stars and binaries currently in the field were still originally formed 
in clusters, which later dissolved \citep[e.g.,][]{2000AJ....120.3139C,2003ARA&A..41...57L,2003AJ....126.1916P,2010ARA&A..48..431P}, 
and binaries may have been affected by dynamical interactions before their release into the field \citep[e.g.,][]{2014MNRAS.441.3703Z}.
%

%
%
\acknowledgments We are grateful to the anonymous referee for constructive comments. We are also grateful to 
Vicky Kalogera for her helpful comments. This work was supported by 
NASA ATP Grant NNX14AP92G and NSF Grant AST1312945. S.C. also acknowledges support from the 
National Aeronautics and Space Administration through Chandra Award Number TM5-16004X/NAS8-03060 
issued by the Chandra X-ray Observatory Center, which is operated by the Smithsonian Astrophysical 
Observatory for and on behalf of the National Aeronautics Space Administration under contract NAS8-03060.

\clearpage
\begin{deluxetable*}{llccccc|cc|cccc|ccccc}
\tabletypesize{\footnotesize}
\tablecolumns{15}
\tablewidth{0pt}
\tablecaption{Initial model parameters.}
\tablehead{
	  \colhead{No.} &
	  \colhead{Name} &
           \colhead{$M$} &
           \colhead{$\rgc$} &
           \colhead{$f_b$} & 
           \colhead{$Z$} &
           \colhead{$r_v$} &
           \multicolumn{2}{c}{BH-formation kick} & 
           \multicolumn{3}{c}{High-mass Binaries} & 
           \multicolumn{2}{c}{IMF}\\
           \cline{8-9}
           \cline{9-12}
           \cline{13-14}
           \colhead{} &
           \colhead{} &
           \colhead{($10^5\,\msun$)} & 
           \colhead{(kpc)} &
           \colhead{} &
           \colhead{} &
           \colhead{(pc)} &
           \colhead{$\frac{\sigma_{\rm{BH}}}{\sigma_{\rm{NS}}}$} & 
           \colhead{{\tt FB}} & 
           \colhead{$f_{b, \rm{high}}$} & 
           \colhead{$q$ range} &
           \colhead{$\frac{dn}{d\log P}$} & 
           \colhead{range} & 
           \colhead{$\alpha_1$} \\ 
}
\startdata
1 & \modelS & $5.2$ & $8$ & $0.05$ & $0.001$ & $2$ & $1$ & y & $0.05$ & $[0.1/m_p, 1]$ & $P^0$ & [0.1, 100] & $2.3$ \\
2 & \modelSRoneZ & $5.2$ & $8$ & $0.05$ & $0.001$ & $2$ & $1$ & y & $0.05$ & $[0.1/m_p, 1]$ & $P^0$ & [0.1, 100] & $2.3$ \\
3 & \modelSRone & $5.2$ & $1$ & $0.05$ & $0.02$ & $2$ & $1$ & y & $0.05$ & $[0.1/m_p, 1]$ & $P^0$ & [0.1, 100] & $2.3$ \\
4 & \modelSRtwo & $5.2$ & $2$ & $0.05$ & $0.007$ & $2$ & $1$ & y & $0.05$ & $[0.1/m_p, 1]$ & $P^0$ & [0.1, 100] & $2.3$ \\
5 & \modelSRthree & $5.2$ & $4$ & $0.05$ & $0.003$ & $2$ & $1$ & y & $0.05$ & $[0.1/m_p, 1]$ & $P^0$ & [0.1, 100] & $2.3$ \\
6 & \modelSRfour & $5.2$ & $20$ & $0.05$ & $0.0003$ & $2$ & $1$ & y & $0.05$ & $[0.1/m_p, 1]$ & $P^0$ & [0.1, 100] & $2.3$ \\
7 & \modelSifb & $4.9$ & $8$ & $0.1$ & $0.001$ & $2$ & $1$ & y & $0.05$ & $[0.1/m_p, 1]$ & $P^0$ & [0.08, 150] & $2.3$ \\
8 & \modelSifbrvone & $4.9$ & $8$ & $0.1$ & $0.001$ & $1$ & $1$ & y & $0.05$ & $[0.1/m_p, 1]$ & $P^0$ & [0.08, 150] & $2.3$ \\
\hline
9 & \modelFzero & $5.2$ & $8$ & $0.05$ & $0.001$ & $2$ & $1$ & y & $0$ & - & - & [0.1, 100] & $2.3$ \\
10 & \modelFone & $5.6$ & $8$ & $0.05$ & $0.001$ & $2$ & $1$ & y & $1$ & $[0.1/m_p, 1]$ & $P^0$ & [0.1, 100] & $2.3$ \\
11 & \modelFDMs & $5.5$ & $8$ & $0.05$ & $0.001$ & $2$ & $1$ & y & 0.7 & $[0.1/m_p, 1]$ & $P^{-0.55}$ & [0.1, 100] & $2.3$ \\
12 & \modelFDq & $5.6$ & $8$ & $0.05$ & $0.001$ & $2$ & $1$ & y & 0.7 & $[0.6, 1]$ & $P^{-0.55}$ & [0.1, 100] & $2.3$ \\
13 & \modelFDqRone & $5.6$ & $1$ & $0.05$ & $0.001$ & $2$ & $1$ & y & 0.7 & $[0.6, 1]$ & $P^{-0.55}$ & [0.1, 100] & $2.3$ \\
14 & \modelFDqRtwo & $5.6$ & $3$ & $0.05$ & $0.001$ & $2$ & $1$ & y & 0.7 & $[0.6, 1]$ & $P^{-0.55}$ & [0.1, 100] & $2.3$ \\
\hline
15 & \modelKone & $5.2$ & $8$ & $0.05$ & $0.001$ & $2$ & $1$ & n & $0.05$ & $[0.1/m_p, 1]$ & $P^0$ & [0.1, 100] & $2.3$ \\
16 & \modelKtwo & $5.2$ & $8$ & $0.05$ & $0.001$ & $2$ & $0.1$ & n & $0.05$ & $[0.1/m_p, 1]$ & $P^0$ & [0.1, 100] & $2.3$ \\
17 & \modelKthree & $5.2$ & $8$ & $0.05$ & $0.001$ & $2$ & $0.01$ & n & $0.05$ & $[0.1/m_p, 1]$ & $P^0$ & [0.1, 100] & $2.3$ \\
18 & \modelSifbrvoneKone & $4.9$ & $8$ & $0.1$ & $0.001$ & $1$ & $1$ & n & $0.05$ & $[0.1/m_p, 1]$ & $P^0$ & [0.08, 150] & $2.3$ \\
19 & \modelSifbrvoneKthree & $4.9$ & $8$ & $0.1$ & $0.001$ & $1$ & $0.01$ & n & $0.05$ & $[0.1/m_p, 1]$ & $P^0$ & [0.08, 150] & $2.3$ \\
20 & \modelKoneRone & $5.2$ & $1$ & $0.05$ & $0.02$ & $2$ & $1$ & n & $0.05$ & $[0.1/m_p, 1]$ & $P^0$ & [0.1, 100] & $2.3$ \\
21 & \modelKoneRtwo & $5.2$ & $2$ & $0.05$ & $0.007$ & $2$ & $1$ & n & $0.05$ & $[0.1/m_p, 1]$ & $P^0$ & [0.1, 100] & $2.3$ \\
22 & \modelKoneRthree & $5.2$ & $4$ & $0.05$ & $0.003$ & $2$ & $1$ & n & $0.05$ & $[0.1/m_p, 1]$ & $P^0$ & [0.1, 100] & $2.3$ \\
23 & \modelKoneRfour & $5.2$ & $20$ & $0.05$ & $0.0003$ & $2$ & $1$ & n & $0.05$ & $[0.1/m_p, 1]$ & $P^0$ & [0.1, 100] & $2.3$ \\
24 & \modelKtwoRone & $5.2$ & $1$ & $0.05$ & $0.02$ & $2$ & $0.1$ & n & $0.05$ & $[0.1/m_p, 1]$ & $P^0$ & [0.1, 100] & $2.3$ \\
25 & \modelKtwoRtwo & $5.2$ & $2$ & $0.05$ & $0.007$ & $2$ & $0.1$ & n & $0.05$ & $[0.1/m_p, 1]$ & $P^0$ & [0.1, 100] & $2.3$ \\
26 & \modelKtwoRthree & $5.2$ & $4$ & $0.05$ & $0.003$ & $2$ & $0.1$ & n & $0.05$ & $[0.1/m_p, 1]$ & $P^0$ & [0.1, 100] & $2.3$ \\
27 & \modelKthreeRone & $5.2$ & $1$ & $0.05$ & $0.02$ & $2$ & $0.01$ & n & $0.05$ & $[0.1/m_p, 1]$ & $P^0$ & [0.1, 100] & $2.3$ \\
28 & \modelKthreeRtwo & $5.2$ & $2$ & $0.05$ & $0.007$ & $2$ & $0.01$ & n & $0.05$ & $[0.1/m_p, 1]$ & $P^0$ & [0.1, 100] & $2.3$ \\
29 & \modelKthreeRthree & $5.2$ & $4$ & $0.05$ & $0.003$ & $2$ & $0.01$ & n & $0.05$ & $[0.1/m_p, 1]$ & $P^0$ & [0.1, 100] & $2.3$ \\
30 & \modelKthreeRfour & $5.2$ & $20$ & $0.05$ & $0.0003$ & $2$ & $0.01$ & n & $0.05$ & $[0.1/m_p, 1]$ & $P^0$ & [0.1, 100] & $2.3$ \\
31 & \modelKoneRoneZ & $5.2$ & $1$ & $0.05$ & $0.001$ & $2$ & $1$ & n & $0.05$ & $[0.1/m_p, 1]$ & $P^0$ & [0.1, 100] & $2.3$ \\
32 & \modelKtwoRoneZ & $5.2$ & $1$ & $0.05$ & $0.001$ & $2$ & $0.1$ & n & $0.05$ & $[0.1/m_p, 1]$ & $P^0$ & [0.1, 100] & $2.3$ \\
33 & \modelKthreeRoneZ & $5.2$ & $1$ & $0.05$ & $0.001$ & $2$ & $0.01$ & n & $0.05$ & $[0.1/m_p, 1]$ & $P^0$ & [0.1, 100] & $2.3$ \\
\hline
34 & \modelFDMsKone & $5.5$ & $8$ & $0.05$ & $0.001$ & $2$ & $1$ & n & 0.7 & $[0.1/m_p, 1]$ & $P^{-0.55}$ & [0.1, 100] & $2.3$ \\
35 & \modelFDMsKtwo & $5.5$ & $8$ & $0.05$ & $0.001$ & $2$ & $0.1$ & n & 0.7 & $[0.1/m_p, 1]$ & $P^{-0.55}$ & [0.1, 100] & $2.3$ \\
36 & \modelFDMsKthree & $5.5$ & $8$ & $0.05$ & $0.001$ & $2$ & $0.01$ & n & 0.7 & $[0.1/m_p, 1]$ & $P^{-0.55}$ & [0.1, 100] & $2.3$ \\
\hline
37 & \modelFDqKone & $5.6$ & $8$ & $0.05$ & $0.001$ & $2$ & $1$ & n & 0.7 & $[0.6, 1]$ & $P^{-0.55}$ & [0.1, 100] & $2.3$ \\
38 & \modelFDqKtwo & $5.6$ & $8$ & $0.05$ & $0.001$ & $2$ & $0.1$ & n & 0.7 & $[0.6, 1]$ & $P^{-0.55}$ & [0.1, 100] & $2.3$ \\
39 & \modelFDqKoneRone & $5.6$ & $1$ & $0.05$ & $0.001$ & $2$ & $0.01$ & n & 0.7 & $[0.6, 1]$ & $P^{-0.55}$ & [0.1, 100] & $2.3$ \\
40 & \modelFDqKoneRtwo & $5.6$ & $3$ & $0.05$ & $0.001$ & $2$ & $0.01$ & n & 0.7 & $[0.6, 1]$ & $P^{-0.55}$ & [0.1, 100] & $2.3$ \\
41 & \modelFDqKtwoRone & $5.6$ & $1$ & $0.05$ & $0.001$ & $2$ & $0.01$ & n & 0.7 & $[0.6, 1]$ & $P^{-0.55}$ & [0.1, 100] & $2.3$ \\
42 & \modelFDqKtwoRtwo & $5.6$ & $3$ & $0.05$ & $0.001$ & $2$ & $0.01$ & n & 0.7 & $[0.6, 1]$ & $P^{-0.55}$ & [0.1, 100] & $2.3$ \\
\hline
43 & \modelIsteep & $3.6$ & $8$ & $0.05$ & $0.001$ & $2$ & $1$ & n & $0.05$ & $[0.1/m_p, 1]$ & $P^0$ & [0.1, 100] & $3.0$ \\
44 & \modelIflat & $16.1$ & $8$ & $0.05$ & $0.001$ & $2$ & $1$ & n & $0.05$ & $[0.1/m_p, 1]$ & $P^0$ & [0.1, 100] & $1.6$ \\
45 & \modelIsteepKone & $3.6$ & $8$ & $0.05$ & $0.001$ & $2$ & $1$ & n & $0.05$ & $[0.1/m_p, 1]$ & $P^0$ & [0.1, 100] & $3.0$ \\
46 & \modelIsteepKtwo & $3.6$ & $8$ & $0.05$ & $0.001$ & $2$ & $0.1$ & n & $0.05$ & $[0.1/m_p, 1]$ & $P^0$ & [0.1, 100] & $3.0$ \\
47 & \modelIsteepKthree & $3.6$ & $8$ & $0.05$ & $0.001$ & $2$ & $0.01$ & n & $0.05$ & $[0.1/m_p, 1]$ & $P^0$ & [0.1, 100] & $3.0$ \\
48 & \modelIflatKone & $16.1$ & $8$ & $0.05$ & $0.001$ & $2$ & $1$ & n & $0.05$ & $[0.1/m_p, 1]$ & $P^0$ & [0.1, 100] & $1.6$ \\
49 & \modelIflatKtwo & $16.1$ & $8$ & $0.05$ & $0.001$ & $2$ & $0.1$ & n & $0.05$ & $[0.1/m_p, 1]$ & $P^0$ & [0.1, 100] & $1.6$ \\
50 & \modelIflatKthree & $16.1$ & $8$ & $0.05$ & $0.001$ & $2$ & $0.01$ & n & $0.05$ & $[0.1/m_p, 1]$ & $P^0$ & [0.1, 100] & $1.6$ \\
\hline
51 & \modelW & $5.2$ & $8$ & $0.05$ & $0.001$ & $2$ & $1$ & y & $0.05$ & $[0.1/m_p, 1]$ & $P^0$ & [0.1, 100] & $2.3$ \\
52 & \modelWKone & $5.2$ & $8$ & $0.05$ & $0.001$ & $2$ & $1$ & n & $0.05$ & $[0.1/m_p, 1]$ & $P^0$ & [0.1, 100] & $2.3$ \\
53 & \modelWKtwo & $5.2$ & $8$ & $0.05$ & $0.001$ & $2$ & $0.1$ & n & $0.05$ & $[0.1/m_p, 1]$ & $P^0$ & [0.1, 100] & $2.3$ \\
54 & \modelWKthree & $5.2$ & $8$ & $0.05$ & $0.001$ & $2$ & $0.01$ & n & $0.05$ & $[0.1/m_p, 1]$ & $P^0$ & [0.1, 100] & $2.3$ \\
\hline
55 & \modelWFDq & $5.6$ & $8$ & $0.05$ & $0.001$ & $2$ & $1$ & y & $0.7$ & $[0.6, 1]$ & $P^{-0.55}$ & [0.1, 100] & $2.3$ \\
56 & \modelWFDqKone & $5.6$ & $8$ & $0.05$ & $0.001$ & $2$ & $1$ & n & $0.7$ & $[0.6, 1]$ & $P^{-0.55}$ & [0.1, 100] & $2.3$ \\
\hline
57 & \modelWIflat & $16.1$ & $8$ & $0.05$ & $0.001$ & $2$ & $1$ & y & $0.05$ & $[0.1/m_p, 1]$ & $P^0$ & [0.1, 100] & $1.6$ \\
58 & \modelWIflatKone & $16.1$ & $8$ & $0.05$ & $0.001$ & $2$ & $1$ & n & $0.05$ & $[0.1/m_p, 1]$ & $P^0$ & [0.1, 100] & $1.6$ \\
59 & \modelWIflatKthree & $16.1$ & $8$ & $0.05$ & $0.001$ & $2$ & $0.01$ & n & $0.05$ & $[0.1/m_p, 1]$ & $P^0$ & [0.1, 100] & $1.6$ \\
\hline
60 & \modelWifb & $5.0$ & $8$ & $0.1$ & $0.001$ & $1$ & $1$ & y & $0.1$ & $[0.1/m_p, 1]$ & $P^0$ & [0.08, 150] & $2.3$ \\
61 & \modelWifbKone & $5.0$ & $8$ & $0.1$ & $0.001$ & $1$ & $1$ & n & $0.1$ & $[0.1/m_p, 1]$ & $P^0$ & [0.08, 150] & $2.3$ \\
62 & \modelWifbKthree & $5.0$ & $8$ & $0.1$ & $0.001$ & $1$ & $0.01$ & n & $0.1$ & $[0.1/m_p, 1]$ & $P^0$ & [0.08, 150] & $2.3$ \\
\enddata
\tablecomments{Each model initially has $N=8\times10^5$ stars. Column {\tt FB} denotes 
whether or not natal kicks for BHs are dependent on fallback. $\alpha_1$ denotes the magnitude of 
the power-law exponent for stars of initial mass $>1\,\msun$. $q$-distribution is always uniform.  
}
\label{T:initial}
\end{deluxetable*}

\clearpage

\begin{deluxetable*}{l|ccccccccccccccccccccc}
\tabletypesize{\footnotesize}
\tablecolumns{17}
\tablewidth{0pt}
\tablecaption{Final structural properties of model clusters.}
\tablehead{
	  \colhead{No.} &
	  \colhead{t} &
           \colhead{$N$} &
           \colhead{$M$} &
           \colhead{$f_b$} &
           \colhead{$f_{b,c}$} &
           \colhead{$r_c$} & 
           \colhead{$\rcobs$} &
           \colhead{$r_h$} & 
           \colhead{$\rhlobs$} &
           \colhead{$\rho_c$} & 
           \colhead{$\sigmacobs$} &
           \colhead{$v_{\sigma,c}$} & 
           \colhead{$v_{\sigma, c, \rm{obs}}$} \\
           \cline{7-10}
           \cline{13-14}
	  \colhead{} &
           \colhead{(Gyr)} &
           \colhead{($10^4$)} &
           \colhead{($10^4\msun$)} &
           \colhead{} &
           \colhead{} & 
           \multicolumn{4}{c}{($\pc$)} &
           \colhead{($10^3\msun/\pc^{3}$)} &
           \colhead{($10^3\lsun/\pc^{2}$)} & 
           \multicolumn{2}{c}{($\kms$)} \\
}
\startdata
1 & 12.0 & 71 & 26 & 0.05 & 0.05 & 2.90 & 3.06 & 7.89 & 5.55 & 17.8 & 1.9 & 9.8 & 3.40 \\ 
2 & $\approx5$ & \multicolumn{13}{c}{Dissolved} \\
3 & 12.0 & 24 & 12 & 0.06 & 0.13 & 0.64 & 0.40 & 2.80 & 1.41 & 44.0 & 31.5 & 10.9 & 4.26 \\ 
4 & 12.0 & 55 & 21 & 0.05 & 0.06 & 2.52 & 2.86 & 6.12 & 4.01 & 9.1 & 2.2 & 9.8 & 3.32 \\ 
5 & 12.0 & 68 & 25 & 0.05 & 0.06 & 2.83 & 2.59 & 7.16 & 4.73 & 12.3 & 2.0 & 9.9 & 3.46 \\ 
6 & 12.0 & 73 & 27 & 0.05 & 0.05 & 3.07 & 4.22 & 8.54 & 6.24 & 21.8 & 1.5 & 9.7 & 3.30 \\ 
%
7 & 12.0 & 76 & 26 & 0.09 & 0.12 & 1.89 & 1.12 & 5.65 & 3.27 & 7.5 & 9.2 & 11.5 & 4.37 \\ 
8 & 12.0 & 71 & 25 & 0.09 & 0.14 & 0.83 & 1.12 & 4.29 & 3.27 & 28.7 & 9.2 & 13.6 & 5.30 \\ 
%
\hline
9 & 12.0 & 71 & 26 & 0.05 & 0.05 & 3.20 & 2.63 & 7.82 & 5.50 & 3.2 & 2.2 & 9.9 & 3.46 \\ 
10 & 12.0 & 73 & 27 & 0.04 & 0.05 & 2.51 & 3.34 & 7.71 & 5.39 & 121.1 & 1.9 & 9.8 & 3.46 \\ 
11 & 12.0 & 73 & 27 & 0.04 & 0.05 & 3.15 & 2.81 & 7.77 & 5.44 & 5.6 & 2.4 & 10.0 & 3.47 \\ 
12 & 12.0 & 73 & 27 & 0.04 & 0.05 & 3.05 & 5.07 & 8.12 & 5.73 & 20.6 & 1.4 & 9.7 & 3.24 \\ 
13 & $\approx4$ & \multicolumn{13}{c}{Dissolved} \\ 
14 & 12.0 & 63 & 24 & 0.04 & 0.05 & 3.05 & 2.24 & 7.54 & 5.25 & 14.0 & 2.4 & 9.2 & 3.23 \\ 
%
\hline
15 & 12.0 & 77 & 28 & 0.04 & 0.11 & 0.36 & 0.22 & 4.98 & 2.47 & 275.1 & 126.6 & 14.2 & 6.08 \\ 
16 & 12.0 & 76 & 27 & 0.05 & 0.07 & 2.00 & 1.42 & 5.71 & 3.48 & 8.3 & 7.6 & 11.7 & 4.38 \\ 
17 & 12.0 & 70 & 26 & 0.05 & 0.05 & 3.09 & 3.73 & 8.20 & 5.90 & 9.1 & 1.3 & 9.7 & 3.30 \\ 
%
18 & 12.0 & 69 & 23 & 0.08 & 0.13 & 0.32 & 0.22 & 4.56 & 2.17 & 371.1 & 121.3 & 14.9 & 6.41 \\ 
19 & 12.0 & 69 & 24 & 0.09 & 0.12 & 1.41 & 1.12 & 5.00 & 2.90 & 68.7 & 122.9 & 12.1 & 4.64 \\ 
%
20 & 12.0 & 30 & 14 & 0.05 & 0.14 & 0.20 & 0.17 & 3.05 & 1.30 & 678.2 & 126.0 & 12.0 & 4.92 \\ 
21 & 12.0 & 66 & 25 & 0.04 & 0.12 & 0.37 & 0.05 & 4.54 & 2.02 & 226.6 & 2121.9 & 13.5 & 5.57 \\ 
22 & 12.0 & 74 & 27 & 0.04 & 0.11 & 0.55 & 0.25 & 4.93 & 2.26 & 96.1 & 88.4 & 13.5 & 5.58 \\ 
23 & 12.0 & 77 & 28 & 0.04 & 0.10 & 0.32 & 0.26 & 5.07 & 2.67 & 384.3 & 92.7 & 14.7 & 6.42 \\ 
24 & 12.0 & 24 & 12 & 0.06 & 0.13 & 0.65 & 0.48 & 2.78 & 1.36 & 36.8 & 35.0 & 11.0 & 4.26 \\ 
25 & 12.0 & 63 & 24 & 0.05 & 0.07 & 1.70 & 1.28 & 4.91 & 2.75 & 4.4 & 8.2 & 11.7 & 4.35 \\ 
26 & 12.0 & 73 & 26 & 0.05 & 0.07 & 2.06 & 1.74 & 5.67 & 3.31 & 3.9 & 5.2 & 11.5 & 4.19 \\ 
27 & 12.0 & 12 & 6 & 0.07 & 0.12 & 0.77 & 1.11 & 2.82 & 1.87 & 79.3 & 5.3 & 7.7 & 2.79 \\ 
28 & 12.0 & 53 & 21 & 0.05 & 0.06 & 2.45 & 2.06 & 6.27 & 4.19 & 17.0 & 2.9 & 9.5 & 3.32 \\ 
29 & 12.0 & 67 & 24 & 0.05 & 0.06 & 2.80 & 3.99 & 7.36 & 5.06 & 15.5 & 1.5 & 9.7 & 3.30 \\ 
30 & 12.0 & 73 & 27 & 0.05 & 0.05 & 3.30 & 2.28 & 8.63 & 6.26 & 5.0 & 2.4 & 9.8 & 3.45 \\ 
31 & 12.0 & 28 & 13 & 0.05 & 0.12 & 0.18 & 0.14 & 2.77 & 1.52 & 1124.2 & 205.1 & 13.7 & 5.78 \\ 
32 & 12.0 & 20 & 10 & 0.06 & 0.09 & 1.02 & 1.44 & 2.89 & 1.92 & 13.9 & 8.1 & 9.7 & 3.39 \\ 
33 & $\approx4$ & \multicolumn{13}{c}{Dissolved} \\ 
%
\hline
34 & 12.0 & 77 & 28 & 0.04 & 0.09 & 0.59 & 0.31 & 5.08 & 2.63 & 68.2 & 66.2 & 13.4 & 5.61 \\ 
35 & 12.0 & 76 & 28 & 0.04 & 0.06 & 2.22 & 1.39 & 6.00 & 3.74 & 3.2 & 7.9 & 11.4 & 4.25 \\ 
36 & 12.0 & 73 & 27 & 0.04 & 0.05 & 3.09 & 2.26 & 7.85 & 5.45 & 7.3 & 2.9 & 10.0 & 3.53 \\ 
%
\hline
37 & 12.0 & 78 & 28 & 0.04 & 0.09 & 0.71 & 0.36 & 5.18 & 2.66 & 42.6 & 58.2 & 13.1 & 5.42 \\ 
38 & 12.0 & 76 & 28 & 0.04 & 0.06 & 2.45 & 1.89 & 6.47 & 4.20 & 6.7 & 4.4 & 11.0 & 4.00 \\ 
39 & 12.0 & 26 & 12 & 0.05 & 0.13 & 0.20 & 0.31 & 2.73 & 1.53 & 616.9 & 55.7 & 12.4 & 5.01 \\ 
40 & 12.0 & 71 & 26 & 0.04 & 0.09 & 0.68 & 0.59 & 4.87 & 2.55 & 58.7 & 26.9 & 13.1 & 5.31 \\ 
41 & 12.0 & 7 & 4 & 0.07 & 0.09 & 0.92 & 0.73 & 2.48 & 1.78 & 10.8 & 6.4 & 7.0 & 2.49 \\ 
42 & 12.0 & 69 & 26 & 0.04 & 0.06 & 2.23 & 1.20 & 5.85 & 3.76 & 8.3 & 8.4 & 11.0 & 4.05 \\ 
%
\hline
43 & 12.0 & 71 & 25 & 0.04 & 0.09 & 0.29 & 0.13 & 4.28 & 1.68 & 384.7 & 335.5 & 15.4 & 6.67 \\ 
44 & $\approx2$ & \multicolumn{13}{c}{Dissolved} \\ 
45 & 12.0 & 74 & 25 & 0.04 & 0.07 & 0.21 & 0.18 & 4.63 & 2.06 & 966.6 & 175.1 & 13.5 & 5.85 \\ 
46 & 12.0 & 72 & 25 & 0.04 & 0.08 & 0.20 & 0.14 & 4.46 & 1.97 & 1107.9 & 260.7 & 13.8 & 5.93 \\ 
47 & 12.0 & 71 & 25 & 0.04 & 0.10 & 0.36 & 0.10 & 4.14 & 1.61 & 286.1 & 646.3 & 15.7 & 6.83 \\ 
48 & $<0.1$ & \multicolumn{13}{c}{Dissolved} \\
49 & $\approx3$ & \multicolumn{13}{c}{Dissolved} \\
50 & $\approx2$ & \multicolumn{13}{c}{Dissolved} \\
%
\hline
51 & 12.0 & 69 & 25 & 0.05 & 0.05 & 3.58 & 4.72 & 8.57 & 6.17 & 3.7 & 1.1 & 9.3 & 3.10 \\ 
52 & 11.0 & 77 & 28 & 0.04 & 0.10 & 0.56 & 0.26 & 4.73 & 2.31 & 89.7 & 112.1 & 14.5 & 6.06 \\ 
53 & 12.0 & 74 & 27 & 0.05 & 0.06 & 2.50 & 1.57 & 6.28 & 4.07 & 2.6 & 5.8 & 11.0 & 4.05 \\ 
54 & 12.0 & 63 & 24 & 0.05 & 0.05 & 3.96 & 7.76 & 10.55 & 7.95 & 10.4 & 0.5 & 8.2 & 2.58 \\ 
%
\hline
55 & 12.0 & 69 & 26 & 0.04 & 0.05 & 3.64 & 2.20 & 8.38 & 5.90 & 2.4 & 2.4 & 9.4 & 3.31 \\ 
56 & 12.0 & 77 & 28 & 0.04 & 0.09 & 0.76 & 0.56 & 5.09 & 2.61 & 34.4 & 35.1 & 13.3 & 5.49 \\ 
%
\hline
%
57 & $\approx0.6$ & \multicolumn{13}{c}{Dissolved} \\
58 & 12.0 & 60 & 24 & 0.04 & 0.04 & 13.95 & 35.76 & 21.29 & 16.29 & 0.0 & 0.1 & 4.8 & 1.34 \\ 
59 & $\approx0.4$ & \multicolumn{13}{c}{Dissolved} \\ 
%
\hline
60 & 12.0 & 67 & 23 & 0.09 & 0.12 & 1.86 & 1.16 & 5.33 & 3.15 & 5.7 & 10.7 & 11.4 & 4.36 \\ 
61 & 12.0 & 64 & 22 & 0.09 & 0.13 & 0.36 & 0.14 & 3.56 & 1.61 & 647.6 & 9.2 & 17.0 & 7.27 \\ 
62 & 12.0 & 65 & 23 & 0.09 & 0.11 & 2.42 & 1.46 & 6.50 & 4.24 & 5.7 & 253.7 & 10.3 & 3.79 \\ 
%
\enddata
\tablecomments{Serial numbers for models are the same as in Table\ 2. Observed structural properties 
are denoted by the subscript ``obs." Definitions for 
observed properties are explained in \S\ref{S:obs_derivation}. The approximate dissolution times 
(\S\ref{S:tdiss}) are listed for dissolved clusters. }
\label{T:clusprop}
\end{deluxetable*}

\clearpage

%
\begin{deluxetable*}{l|cccccc|cc|cc|cc}
\tabletypesize{\footnotesize}
\tablecolumns{13}
\tablewidth{0pt}
\tablecaption{Properties of Black Holes.}
\tablehead{
	  \colhead{No.} &
	  \colhead{$\nbh$} &
	  \colhead{$\nbh_{,\rm{esc}}$} &
           \colhead{$\nbhbh$} &
           \colhead{$\nbhbh_{,\rm{esc}}$} &
           \colhead{$\nbhnbh$} &
           \colhead{$\nbhnbh_{,\rm{esc}}$} &
           \multicolumn{4}{c}{$N_{\rm{merge}}$} &
           \multicolumn{2}{c}{$M_{\rm{tot}}$} \\
           \colhead{} &
           \colhead{} &
           \colhead{} &
           \colhead{} &
           \colhead{} &
           \colhead{} &
           \colhead{} &
           \multicolumn{2}{c}{$z<1$} &
           \multicolumn{2}{c}{$z<2$} &
           \colhead{$z<1$} &
           \colhead{$z<2$} \\
           \cline{8-11}
           \cline{12-13}
           \colhead{} &
           \colhead{} &
           \colhead{} &
           \colhead{} &
           \colhead{} &
           \colhead{} &
           \colhead{} &
           \colhead{in} &
           \colhead{out} &
           \colhead{in} & 
           \colhead{out} &
           \multicolumn{2}{c}{($M_\odot$)} \\
}
\startdata
1 & 464 & 1260 & 4 & 130 & 2 & 5 & 0 & 32 & 0 & 63 & $33.2_{23.8}^{47.6}$ & $44.2_{24.8}^{49.3}$ \\
2 & 601 & 1001 & 2 & 97 & 1 & 2 & 0 & 17 & 0 & 46 & $34.7_{27.5}^{46.7}$ & $44.4_{28.5}^{49.1}$ \\
3 & 165 & 1259 & 4 & 65 & 10 & 5 & 0 & 37 & 0 & 59 & $19.8_{8.8}^{21.0}$ & $19.9_{10.4}^{21.5}$ \\
4 & 327 & 1213 & 4 & 132 & 2 & 1 & 0 & 42 & 0 & 83 & $26.7_{21.8}^{30.5}$ & $28.7_{22.4}^{31.9}$ \\
5 & 393 & 1227 & 3 & 131 & 1 & 3 & 0 & 40 & 0 & 77 & $30.7_{24.0}^{37.4}$ & $31.5_{24.0}^{37.4}$ \\
6 & 579 & 1200 & 4 & 146 & 1 & 6 & 0 & 45 & 0 & 80 & $32.0_{23.6}^{44.9}$ & $36.7_{24.0}^{50.5}$ \\
7 & 139 & 1318 & 4 & 65 & 4 & 0 & 0 & 22 & 0 & 46 & $29.4_{20.0}^{37.2}$ & $32.2_{21.5}^{38.6}$ \\
8 & 49 & 1458 & 6 & 103 & 4 & 11 & 0 & 32 & 0 & 55 & $21.6_{15.7}^{34.7}$ & $24.9_{15.7}^{36.1}$ \\
%
\hline
9 & 458 & 1265 & 5 & 130 & 4 & 2 & 0 & 35 & 0 & 66 & $32.9_{26.7}^{46.9}$ & $40.8_{27.2}^{49.7}$ \\
10 & 467 & 1286 & 3 & 156 & 1 & 37 & 0 & 39 & 0 & 68 & $35.4_{23.5}^{48.3}$ & $42.3_{21.2}^{50.5}$ \\
11 & 437 & 1298 & 4 & 153 & 1 & 29 & 0 & 32 & 0 & 64 & $35.8_{25.9}^{47.8}$ & $41.8_{26.3}^{50.7}$ \\
12 & 503 & 1367 & 2 & 159 & 2 & 23 & 0 & 42 & 0 & 66 & $35.7_{26.9}^{50.8}$ & $37.7_{23.9}^{51.9}$ \\
13 & 686 & 1083 & 3 & 116 & 3 & 30 & 0 & 21 & 0 & 56 & $38.2_{25.2}^{47.1}$ & $41.0_{23.0}^{50.8}$ \\
14 & 462 & 1346 & 3 & 151 & 4 & 32 & 0 & 44 & 0 & 75 & $32.1_{24.8}^{50.9}$ & $39.5_{24.3}^{52.0}$ \\
%
\hline
15 & 6 & 1757 & 1 & 1 & 2 & 3 & 0 & 0 & 0 & 0 & $0.0_{0.0}^{0.0}$ & $0.0_{0.0}^{0.0}$ \\
16 & 241 & 1465 & 3 & 89 & 2 & 0 & 0 & 25 & 0 & 49 & $28.1_{20.1}^{45.6}$ & $35.4_{20.4}^{49.7}$ \\
17 & 759 & 967 & 3 & 138 & 5 & 2 & 0 & 50 & 0 & 71 & $34.9_{24.6}^{49.5}$ & $40.5_{24.7}^{50.2}$ \\
18 & 1 & 1521 & 0 & 0 & 1 & 1 & 1 & 0 & 1 & 0 & $26.9_{26.9}^{26.9}$ & $26.9_{26.9}^{26.9}$ \\
19 & 282 & 1218 & 4 & 172 & 6 & 11 & 0 & 52 & 0 & 102 & $19.6_{13.4}^{29.6}$ & $22.8_{14.1}^{35.1}$ \\
20 & 0 & 1512 & 0 & 2 & 0 & 3 & 0 & 0 & 0 & 0 & $0.0_{0.0}^{0.0}$ & $0.0_{0.0}^{0.0}$ \\
21 & 3 & 1637 & 1 & 0 & 0 & 2 & 1 & 0 & 1 & 0 & $11.6_{11.6}^{11.6}$ & $11.6_{11.6}^{11.6}$ \\
22 & 3 & 1689 & 1 & 2 & 0 & 7 & 0 & 0 & 0 & 0 & $0.0_{0.0}^{0.0}$ & $0.0_{0.0}^{0.0}$ \\
23 & 3 & 1797 & 0 & 2 & 1 & 7 & 1 & 0 & 1 & 0 & $37.5_{37.5}^{37.5}$ & $37.5_{37.5}^{37.5}$ \\
24 & 132 & 1219 & 2 & 69 & 9 & 6 & 0 & 41 & 0 & 62 & $16.1_{10.7}^{21.0}$ & $17.6_{11.3}^{21.3}$ \\
25 & 179 & 1345 & 2 & 80 & 3 & 2 & 0 & 25 & 0 & 54 & $25.6_{19.0}^{30.7}$ & $27.0_{20.0}^{31.5}$ \\
26 & 239 & 1358 & 5 & 75 & 1 & 3 & 0 & 27 & 0 & 49 & $27.5_{19.3}^{37.4}$ & $31.4_{19.2}^{37.5}$ \\
27 & 398 & 980 & 4 & 141 & 3 & 4 & 0 & 75 & 0 & 118 & $16.0_{12.0}^{20.0}$ & $17.1_{12.1}^{20.3}$ \\
28 & 561 & 972 & 4 & 141 & 3 & 1 & 0 & 53 & 0 & 89 & $26.7_{20.0}^{31.4}$ & $27.4_{21.8}^{31.5}$ \\
29 & 661 & 944 & 2 & 130 & 3 & 1 & 0 & 48 & 0 & 79 & $29.0_{21.9}^{36.9}$ & $31.2_{22.7}^{37.4}$ \\
30 & 741 & 1046 & 7 & 145 & 1 & 5 & 0 & 43 & 0 & 76 & $30.3_{23.7}^{48.0}$ & $39.8_{24.0}^{49.5}$ \\
31 & 1 & 1752 & 0 & 2 & 1 & 6 & 0 & 0 & 0 & 0 & $0.0_{0.0}^{0.0}$ & $0.0_{0.0}^{0.0}$ \\
32 & 125 & 1418 & 3 & 81 & 1 & 4 & 0 & 36 & 0 & 51 & $27.9_{19.3}^{48.0}$ & $30.0_{19.7}^{48.6}$ \\
33 & 881 & 671 & 3 & 94 & 3 & 1 & 0 & 20 & 0 & 51 & $36.2_{27.1}^{48.0}$ & $43.6_{27.7}^{50.9}$ \\
%
\hline
34 & 18 & 1732 & 2 & 20 & 5 & 41 & 0 & 3 & 0 & 3 & $23.3_{16.9}^{35.9}$ & $23.3_{16.9}^{35.9}$ \\
35 & 279 & 1432 & 3 & 158 & 6 & 17 & 0 & 17 & 0 & 37 & $30.0_{23.5}^{44.1}$ & $38.1_{23.6}^{51.1}$ \\
36 & 673 & 1020 & 3 & 190 & 3 & 13 & 0 & 46 & 1 & 74 & $36.3_{23.5}^{50.9}$ & $40.5_{23.7}^{51.1}$ \\
%
\hline
37 & 31 & 1794 & 2 & 25 & 4 & 52 & 0 & 5 & 0 & 5 & $22.9_{16.3}^{26.3}$ & $22.9_{16.3}^{26.3}$ \\
38 & 338 & 1498 & 3 & 215 & 3 & 18 & 0 & 32 & 1 & 55 & $29.7_{21.2}^{41.3}$ & $32.4_{18.4}^{50.8}$ \\
39 & 13 & 1723 & 0 & 30 & 5 & 57 & 0 & 7 & 0 & 7 & $19.6_{13.2}^{21.9}$ & $19.6_{13.2}^{21.9}$ \\
40 & 31 & 1795 & 3 & 28 & 5 & 61 & 0 & 3 & 0 & 3 & $23.7_{22.6}^{25.9}$ & $23.7_{22.6}^{25.9}$ \\
41 & 176 & 1422 & 2 & 214 & 0 & 17 & 0 & 26 & 2 & 49 & $26.9_{9.5}^{43.2}$ & $28.3_{11.0}^{46.2}$ \\
42 & 315 & 1449 & 5 & 208 & 6 & 14 & 0 & 34 & 0 & 65 & $25.4_{19.3}^{46.9}$ & $29.0_{19.0}^{49.6}$ \\
%
\hline
43 & 1 & 165 & 0 & 13 & 1 & 2 & 0 & 2 & 0 & 4 & $25.9_{23.5}^{28.3}$ & $33.1_{23.7}^{46.7}$ \\
44 & 1430 & 5377 & 2 & 185 & 5 & 25 & 0 & 38 & 0 & 105 & $50.0_{21.8}^{53.3}$ & $50.0_{23.7}^{53.4}$ \\
45 & 0 & 169 & 0 & 0 & 0 & 1 & 0 & 0 & 0 & 0 & $0.0_{0.0}^{0.0}$ & $0.0_{0.0}^{0.0}$ \\
46 & 0 & 165 & 0 & 9 & 0 & 2 & 1 & 1 & 1 & 5 & $26.0_{25.0}^{26.9}$ & $39.4_{25.2}^{49.3}$ \\
47 & 4 & 164 & 1 & 24 & 0 & 3 & 0 & 4 & 0 & 12 & $19.1_{15.7}^{24.7}$ & $23.7_{15.8}^{47.9}$ \\
48 & 61 & 19012 & 0 & 30 & 0 & 45 & 0 & 0 & 0 & 0 & $0.0_{0.0}^{0.0}$ & $0.0_{0.0}^{0.0}$ \\
49 & 1452 & 7774 & 5 & 148 & 4 & 10 & 1 & 48 & 1 & 78 & $49.2_{18.2}^{53.8}$ & $49.6_{19.9}^{53.9}$ \\
50 & 1557 & 2502 & 4 & 202 & 4 & 9 & 0 & 38 & 0 & 106 & $49.9_{46.0}^{54.4}$ & $50.0_{43.4}^{54.5}$ \\
%
\hline
51 & 393 & 1341 & 2 & 139 & 3 & 10 & 0 & 36 & 0 & 69 & $44.3_{32.0}^{55.3}$ & $49.3_{32.2}^{55.4}$ \\
52 & 1 & 1764 & 0 & 1 & 1 & 12 & 0 & 0 & 0 & 0 & $0.0_{0.0}^{0.0}$ & $0.0_{0.0}^{0.0}$ \\
53 & 165 & 1563 & 3 & 103 & 1 & 3 & 0 & 29 & 0 & 52 & $43.1_{32.0}^{63.3}$ & $48.6_{33.1}^{58.0}$ \\
54 & 653 & 1068 & 1 & 162 & 2 & 2 & 0 & 31 & 0 & 59 & $49.7_{40.2}^{57.2}$ & $51.0_{40.5}^{58.7}$ \\
%
\hline
55 & 310 & 1761 & 3 & 282 & 3 & 103 & 0 & 24 & 0 & 55 & $48.0_{36.4}^{54.2}$ & $49.7_{27.3}^{74.9}$ \\
56 & 21 & 1916 & 1 & 51 & 5 & 160 & 0 & 1 & 0 & 2 & $60.2_{60.2}^{60.2}$ & $67.6_{60.6}^{74.6}$ \\
%
\hline
57 & 1500 & 8638 & 9 & 248 & 2 & 75 & 0 & 28 & 1 & 94 & $54.6_{19.1}^{56.4}$ & $54.8_{23.0}^{57.6}$ \\
58 & 39 & 19134 & 0 & 43 & 17 & 143 & 0 & 0 & 0 & 0 & $0.0_{0.0}^{0.0}$ & $0.0_{0.0}^{0.0}$ \\
59 & 1917 & 3554 & 11 & 322 & 4 & 13 & 0 & 14 & 0 & 117 & $77.4_{35.1}^{82.0}$ & $71.0_{19.8}^{84.7}$ \\
%
\hline
60 & 54 & 1469 & 1 & 141 & 3 & 7 & 0 & 33 & 0 & 59 & $33.6_{30.3}^{45.7}$ & $36.2_{30.4}^{51.4}$ \\
61 & 1 & 1519 & 0 & 1 & 1 & 25 & 3 & 0 & 3 & 1 & $68.3_{59.7}^{74.6}$ & $71.6_{59.9}^{79.5}$ \\
62 & 289 & 1237 & 3 & 175 & 7 & 10 & 0 & 31 & 0 & 69 & $34.3_{24.3}^{43.0}$ & $36.2_{27.5}^{48.3}$ \\
%
%
\enddata
\tablecomments{Serial numbers for models are the same as in Table\ 2. Columns ``in" and 
``out" denote BH-BH mergers inside and outside clusters. Number of BH-BH mergers and their total masses 
in the source frame are listed for $z<1$ and $z<2$ assuming $z=0\equiv t=12\,\gyr$. The median and $2\sigma$ ranges in total 
mass are listed. }
\label{T:bbh}
\end{deluxetable*}

\clearpage
\clearpage

\end{document}